\documentclass[11pt]{article}

\usepackage[subpreambles=false]{standalone}
\usepackage[dvipsnames]{xcolor, colortbl}
\usepackage[utf8]{inputenc}
\usepackage[english]{babel}
\usepackage{multicol}
\usepackage{times}
\usepackage{graphicx}
\usepackage[margin=1in]{geometry}
\usepackage{comment}

\usepackage{fnpct}
\usepackage{xcolor}
\PassOptionsToPackage{hyphens}{url}
\usepackage{hyperref}
\hypersetup{
    breaklinks=true,
}

\usepackage{booktabs}
\usepackage{changepage}

\usepackage[anythingbreaks]{breakurl} 
\usepackage{eurosym}
\usepackage{xspace}
\usepackage{caption}
\newcommand{\carbon}{\texorpdfstring{CO\textsubscript{2}}{CO2}}
\usepackage{color}

\newcommand{\et}{et al.~} 
\newcommand{\cd}{CO$_2$\xspace}

\newcommand{\Gap}{\texorpdfstring{\hfill}{}}

\definecolor{ccai-yellow}{HTML}{ffea94}
\definecolor{ccai-yellow-darker}{HTML}{dbaf00}
\definecolor{ccai-green}{HTML}{007849}
\definecolor{ccai-blue}{HTML}{0375B4}
\definecolor{ccai-blue-darker}{HTML}{035582}
\definecolor{ccai-blue-lightest}{HTML}{cdedfe}

\newcommand{\Rec}{\textbf{\texorpdfstring{{\small\emph{\color{ccai-blue}{\fbox{High Leverage}}}}}{}}}
\newcommand{\HighRisk}{\textbf{\texorpdfstring{{\small\emph{\color{ccai-yellow-darker}{\fbox{Uncertain Impact}}}}}{}}}
\newcommand{\Longterm}{\textbf{\texorpdfstring{{\small\emph{\color{ccai-green}{\fbox{Long-term}}}}}{}}}

\usepackage[sectionbib,numbers,sort&compress]{natbib}

\title{Tackling Climate Change with Machine Learning}
\author{David Rolnick$^1$\footnote{D.R. conceived and edited this work, with P.L.D., L.H.K., and K.K.  Authors P.L.D., L.H.K., K.K., A.L., K.S., A.S.R., N.M-D., N.J., A.W-B., A.L., T.M., and E.D.S. researched and wrote individual sections. S.K.M., K.P.K., C.G., A.Y.N., D.H., J.C.P., F.C., J.C., and Y.B. contributed expert advice. Correspondence to \url{drolnick@seas.upenn.edu}.}
, Priya L.~Donti$^2$, Lynn H.~Kaack$^3$, Kelly Kochanski$^4$, Alexandre Lacoste$^5$,\\
Kris Sankaran$^{6,7}$,
Andrew Slavin Ross$^9$, Nikola Milojevic-Dupont$^{10,11}$, Natasha Jaques$^{12}$,\\Anna Waldman-Brown$^{12}$, Alexandra Luccioni$^{6,7}$, Tegan Maharaj$^{6,8}$, Evan D.~Sherwin$^2$,\\S.~Karthik Mukkavilli$^{6,7}$, Konrad P.~K\"ording$^1$, Carla Gomes$^{13}$, Andrew Y.~Ng$^{14}$,\\ Demis Hassabis$^{15}$,
John C.~Platt$^{16}$, Felix Creutzig$^{10,11}$, Jennifer Chayes$^{17}$, Yoshua Bengio$^{6,7}$\vspace{0.1in}\\
\small{$^1$University of Pennsylvania, $^2$Carnegie Mellon University, $^3$ETH Z\"urich, $^4$University of Colorado Boulder,}\\\small{ $^5$Element AI, $^6$Mila, $^7$Universit\'e de Montr\'
eal, $^8$\'Ecole Polytechnique de Montr\'eal, $^9$Harvard University,}\\\small{$^{10}$Mercator Research Institute on Global Commons and Climate Change, $^{11}$Technische Universit\"at Berlin,}\\\small{$^{12}$Massachusetts Institute of Technology, $^{13}$Cornell University, $^{14}$Stanford University, }\\\small{$^{15}$DeepMind, $^{16}$Google AI, $^{17}$Microsoft Research}
}
\date{}

\begin{document}
 
    \maketitle

\begin{abstract}
Climate change is one of the greatest challenges facing humanity, and we, as machine learning experts, may wonder how we can help. Here we describe how machine learning can be a powerful tool in reducing greenhouse gas emissions and helping society adapt to a changing climate. From smart grids to disaster management, we identify high impact problems where existing gaps can be filled by machine learning, in collaboration with other fields. Our recommendations encompass exciting research questions as well as promising business opportunities. We call on the machine learning community to join the global effort against climate change.
\vskip .5in
\end{abstract}

\part*{Introduction}
The effects of climate change are increasingly visible.\footnote{For a layman's introduction to the topic of climate change, see \cite{romm2018climate, archer2010climate}.} Storms, droughts, fires, and flooding have become stronger and more frequent \cite{field2012managing}. Global ecosystems are changing, including the natural resources and agriculture on which humanity depends. The 2018 intergovernmental report on climate change estimated that the world will face catastrophic consequences unless global greenhouse gas emissions are eliminated within thirty years \cite{ipcc_global_2018}. Yet year after year, these emissions rise.

Addressing climate change involves mitigation (reducing emissions) and adaptation (preparing for unavoidable consequences). Both are multifaceted issues. Mitigation of greenhouse gas (GHG) emissions requires changes to electricity systems, transportation, buildings, industry, and land use. Adaptation requires planning for resilience and disaster management, given an understanding of climate and extreme events. Such a diversity of problems can be seen as an opportunity: there are many ways to have an impact.

In recent years, machine learning (ML) has been recognized as a broadly powerful tool for technological progress. Despite the growth of movements applying ML and AI to problems of societal and global good,\footnote{See the AI for social good movement (e.g.~\cite{hager2019artificial, berendt2019ai}), ML for the developing world~\cite{de2018machine}, the computational sustainability movement (e.g.~\cite{kelling2018computational, joppa2017case, lassig2016computational, gomes2009computational, dietterich2009machine}, the American Meteorological Society's Committee on AI Applications to Environmental Science, and the field of Climate Informatics (\url{www.climateinformatics.org}) \cite{Monteleoni2013chapter}, as well as the relevant survey papers \cite{faghmous2014big, kaack2019challenges, ford2016opinion}.} there remains the need for a concerted effort to identify how these tools may best be applied to tackle climate change. Many ML practitioners wish to act, but are uncertain how. On the other side, many fields have begun actively seeking input from the ML community.

This paper aims to provide an overview of where machine learning can be applied with high impact in the fight against climate change, through either effective engineering or innovative research. The strategies we highlight include climate mitigation and adaptation, as well as meta-level tools that enable other strategies. In order to maximize the relevance of our recommendations, we have consulted experts across many fields (see \hyperref[sec:acknowledgments]{{\small{Acknowledgments}}}) in the preparation of this paper.

\begin{table}
\begin{small}
\begin{center}
\begin{tabular}{l l l l l l l l l l l l}  \toprule
     \multicolumn{2}{l}{ }
         & \small{\rotatebox{90}{\parbox{2.2cm}{Causal\\inference}}}
         & \small{\rotatebox{90}{\parbox{2.2cm}{Computer\\vision}}}
         & \small{\rotatebox{90}{\parbox{2.2cm}{Interpretable\\models}}}
         & \small{\rotatebox{90}{NLP}}
         & \small{\rotatebox{90}{\parbox{2.2cm}{RL \& Control}}}
         & \small{\rotatebox{90}{\parbox{2.2cm}{Time-series analysis}}}
         & \small{\rotatebox{90}{\parbox{2.2cm}{Transfer\\learning}}}
         & \small{\rotatebox{90}{\parbox{2.2cm}{Uncertainty\\quantification}}}
         & \small{\rotatebox{90}{\parbox{2.2cm}{Unsupervised\\learning}}}
    \\ \midrule
    \rowcolor{ccai-blue-lightest}
    \multicolumn{2}{l}{1 \hyperref[sec:electricity-systems]{Electricity systems}} 
        & 
        &  
        & 
        & 
        & 
        & 
        & 
        & 
        & \\
    & \hyperref[sec:electricity-lowCarbon]{Enabling low-carbon electricity}
        & 
        & $\bullet$
        & $\bullet$
        & 
        & $\bullet$
        & $\bullet$
        & 
        & $\bullet$
        & $\bullet$\\
    & \hyperref[sec:electricity-currentSystemImpact]{Reducing current-system impacts}
        & 
        & $\bullet$
        & 
        & 
        & 
        & $\bullet$
        & 
        & $\bullet$
        & $\bullet$\\
    & \hyperref[sec:electricity-developing]{Ensuring global impact}
        & 
        & $\bullet$
        & 
        & 
        & 
        & 
        & $\bullet$ 
        & 
        & $\bullet$\\
    \rowcolor{ccai-blue-lightest}
    \multicolumn{2}{l}{2 \hyperref[sec:transportation]{Transportation}} 
        & 
        & 
        &
        & 
        & 
        & 
        & 
        & 
        & \\
    & \hyperref[sec:TReducing]{Reducing transport activity}
        & 
        & $\bullet$
        & 
        & 
        & 
        & $\bullet$
        & 
        & $\bullet$
        & $\bullet$\\
   & \hyperref[sec:TEfficient]{Improving vehicle efficiency}
        & 
        & $\bullet$
        & 
        & 
        & $\bullet$
        & 
        & 
        & 
        & \\
   & \hyperref[sec:TFuels]{Alternative fuels \& electrification}
        & 
        & 
        & 
        & 
        & $\bullet$
        & 
        & 
        & 
        & $\bullet$ \\
   & \hyperref[sec:modalshift]{Modal shift}
        & $\bullet$
        & $\bullet$
        & 
        & 
        & 
        & $\bullet$
        & 
        & $\bullet$
        & \\
    \rowcolor{ccai-blue-lightest}
    \multicolumn{2}{l}{3 \hyperref[sec:buildings-cities]{Buildings and cities}} 
        & 
        & 
        & 
        & 
        & 
        & 
        & 
        & 
        & \\
    & \hyperref[sec:indv]{Optimizing buildings}
        & $\bullet$
        & 
        & 
        & 
        & $\bullet$
        & $\bullet$
        & $\bullet$
        & 
        & \\
    & \hyperref[sec:distr]{Urban planning}
        & 
        & $\bullet$
        & 
        & 
        & 
        & $\bullet$
        & $\bullet$
        & 
        & $\bullet$\\
    & \hyperref[sec:cities]{The future of cities}
        & 
        & 
        & 
        & $\bullet$
        & 
        & 
        & $\bullet$
        & $\bullet$
        & $\bullet$\\
    \rowcolor{ccai-blue-lightest}
    \multicolumn{2}{l}{4 \hyperref[sec:industry]{Industry}} 
        & 
        & 
        & 
        & 
        & 
        & 
        & 
        & 
        & \\
    & \hyperref[sec:supplychains]{Optimizing supply chains}
        & 
        & $\bullet$ 
        & 
        & 
        & $\bullet$ 
        & $\bullet$ 
        & 
        & 
        & \\
    & \hyperref[sec:materialsandconstruction]{Improving materials}
        & 
        & 
        & 
        & 
        & 
        & 
        & 
        & 
        & $\bullet$ \\
    & \hyperref[sec:demandresponse]{Production \& energy}
        & 
        & $\bullet$
        & $\bullet$ 
        & 
        & $\bullet$
        & 
        & 
        & 
        & \\
    \rowcolor{ccai-blue-lightest}
    \multicolumn{2}{l}{5 \hyperref[sec:afolu]{Farms \& forests}} 
        & 
        & 
        & 
        & 
        & 
        & 
        & 
        & 
        & \\
    & \hyperref[sec:emissions-detection]{Remote sensing of emissions}
        & 
        & $\bullet$
        & 
        & 
        & 
        & 
        & 
        & 
        & \\
    & \hyperref[sec:agriculture]{Precision agriculture}
        & 
        & $\bullet$
        & 
        & 
        & $\bullet$
        & $\bullet$
        & 
        & 
        & \\
    & \hyperref[sec:peatlands]{Monitoring peatlands}
        & 
        & $\bullet$
        & 
        & 
        & 
        & 
        & 
        & 
        & \\
    & \hyperref[sec:forests]{Managing forests}
        & 
        & $\bullet$
        & 
        & 
        & $\bullet$ 
        & $\bullet$ 
        & 
        & 
        & \\
    \rowcolor{ccai-blue-lightest}
    \multicolumn{2}{l}{6 \hyperref[sec:ccs]{Carbon dioxide removal}}
        & 
        & 
        & 
        & 
        & 
        & 
        & 
        & 
        & \\
    & \hyperref[sec:ccs]{Direct air capture}
        & 
        & 
        & 
        & 
        & 
        & 
        & 
        & 
        & $\bullet$\\
    & \hyperref[subsubsec: sequestrativervin]{Sequestering~\cd}
        & 
        & $\bullet$
        & 
        & 
        & 
        & 
        & 
        & $\bullet$
        & $\bullet$\\
    \rowcolor{ccai-blue-lightest}
    \multicolumn{2}{l}{7 \hyperref[sec: climate prediction]{Climate prediction}} 
        & 
        & 
        & 
        & 
        & 
        & 
        & 
        & 
        & \\
    & \hyperref[sec:climate-models-params]{Uniting data, ML \& climate science}
        & 
        & $\bullet$
        & $\bullet$
        & 
        & 
        & $\bullet$
        & 
        & $\bullet$
        & \\
    & \hyperref[sec:models-extreme-events]{Forecasting extreme events}
        & 
        & $\bullet$
        & $\bullet$
        & 
        & 
        & $\bullet$
        & 
        & $\bullet$
        & \\
    \rowcolor{ccai-blue-lightest}
    \multicolumn{2}{l}{8 \hyperref[sec:societal-impacts]{Societal impacts}} 
        & 
        & 
        & 
        & 
        & 
        & 
        & 
        & 
        & \\
    & \hyperref[subsub:ecology]{Ecology}
        & 
        & $\bullet$
        & 
        & 
        & 
        & 
        & $\bullet$
        & 
        & \\
    & \hyperref[subsub:infrastructure]{Infrastructure}
        & 
        & 
        & 
        & 
        & $\bullet$
        & $\bullet$
        & 
        & $\bullet$
        & \\
    & \hyperref[subsub:social_systems]{Social systems}
        & 
        & $\bullet$
        & 
        & 
        & 
        & $\bullet$
        & 
        & 
        & $\bullet$\\
    & \hyperref[subsub:crisis]{Crisis}
        & 
        & $\bullet$
        & 
        & $\bullet$
        & 
        & 
        & 
        & 
        & \\
    \rowcolor{ccai-blue-lightest}
    \multicolumn{2}{l}{9 \hyperref[sec:geoengineering]{Solar geoengineering}} 
        & 
        & 
        & 
        & 
        & 
        & 
        & 
        & 
        & \\
    & \hyperref[subsub:better-aerosols]{Understanding \& improving aerosols}
        & 
        & 
        & 
        & 
        & 
        & $\bullet$
        & 
        & $\bullet$
        & \\
    & \hyperref[subsub:planetary-control]{Engineering a planetary control system}
        & 
        & 
        & 
        & 
        & $\bullet$
        & 
        & 
        & $\bullet$
        & \\
    & \hyperref[subsub:impact-models]{Modeling impacts}
        & 
        & 
        & 
        & 
        & 
        & $\bullet$
        & 
        & $\bullet$
        & \\
    \rowcolor{ccai-blue-lightest}
    \multicolumn{2}{l}{10 \hyperref[sec:tools-individuals]{Individual action}} 
        & 
        & 
        & 
        & 
        & 
        & 
        & 
        & 
        & \\
    & \hyperref[sec:personal_carbon_footprint]{Understanding personal footprint}
        & $\bullet$
        & 
        & 
        & $\bullet$
        & $\bullet$
        & $\bullet$
        & 
        & 
        & \\
    & \hyperref[sec:behavior_change]{Facilitating behavior change}
        & 
        & 
        & 
        & $\bullet$
        & 
        & 
        & 
        & 
        & $\bullet$\\
    \rowcolor{ccai-blue-lightest}
    \multicolumn{2}{l}{11 \hyperref[sec:toolsforsociety]{Collective decisions}} 
        & 
        & 
        & 
        & 
        & 
        & 
        & 
        & 
        &  \\
    & \hyperref[sec:coordination]{Modeling social interactions}
        & 
        & 
        & $\bullet$ 
        & 
        & $\bullet$ 
        & 
        & 
        & 
        & \\
    & \hyperref[sec:decisionmaking]{Informing policy}
        & $\bullet$ 
        & $\bullet$ 
        & 
        & $\bullet$
        & 
        & 
        & 
        & $\bullet$
        & $\bullet$\\
    & \hyperref[subsec:markets]{Designing markets}
        & 
        & 
        & 
        & 
        & $\bullet$
        & $\bullet$
        & 
        & 
        & $\bullet$\\
    \rowcolor{ccai-blue-lightest}
    \multicolumn{2}{l}{12 \hyperref[sec:education]{Education}} 
        & 
        & 
        & 
        & $\bullet$
        & $\bullet$
        & 
        & 
        & 
        & \\
    \rowcolor{ccai-blue-lightest}
    \multicolumn{2}{l}{13 \hyperref[sec:finance]{Finance}} 
        & 
        & 
        & 
        & $\bullet$
        & 
        & $\bullet$
        & 
        & $\bullet$
        & \\
    \bottomrule
\end{tabular}
\caption{Climate change solution domains, corresponding to sections of this paper, matched with selected areas of ML that are relevant to each. }
\label{tab:summary}
\end{center}
\end{small}
\end{table}

\subsection*{Who is this paper written for?}

We believe that our recommendations will prove valuable to several different audiences (detailed below). In our writing, we have assumed some familiarity with basic terminology in machine learning, but do not assume any prior familiarity with application domains (such as agriculture or electric grids).\\

\textbf{Researchers and engineers:}
We identify many problems that require conceptual innovation and can advance the field of ML, as well as being highly impactful. For example, we highlight how climate models afford an exciting domain for interpretable ML (see \S\ref{sec: climate prediction}).
We encourage researchers and engineers across fields to use their expertise in solving urgent problems relevant to society.\\

\textbf{Entrepreneurs and investors:} We identify many problems where existing ML techniques could have a major impact without further research, and where the missing piece is deployment. We realize that some of the recommendations we offer here will make valuable startups and nonprofits. For example, we highlight techniques for providing fine-grained solar forecasts for power companies (see \S\ref{sec:electricity-lowCarbon}), tools for helping reduce personal energy consumption (see \S\ref{sec:behavior_change}), and predictions for the financial impacts of climate change (see \S\ref{sec:finance}). We encourage entrepreneurs and investors to fill what is currently a wide-open space.\\

\textbf{Corporate leaders:} We identify problems where ML can lead to massive efficiency gains if adopted at scale by corporate players. For example, we highlight means of optimizing supply chains to reduce waste (see \S\ref{sec:supplychains}) and software/hardware tools for precision agriculture (see \S\ref{sec:agriculture}). We encourage corporate leaders to take advantage of opportunities offered by ML to benefit both the world and the bottom line.\\

\textbf{Local and national governments:} We identify problems where ML can improve public services, help gather data for decision-making, and guide plans for future development. For example, we highlight intelligent transportation systems (see \S\ref{sec:modalshift}), techniques for automatically assessing the energy consumption of buildings in cities (see \S\ref{sec:indv}),
and tools for improving disaster management (see \S\ref{subsub:crisis}). We encourage governments to consult ML experts while planning infrastructure and development, as this can lead to better, more cost-effective outcomes. We further encourage public entities to release data that may be relevant to climate change mitigation and adaptation goals.\\

\subsection*{How to read this paper} \label{sub:howtoread}
The paper is broken into sections according to application domain (see Table \ref{tab:summary}). To help the reader, we have also included the following flags at the level of individual strategies.
\begin{itemize}
\item \textbf{\Rec} $\,$ denotes bottlenecks that domain experts have identified in climate change mitigation or adaptation and that we believe to be particularly well-suited to tools from ML. These areas may be especially fruitful for ML practitioners wishing to have an outsized impact, though applications not marked with this flag are also valuable and should be pursued.
\item \textbf{\Longterm} $\,$ denotes applications that will have their primary impact after 2040. While extremely important, these may in some cases be less pressing than those which can help act on climate change in the near term.
\item \textbf{\HighRisk} $\,$ denotes applications where the impact on GHG emissions is uncertain (for example, the \emph{Jevons paradox} may apply\footnote{The Jevons paradox in economics refers to a situation where increased efficiency nonetheless results in higher overall demand. For example, autonomous vehicles could cause people to drive far more, so that overall GHG emissions could increase even if each ride is more efficient. In such cases, it becomes especially important to make use of specific policies, such as carbon pricing, to direct new technologies and the ML behind them. See also the literature on rebound effects and induced demand.}) or where there is  potential for undesirable side effects (\emph{negative externalities}).
\end{itemize}

These flags should not be taken as definitive; they represent our understanding of more rigorous analyses within the domains we consider, combined with our subjective evaluation of the potential role of ML in these various applications.

Despite the length of the paper, we cannot cover everything. There will certainly be many applications that we have not considered, or that we have erroneously dismissed. We look forward to seeing where future work leads.

\subsection*{A call for collaboration}

All of the problems we highlight in this paper require collaboration across fields. As the language used to refer to problems often varies between disciplines, we have provided keywords and background reading within each section of the paper. Finding collaborators and relevant data can sometimes be difficult; for additional resources, please visit the website that accompanies this paper: \url{https://www.climatechange.ai/}.

Collaboration makes it easier to develop effective strategies. Working with domain experts reduces the chance of using powerful tools when simple tools will do the job, of working on a problem that isn't actually relevant to practitioners, of overly simplifying a complex issue,
or of failing to anticipate risks.

Collaboration can also help ensure that new work reaches the audience that will use it. To be impactful, ML code should be accessible and published using a language and a platform that are already popular with the intended users. For maximal impact, new code can be integrated into an existing, widely used tool.

We emphasize that machine learning is not a silver bullet. The applications we highlight are impactful, but no one solution will ``fix'' climate change. There are also many areas of action where ML is inapplicable, and we omit these entirely. Furthermore, technology alone is not enough -- technologies that would address climate change have been available for years, but have largely not been adopted at scale by society. While we hope that ML will be useful in reducing the costs associated with climate action, humanity also must decide to act.

    \newpage
    
    \part*{Mitigation}
    \addcontentsline{toc}{part}{Mitigation}

\section{Electricity Systems\texorpdfstring{\hfill\textit{by Priya L.~Donti}}{}}
\label{sec:electricity-systems}
AI has been called the new electricity, given its potential to transform entire industries \cite{youtube2017andrew}. Interestingly, electricity itself is one of the industries that AI is poised to transform. Many electricity systems are awash in data, and the industry has begun to envision next-generation systems (\emph{smart grids}) driven by AI and ML \cite{ramchurn2012putting, perera2014machine, victor2019how}.

Electricity systems\footnote{Throughout this section, we use the term ``electricity systems'' to refer to the procurement of fuels and raw materials for electric grid components; the generation and storage of electricity; and the delivery of electricity to end-use consumers. For primers on these topics, see \cite{ipcc2014energy,federal2015energy, von2006electric, kirschen2004fundamentals, wood2013power}.} are responsible for about a quarter of human-caused greenhouse gas emissions each year \cite{ipcc2014summary}. Moreover, as buildings, transportation, and other sectors seek to replace GHG-emitting fuels (\S\ref{sec:transportation}-\ref{sec:buildings-cities}), demand for low-carbon electricity will grow. To reduce emissions from electricity systems, society must
\begin{itemize}
    \item Rapidly transition to low-carbon\footnote{We use the term ``low-carbon'' here instead of ``renewable'' because of this paper's explicit focus on climate change goals. Renewable energy is produced from inexhaustible or easily replenished energy sources such as the sun, wind, or water, but need not necessarily be carbon-free (as in the case of some biomass \cite{creutzig2016economic}). Similarly, not all low-carbon energy is renewable (as in the case of nuclear energy).} electricity sources (such as solar, wind, hydro, and nuclear) and phase out carbon-emitting sources (such as coal, natural gas, and other fossil fuels).
    \item Reduce emissions from existing \carbon-emitting power plants, since the transition to low-carbon power will not happen overnight.
    \item Implement these changes across all countries and contexts, as electricity systems are everywhere.
\end{itemize}

\begin{figure}
    \centering
    \includegraphics[width=.99\textwidth]{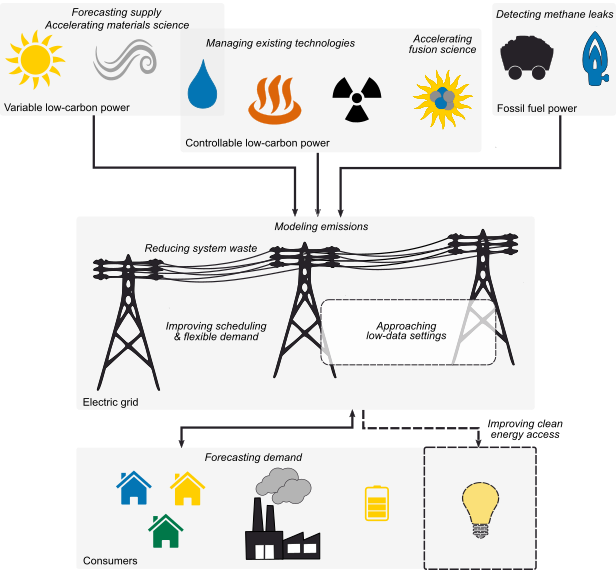}
    \caption{Selected opportunities to reduce GHG emissions from electricity systems using machine learning.}
    \label{fig:electricitySystems}
\end{figure}

ML can contribute on all fronts by informing the research, deployment, and operation of electricity system technologies (Fig.~\ref{fig:electricitySystems}). Such contributions include accelerating the development of clean energy technologies, improving forecasts of demand and clean energy, improving electricity system optimization and management, and enhancing system monitoring. These contributions require a variety of ML paradigms and techniques, as well as close collaborations with the electricity industry and other experts to integrate insights from operations research, electrical engineering, physics, chemistry, the social sciences, and other fields.

\subsection{Enabling low-carbon electricity}
\label{sec:electricity-lowCarbon}
Low-carbon electricity sources are essential to tackling climate change. These sources come in two forms:~variable and controllable. Variable sources fluctuate based on external factors;~for instance, solar panels produce power only when the sun is shining, and wind turbines only when the wind is blowing. On the other hand, controllable sources such as nuclear or geothermal plants can be turned on and off (though not instantaneously\footnote{Nuclear power plants are often viewed as inflexible since they can take hours or days to turn on or off, and are often left on (at full capacity) to operate as \emph{baseload}. That said, nuclear power plants may have some flexibility to change their power generation for load-following and other electric grid services, as in the case of France \cite{lokhov2011technical}.}). These two types of sources affect electricity systems differently, and so present distinct opportunities for ML techniques.

\subsubsection{Variable sources}
\label{sec:electricity-variable}
Most electricity is delivered to consumers using a physical network called the electric grid, where the power generated must equal the power consumed at every moment. This implies that for every solar panel, wind turbine, or other variable electricity generator, there is some mix of natural gas plants, storage, or other controllable sources ready to buffer changes in its output (e.g.~when unexpected clouds block the sun or the wind blows less strongly than predicted). Today, this buffer is often provided by coal and natural gas plants run in a CO$_2$-emitting standby mode called \emph{spinning reserve}. In the future, this role is expected to be played by energy storage technologies such as batteries (\S\ref{sec:transport-evs}), pumped hydro, or power-to-gas~\cite{evans2012assessment}\footnote{It is worth noting that in systems with many fossil fuel plants, storage may increase emissions depending on how it is operated \cite{hittinger2015bulk, babacan2018unintended}.}.
ML can both reduce emissions from today's standby generators and enable the transition to carbon-free systems by helping improve necessary technologies (namely forecasting, scheduling, and control) and by helping create advanced electricity markets that accommodate both variable electricity and flexible demand.

\paragraph{Forecasting supply and demand}\Gap\textbf{\Rec}\mbox{}\\
Since variable generation and electricity demand both fluctuate, they must be forecast ahead of time to inform real-time electricity scheduling and longer-term system planning. Better short-term forecasts can allow system operators to reduce their reliance on polluting standby plants and to proactively manage increasing amounts of variable sources. Better long-term forecasts can help system operators (and investors) determine where and when variable plants should be built. While many system operators today use basic forecasting techniques, forecasts will need to become increasingly accurate, span multiple horizons in time and space, and better quantify uncertainty to support these use cases. ML can help on all these fronts.

To date, many ML methods have been used to forecast electricity supply and demand. These methods have employed historical data, physical model outputs, images, and even video data to create short- to medium-term forecasts of solar power \cite{mathe2019pvnet, das2018forecasting, voyant2017machine, wan2015photovoltaic, sun2018solar, nationalgrid2018powerswarm, alzahrani2017solar, li2016machine, sharma2011predicting}, wind power \cite{foley2012current, deepmind2019machine, wan2014probabilistic, liu2014short, pinson2003wind}, ``run-of-the-river'' hydro power \cite{perera2014machine}, demand \cite{hong2016probabilistic, soliman2010electrical, alfares2002electric, hippert2001neural}, or more than one of these \cite{juban2016multiple, wytock2013sparse} at aggregate spatial scales. These methods span various types of supervised machine learning, fuzzy logic, and hybrid physical models, and take different approaches to quantifying (or not quantifying) uncertainty. At a more spatially granular level, some work has attempted to understand specific categories of demand, for instance by clustering households \cite{kell2018segmenting, beckel2013automatic} or by disaggregating electricity signals using game theory, optimization, regression, and/or online learning \cite{anderson2018disaggregation, kara2018disaggregating, ledva2018real}. 

While much of this previous work has used domain-agnostic techniques, ML algorithms of the future will need to incorporate domain-specific insights. For instance, since weather fundamentally drives both variable generation and electricity demand, ML algorithms forecasting these quantities should draw from innovations in climate modeling and weather forecasting (\S\ref{sec: climate prediction}) and in hybrid physics-plus-ML modeling techniques \cite{wan2015photovoltaic, das2018forecasting, voyant2017machine}. Such techniques can help improve short- to medium-term forecasts, and are also necessary for ML to contribute to longer-term (e.g.~year-scale) forecasts since weather distributions shift over time \cite{kaack2017empirical}. In addition to incorporating system physics, ML models should also directly optimize for system goals \cite{donti2017task, elmachtoub2017smart, wilder2018melding}. For instance, the authors of \cite{donti2017task} use a deep neural network to produce demand forecasts that optimize for electricity scheduling costs rather than forecast accuracy; this notion could be extended to produce forecasts that minimize GHG emissions. In non-automated settings where power system control engineers (partially) determine how much power each generator should produce, interpretable ML and automated visualization techniques could help engineers better understand forecasts and thus improve how they schedule low-carbon generators. More broadly, understanding the domain value of improved forecasts is an interesting challenge. For example, previous work has characterized the benefits of specific solar forecast improvements in a region of the 
United States \cite{martinez2016value}; further study in different contexts and for different types of improvements could help better direct ML work in the forecasting space.

\paragraph{Improving scheduling and flexible demand}\mbox{}\\\label{sec:dispatchDR}When balancing electricity systems, system operators use a process called
\emph{scheduling and dispatch} to determine how much power every controllable generator should produce. This process is slow and complex, as it is governed by NP-hard optimization problems such as \emph{unit commitment} and \emph{optimal power flow} that must be coordinated across multiple time scales (from sub-second to days ahead). 
Further, scheduling will become even more complex as electricity systems include more storage, variable generators, and \emph{flexible demand}, since operators will need to manage even more system components while simultaneously solving scheduling problems more quickly to account for real-time variations in electricity production. Scheduling processes must therefore improve significantly for operators to manage systems with a high reliance on variable sources.

ML can help improve the existing (centralized) process of scheduling and dispatch by speeding up power system optimization problems and improving the quality of optimization solutions. A great deal of work primarily in optimization, but also using techniques such as neural networks, genetic algorithms, and fuzzy logic~\cite{pandya2008survey}, has focused on improving the tractability of power system optimization problems. ML could also be used to approximate or simplify existing optimization problems \cite{guha2019machine, bertsimas2019online, zamzam2019learning}, to find good starting points for optimization \cite{jamei2019meta}, or to learn from the actions of power system control engineers \cite{donnot2017introducing}.
Dynamic scheduling \cite{essl2017machine, moehle2019dynamic} and safe reinforcement learning could also be used to balance the electric grid in real time; in fact, some electricity system operators have started to pilot similar methods at small, test case-based scales. 

While many modern electricity systems are centrally coordinated, recent work has examined how to (at least partially) \emph{decentralize} scheduling and dispatch using energy storage, flexible demand, low-carbon generators, and other resources connected to the electric grid.
One strategy is to explicitly design local control algorithms; for instance, recent work has controlled energy storage and \emph{solar inverters} using supervised learning techniques trained on historical optimization data \cite{dobbe2019towards, karagiannopoulos2019data, karagiannopoulos2019data2, dobbe2017fully}.
Another strategy is to let storage, demand, and generation respond to real-time
prices\footnote{For discussions and examples of different types of advanced electricity markets, see \cite{lian2018transactive, lian2018transactive2, zhang2017machine, camusenergy}.} that reflect (for example) how emissions-intensive electricity currently is. 
In this case, ML can help both to design real-time prices and to respond to these prices.
Previous work has used dynamic programming to set real-time electricity prices \cite{borgs2014optimal} and reinforcement learning to set real-time prices in more general settings \cite{maestre}; similar techniques could be applied to create prices that instead optimize for GHG emissions. Techniques such as agent-based models \cite{ramchurn2011agent2, ramchurn2011agent1, deindl2008load, ygge1999homebots}, online optimization \cite{buchbinder2011online}, and dynamic programming \cite{salas2017benchmarking} can then help maximize profits for decentralized storage, demand, and generation, given real-time prices.
In general, much more work is needed to test and scale existing decentralized solutions; barring deployment on real systems, platforms such as PowerTAC~\cite{powertac} can provide large-scale simulated electricity markets on which to perform these tests.

\paragraph{Accelerating materials science}\Gap\textbf{\Rec\Longterm}\mbox{}\\\label{sec:electricity-materials}Scientists are working to develop new materials that can better store or otherwise harness energy from variable natural resources. For instance, creating \emph{solar fuels} (synthetic fuels produced from sunlight or solar heat) could allow us to capture solar energy when the sun is shining and then store this energy for later use. However, the process of discovering new materials can be slow and imprecise; the physics behind materials are not completely understood, so human experts often manually apply heuristics to understand a proposed material's physical properties \cite{butler2018machine, bai2018phase}.
ML can automate this process by combining existing heuristics with experimental data, physics, and reasoning to apply and even extend existing physical knowledge. For instance, recent work has used tools from ML, AI, optimization, and physics to figure out a proposed material's crystal structure, with the goal of accelerating materials discovery for solar fuels \cite{gomes2019crystal,bai2018phase,suram2016automated}. Other work seeking to improve battery storage technologies has combined first-principles physics calculations with support-vector regression to design conducting solids for lithium-ion batteries \cite{fujimura2013accelerated}. (Additional applications of ML to batteries are discussed in \S\ref{sec:TFuels}.)

More generally in materials science, ML techniques including supervised learning, active learning, and generative models have been used to help synthesize, characterize, model, and design materials, as described in reviews \cite{butler2018machine, liu2017materials} and more recent work \cite{gomez2018automatic}. As discussed in \cite{butler2018machine}, novel challenges for ML in materials science include coping with moderately sized datasets and inferring physical principles from trained models \cite{umehara2019analyzing}. In addition to advancing technology, ML can inform policy for accelerated materials science; for instance, previous work has applied natural language processing to patent data to understand the solar panel innovation process \cite{venugopalan2015topic}. We note that while our focus here has been on electricity system applications, ML for accelerated science may also have significant impacts outside electricity systems, e.g.~by helping design alternatives to cement (\S\ref{sec:materialsandconstruction}) or create better CO$_2$ sorbents (\S\ref{subsec:dac}).

\paragraph{Additional applications}\mbox{}\\
There are many additional opportunities for ML to advance variable power generation. For instance, it is important to ensure that low-carbon variable generators produce energy as efficiently and profitably as possible. Prior work has attempted to maximize electricity production by controlling movable solar panels \cite{abel2018bandit, abdelrahman2016bayesian} or wind turbine blades \cite{kolter2012design} using reinforcement learning or Bayesian optimization. Other work has used graphical models to detect faults in rooftop solar panels \cite{iyengar2018solarclique} and genetic algorithms to optimally place wind turbines within a wind farm \cite{dilkina2015method}.
ML can also help control batteries located at solar and wind farms to increase these farms' profits, for instance by storing their electricity when prices are low and then selling it when prices are high; prior work has used ML to forecast electricity prices \cite{weron2014electricity, lago2018forecasting} or reinforcement learning to control batteries based on current and historical prices \cite{wang2018energy}.

ML can also help integrate rooftop solar panels into the electric grid, particularly in the United States and Europe. Rooftop solar panels are connected to a part of the electric grid called the distribution grid, which traditionally did not have many sensors because it was only used to  deliver electricity ``one-way'' from centralized power plants to consumers. However, rooftop solar and other \emph{distributed energy resources} have created a ``two-way'' flow of electricity on distribution grids. Since the locations and sizes of rooftop solar panels are often unknown to electricity system operators, previous work has used computer vision techniques on satellite imagery to generate size and location data for rooftop solar panels \cite{malof2016automatic, yu2018deepsolar}. Further, to ensure that the distribution system runs smoothly, recent work has employed techniques such as matrix completion and deep neural networks to estimate the state of the system when there are few sensors \cite{donti2018matrix, jiang2016short, pertl2016voltage}.

\subsubsection{Controllable sources}
\label{sec:electricity-controllable}
Controllable low-carbon electricity sources can help achieve climate change goals while requiring very few changes to how the electric grid is run (since today's fossil fuel power plants are also controllable). ML can support existing controllable technologies while accelerating the development of new technologies such as nuclear fusion power plants.

\paragraph{Managing existing technologies}
\mbox{}\\ 
Many controllable low-carbon technologies are already commercially available; these technologies include geothermal, nuclear fission, and (in some cases\footnote{Dam-based hydropower may produce methane, primarily due to biomass that decomposes when a hydro reservoir floods, but the amount produced varies between power plants \cite{steinhurst2012hydropower}.}) dam-based hydropower. 
ML can provide valuable input in planning where these technologies should be deployed and can also help maintain already-operating power plants.
For instance, recent work has proposed to use ML to identify and manage sites for geothermal energy, using satellite imagery and seismic data \cite{eere2019geothermal}.
Previous work has also used multi-objective optimization to place hydropower dams in a way that satisfies both energy and ecological objectives \cite{wu2018efficiently}.
Finally, ML can help maintain nuclear fission reactors (i.e., nuclear power plants) by detecting cracks and anomalies from image and video data \cite{chen2018nb} or by preemptively detecting faults from high-dimensional sensor and simulation data \cite{caliva2018deep}. (The authors of \cite{touran2017computational} speculate that ML and high performance computing could also be used to help simulate nuclear waste disposal options or even design next-generation nuclear reactors.)

\paragraph{Accelerating fusion science}\Gap
\textbf{\Rec\Longterm}\mbox{}\\
Nuclear fusion reactors \cite{nature2016nuclear} have the potential to produce safe and carbon-free electricity using a virtually limitless hydrogen fuel supply, but currently consume more energy than they produce \cite{cowley2016quest}. While considerable scientific and engineering research is still needed, ML can help accelerate this work by guiding experimental design and monitoring physical processes. Fusion reactors require intelligent experimental design because they have a large number of tunable parameters; ML can help prioritize which parameter configurations should be explored during physical experiments. For instance, Google and TAE Technologies have developed a human-in-the-loop experimental design algorithm enabling rapid parameter exploration for TAE's reactor \cite{baltz2017achievement}.

Physically monitoring fusion reactors is also an important application for ML.
Modern reactors attempt to super-heat hydrogen into a plasma state and then stabilize it, but during this process, the plasma may experience rapid instabilities that damage the reactor. Prior work has tried to preemptively detect disruptions for \emph{tokamak} reactors, using supervised learning methods such as support-vector machines, adaptive fuzzy logic, decision trees, and deep learning \cite{ cannas2003disruption, murari2008prototype, vega2013results, windsor2005cross, wroblewski1997tokamak, katesharbeck2019predicting} on previous disruption data. While many of these methods are tuned to work on individual reactors, recent work has shown that deep learning may enable insights that generalize to multiple reactors \cite{katesharbeck2019predicting}. More generally, rather than simply detecting disruptions, scientists need to understand how plasma's state evolves over time, e.g.~by finding the solutions of time-dependent magnetohydrodynamic equations \cite{barton2015simultaneous}; speculatively, ML could help characterize this evolution and even help steer plasma into safe states through reactor control. ML models for such fusion applications would likely employ a combination of simulated\footnote{Plasma simulation frameworks for tokamak reactors include RAPTOR \cite{felici2011real, felici2012non}, ASTRA \cite{pereverzev2002astra}, CRONOS \cite{artaud2010cronos}, PTRANSP \cite{budny2008predictions}, and IPS \cite{ips}.} 
and experimental data, and would need to account for the different physical characteristics, data volumes, and simulator speeds or accuracies associated with different reactor types.

\subsection{Reducing current-system impacts}
\label{sec:electricity-currentSystemImpact}
While switching to low-carbon electricity sources will be essential, in the meantime, it will also be important to mitigate emissions from the electricity system as it currently stands. Some methods for mitigating current-system impacts include cutting emissions from fossil fuels, reducing waste from electricity delivery, and flexibly managing demand to minimize its emissions impacts.

\paragraph{Reducing life-cycle fossil fuel emissions}\Gap\textbf{\Rec\HighRisk}\mbox{}\\\label{sec:electricity-methane}Reducing emissions from fossil fuels is a necessary stopgap while society transitions towards low-carbon electricity. In particular, ML can help prevent the leakage of methane (an extremely potent greenhouse gas) from natural gas pipelines and compressor stations. Previous and ongoing work has used sensor and/or satellite data to proactively suggest pipeline maintenance \cite{zukhrufany2018utilization, edward2018oil} or detect existing leaks \cite{wan2012hierarchical, swri2016swri, bluefieldtechnologies}, and there is a great deal of opportunity in this space to improve and scale existing strategies. In addition to leak detection, ML can help reduce emissions from freight transportation of solid fuels (\S\ref{sec:transportation}), identify and manage storage sites for \carbon~sequestered from power plant flue gas (\S\ref{subsubsec: sequestrativervin}), and optimize power plant parameters to reduce \carbon~emissions.
In all these cases, projects should be pursued with great care so as not to impede or prolong the transition to a low-carbon electricity system; ideally, projects should be preceded by system impact analyses to ensure that they will indeed decrease GHG emissions.

\paragraph{Reducing system waste}
\mbox{}\\ As electricity gets transported from generators to consumers, some of it gets lost as resistive heat on electricity lines. While some of these losses are unavoidable, others can be significantly mitigated to reduce waste and emissions. ML can help prevent avoidable losses through predictive maintenance, i.e.,~by suggesting proactive electricity grid upgrades. Prior work has performed predictive maintenance using LSTMs \cite{bhattacharya2017deep}, bipartite ranking \cite{rudin2012machine}, and neural network-plus-clustering techniques \cite{nguyen2018automatic} on electric grid data, and future work will need to improve and/or localize these approaches to different contexts.

\paragraph{Modeling emissions}
\mbox{}\\\label{subsec:electricity-systems-modeling-emissions}Flexibly managing household, commercial, industrial, and electric vehicle demand (as well as energy storage) can help minimize electricity-based emissions (\S\ref{sec:transportation},~\ref{sec:buildings-cities},~\ref{sec:industry},~\ref{sec:tools-individuals}), but doing so involves understanding what the emissions on the electric grid actually are at any moment. Specifically, \emph{marginal emissions factors} capture the emissions effects of small changes in demand at any given time. To inform consumers about marginal emissions factors, WattTime \cite{watttime} estimates these factors in real time for the United States using regression-based techniques, and the electricityMap project \cite{electricitymap} provides multi-day forecasts for Europe using ensemble models on electricity and weather data. 
Great Britain's National Grid ESO also uses ensemble models to forecast \emph{average} emissions factors, which measure the aggregate emissions intensity of all power plants \cite{carbonintensityapi}. There is still much room to improve the performance of these methods, as well as to forecast related quantities such as electricity curtailments (i.e.~the wasting of usually low-carbon electricity for grid balancing purposes). As most existing methods produce point estimates, it would also be important to quantify the uncertainty of these estimates to ensure that load-shifting techniques indeed decrease (rather than increase) emissions.

\subsection{Ensuring global impact}
\label{sec:electricity-developing}

Much of the discussion around electricity systems often focuses on settings such as the United States with near universal electricity access and relatively abundant data. However, many places that do not share these attributes are still integral to tackling climate change \cite{ipcc2014summary} and warrant serious consideration. To ensure global impact, ML can help improve electricity access and translate electricity system insights from high-data to low-data contexts.

\paragraph{Improving clean energy access}\mbox{}\\ 
Improving access to clean electricity can address climate change while simultaneously improving social and economic development \cite{khandker2009welfare1, khandker2009welfare2}. Specifically, clean electricity provided via electric grids, \emph{microgrids}, or off-grid methods can displace diesel generators, wood-burning stoves, and other carbon-emitting energy sources. Figuring out what clean electrification methods are best for different areas can require intensive, boots-on-the-ground surveying work, but ML can help provide input to this process in a scalable manner. For instance, previous work has used image processing, clustering, and optimization techniques on satellite imagery to inform electrification initiatives \cite{ellman2015reference}. ML and statistics can also help operate rural microgrids through accurate forecasts of demand and power production \cite{cenek2018climate, otieno2018forecasting}, since small microgrids are even harder to balance than country-scale electric grids. Generating data to aid energy access policy and better managing energy access strategies are therefore two areas in which ML may have promising applications.

\paragraph{Approaching low-data settings}\Gap\textbf{\Rec}\mbox{}\\
While ML methods have often been applied to grids with widespread sensors, system operators in many countries do not collect or share system data. Although these data availability practices may evolve, it may meanwhile be beneficial to use ML techniques such as transfer learning to translate insights from high-data to low-data settings (especially since all electric grids share the same underlying system physics). Developing data-efficient ML techniques will likely also be useful in low-data settings; for instance, in \cite{ren2018learning}, the authors enforce physical or other domain-specific constraints on weakly supervised ML models, allowing these models to learn from very little labeled data. 

ML can also help generate information within low-data settings. For instance, recent work has estimated the layout of electricity grids in regions where they are not explicitly mapped, using computer vision on satellite imagery along with graph search techniques \cite{facebook2019predictive}. Companies have also proposed to use satellite imagery to measure power plant CO$_2$ emissions \cite{carbontracker} (also see \S\ref{sec:emissions-detection}). Other recent work has modeled electricity consumption using regression-based techniques on cellular network data \cite{bogomolov2016energy}, which may prove useful in settings with many cellular towers but few electric grid sensors. Although low-data settings are generally underexplored by the ML community, electricity systems research in these settings presents opportunities for both innovative ML and climate change mitigation. 

\subsection{Discussion}
Data-driven and critical to climate change, electricity systems hold many opportunities for ML. 
At the same time, applications in this space hold many potential pitfalls; for instance, innovations that seek to reduce GHG emissions in the oil and gas industries could actually \emph{increase} emissions by making them cheaper to emit \cite{victor2019how}. Given these domain-specific nuances, working in this area requires close collaborations with electricity system decision-makers and with practitioners in fields including electrical engineering, the natural sciences, and the social sciences. Interpretable ML may enable stakeholders outside ML to better understand and apply models in real-world settings. Similarly, it will be important to develop hybrid ML models that explicitly account for system physics (see e.g.~\cite{chen2018neural, schenck2018spnets, de2018end, ren2018learning}), directly optimize for domain-specific goals \cite{donti2017task, elmachtoub2017smart, wilder2018melding}, or otherwise incorporate or scale existing domain knowledge. Finally, since most modern electric grids are not data-abundant (although they may be data-driven), understanding how to apply data-driven insights to these grids may be the next grand challenge for ML in electricity systems.

    \newpage

\section{Transportation\texorpdfstring{\hfill\textit{by Lynn H.~Kaack}}{}}
\label{sec:transportation}

Transportation systems form a complex web that is fundamental to an active and prosperous society. Globally, the transportation sector accounts for about a quarter of energy-related CO$_2$ emissions \cite{ipcc_global_2018}. 
In contrast to the electricity sector, however, transportation has not made significant progress to lower its CO$_2$ emissions \cite{Creutzig911} and much of the sector is regarded as hard to decarbonize \cite{Daviseaas9793}. This is because of the high energy density of fuels required for many types of vehicles, which constrains low-carbon alternatives, and because transport policies directly impact end-users and are thus more likely to be controversial.

Passenger and freight transportation are each responsible for about half of transport GHG emissions \cite{AR5_transport}. 
Both freight and passengers can travel by road, by rail, by water, or by air (referred to as \emph{transport modes}). Different modes carry vastly different carbon emission intensities.\footnote{Carbon intensity is measured in grams of CO$_2$-equivalent per person-km or per ton-km, respectively.}
At present, more than two-thirds of transportation emissions are from road travel \cite{AR5_transport}, but air travel has the highest emission intensity and is responsible for an increasingly large share.
Strategies to reduce GHG emissions\footnote{For general resources on how to decarbonize the transportation sector, see the AR5 chapter on transportation \cite{AR5_transport}, and \cite{figueroa2014energy, iea2017future, kaack2018decarbonizing}.} from transportation include \cite{AR5_transport}: 
\begin{itemize}
    \item Reducing transport activity.
    \item Improving vehicle efficiency.
    \item Alternative fuels and electrification.
    \item Modal shift (shifting to lower-carbon options, like rail).
\end{itemize}

\begin{figure}[tb]
    \centering
    \includegraphics[width=\textwidth]{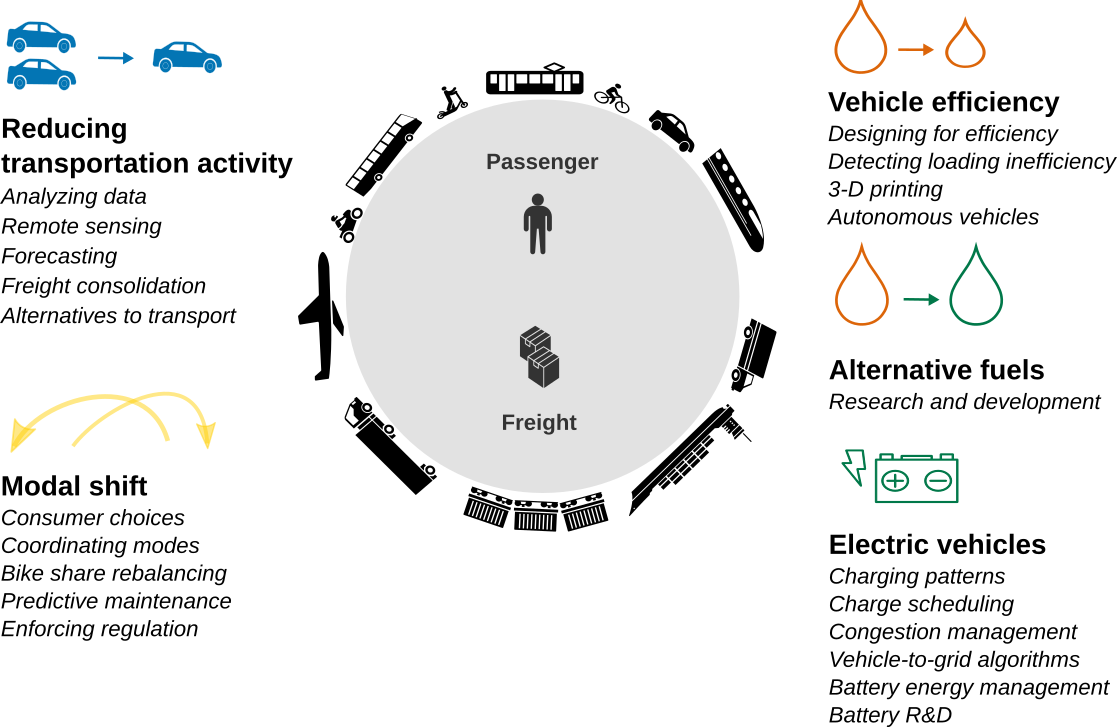}
    \caption{Selected strategies to mitigate GHG emissions from transportation using machine learning.}
    \label{fig:transport}
\end{figure}
Each of these mitigation strategies offers opportunities for ML (Fig.~\ref{fig:transport}).  
While many of us probably think of autonomous vehicles and ride-sharing when we think of transport and ML, these technologies have uncertain impacts on GHG emissions \cite{WADUD20161}, potentially even increasing them. 
We discuss these disruptive technologies in \S\ref{sec:TReducing} but show that ML can play a role for decarbonizing transportation that goes much further. 
ML can improve vehicle engineering, enable intelligent infrastructure, and provide policy-relevant information. 
Many interventions that reduce GHG emissions in the transportation sector require changes in planning, maintenance, and operations of transportation systems, even though the GHG reduction potential of those measures might not be immediately apparent. ML can help in implementing such interventions, for example by providing better demand forecasts. Typically, ML strategies are most effective in tandem with strong public policies.  
While we do not cover all ML applications in the transportation sector, we aim to include those areas that can conceivably reduce GHG emissions. 

\subsection{Reducing transport activity}\label{sec:TReducing}

A colossal amount of transport occurs each day across the world, but much of this mileage occurs inefficiently, resulting in needless GHG emissions. With the help of ML, the number of vehicle-miles traveled can be reduced by making long trips less necessary, increasing loading, and optimizing vehicle routing. Here, we discuss the first two in depth -- for a discussion of ML and routing, see for example \cite{ZENG2017458}. 

\paragraph*{Understanding transportation data}\mbox{}\\\label{sec:transport-data}Many areas of transportation lack data, and decision-makers often design infrastructure and policy with uncertain information. In recent years, new types of sensors have become available, and ML can turn this raw data into useful information.
Traditionally, traffic is monitored with ground-based counters that are installed on selected roads. A variety of technologies are used, such as inductive loop detectors or pneumatic tubes. Traffic is sometimes monitored with video systems, in particular when counting pedestrians and cyclists, which can be automated with computer vision \cite{Zaki2016361}.
Since counts on most roads are often available only over short time frames, these roads are modeled by looking at known traffic patterns for similar roads. ML methods, such as SVMs and neural networks, have made it easier to classify roads with similar traffic patterns \cite{krile2016assessing, tsapakis2015use, gastaldi2013annual}. 
As ground-based counters require costly installation and maintenance, many countries do not have such systems. Vehicles can also be detected in high-resolution satellite images with high accuracy \cite{sommer2017fast, jiang2015deep, mundhenk2016large, deng2017toward}, and image counts can serve to estimate average vehicle traffic \cite{kaack2019truck}. 
Similarly, ML methods can help with imputing missing data for precise bottom-up estimation of GHG emissions \cite{NOCERA2018125} and they are also applied in simulation models of vehicle emissions \cite{doi:10.1080/15568318.2017.1346732}.

\paragraph*{Modeling demand}\Gap\textbf{\Rec}\mbox{}\\Modeling demand and planning new infrastructure can significantly shape how long trips are and which transport modes are chosen by passengers and shippers -- for example, discouraging sprawl and creating new transportation links can both reduce GHG emissions. 
ML can provide information about mobility patterns, which is directly necessary for agent-based travel demand models, one of the main transport planning tools \cite{feygin7932990}.
For example, ML makes it possible to estimate origin-destination demand from traffic counts \cite{MA201896}, and it offers new methods for spatio-temporal road traffic forecasting -- which do not always outperform other statistical methods \cite{doi:10.1080/01441647.2018.1442887} 
but may transfer well between areas \cite{Tang2018spatial}.
Also, short-term forecasting of public transit ridership can improve with ML; see for example \cite{Dai:2018aa, NOURSALEHI2018277}. 
ML is particularly relevant for deducing information from novel data -- for example, learning about the behavior of public transit users from smart card data \cite{Manley2018, Ghaemi_2017}.
Also, mobile phone sensors provide new means to understand personal travel demand and the urban topology, such as walking route choices \cite{doi:10.1177/0265813516659286}. 
Similarly, ML-based modeling of demand can help mitigate climate change by improving operational efficiency of modes that emit significant CO$_2$, such as aviation. ML can help predict runway demand and aircraft taxi time in order to reduce the excess fuel burned in the air and on the ground due to congestion in airports \cite{JACQUILLAT2018168, lee2015taxi}.

\paragraph*{Shared mobility}\Gap\textbf{\HighRisk}\mbox{}\\In the passenger sector, shared mobility (such as on-demand ride services or vehicle-sharing\footnote{In this section, we discuss shared cars; see \S\ref{sec:modalshift} for bike shares and electric scooters.}), is undoubtedly disrupting the way people travel and think about vehicle ownership, and ML plays an integral part in optimizing these services (e.g.~\cite{8317908}). 
However, it is largely unclear what the impact of this development will be. For example, shared cars can actually cause more people to travel by car, as opposed to using public transportation. Similarly, on-demand taxi services add mileage when traveling without a customer, possibly negating any GHG emission savings \cite{suatmadi2019demand}. 
On the other hand, shared mobility can lead to higher utilization of each vehicle, which means a more efficient use of materials \cite{Hertwich_2019}. The use of newer and more efficient vehicles, ideally electric ones, could increase with vehicle-sharing concepts, reducing GHG emissions. Some of the issues raised above could also perhaps be overcome by making taxis autonomous. Such vehicles also might integrate better with public transportation, and offer new concepts for pooled rides, which substantially reduce the emissions per person-mile.

ML methods can help to understand the energy impact of shared mobility concepts.
For example, they can be used to predict if a customer decides to share a ride with other passengers from an on-demand ride service \cite{CHEN201751}. For decision-makers it is important to have access to timely location-specific empirical analysis to understand if a ride share service is taking away customers from low-carbon transit modes and increasing the use of cars. Some local governments are beginning to require data-sharing from these providers (see \S\ref{sec:data_pol}).

Car-sharing services using autonomous vehicles could yield GHG emission savings when they encourage people to use public transit for part of the journey \cite{2017-01-1276} or with autonomous electric vehicles \cite{kang2016autonous}. However, using autonomous shared vehicles alone could increase the total vehicle-miles traveled and therefore do not necessarily lead to lower emissions as long as the vehicles have internal combustion engines (or electrical engines on a ``dirty'' electrical grid) \cite{lu2018multiagent, CHEN2016243}.
We see the intersection of shared mobility, autonomous and electric vehicles, and smart public transit as a path where ML can make a contribution to shaping future mobility. See also \S\ref{sec:TEfficient} for more on autonomous vehicles.

When designing and promoting new mobility services, it is important that industry and public policy prioritize lowering GHG emissions. Misaligned incentives in the early stages of technological development could result in the lock-in to a service with high GHG emissions \cite{AXSEN20191, 10.2307/2234208}.

\paragraph*{Freight routing and consolidation}\Gap\textbf{\Rec}\mbox{}\\Bundling shipments together, which is referred to as freight consolidation, dramatically reduces the number of trips (and therefore the GHG emissions). The same is true for changing routing so that trucks do not have to return empty. As rail and water modes require much larger loads than trucks, consolidation also enables shipments to use these modes for part of the journey \cite{kaack2018decarbonizing}. Freight consolidation and routing decisions are often taken by third-party \emph{logistics service providers} and other freight forwarders, such as in the less-than-truckload market, which deals with shipments of smaller sizes. ML offers opportunities to optimize this complex interaction of shipment sizes, modes, origin-destination pairs, and service requirements. Many problem settings are addressed with methods from the field of operations research. There is evidence that ML can improve upon these methods, in particular mixed-integer linear programming \cite{2018arXiv181106128B}. Other proposed and deployed applications of ML include predicting arrival times or demand, identifying and planning around transportation disruptions \cite{DHL}, and clustering suppliers by their geographical location and common shipping destinations. Proposed planning approaches include designing allocation algorithms and freight auctions, and ML has for example been shown to help pick good algorithms and parameters to solve auction markets \cite{Sandholm_very-large-scalegeneralized}.

\paragraph*{Alternatives to transport}\Gap\textbf{\HighRisk}\mbox{}\\Disruptive technologies that are based on ML could replace or reduce transportation demand. 
For example, additive manufacturing (AM, or 3-D printing) has (limited) potential to reduce freight transport by producing lighter goods and enabling production closer to the consumer \cite{kaack2018decarbonizing}. ML can be a valuable tool for improving AM processes \cite{PhysRevLett.120.145301}. ML can also help to improve virtual communication \cite{Shiarlis}. If passenger trips are replaced by telepresence, travel demand can be reduced, as has been shown for example in public agencies \cite{ARNFALK2016101} and for scientific teams \cite{Marlow4841}. However, it is uncertain to what extent virtual meetings replace physical travel, or if they may actually give rise to more face-to-face meetings \cite{doi:10.1080/17450101.2014.902655}.

\subsection{Improving vehicle efficiency}\label{sec:TEfficient}
Most vehicles are not very efficient compared to what is technically possible: for example, aircraft carbon intensity is expected to decline by more than a third with respect to 2012, simply by virtue of newer models replacing aging jets \cite{schaefer2015}. Both the design of the vehicle and the way it is operated can increase the fuel economy. Here, we discuss how ML can help design more efficient vehicles and the impacts that autonomous driving may have on GHG emissions. Encouraging drivers to adopt more efficient vehicles is also a priority; while we do not focus on this here, ML plays a role in studying consumer preferences in vehicle markets \cite{burnap2016improving}.

\paragraph*{Designing for efficiency}\Gap\mbox{}\\There are many ways to reduce the energy a vehicle uses -- such as more efficient engines, improved aerodynamics, hybrid electric engines, and reducing the vehicle's weight or tire resistance.
These different strategies require a broad range of engineering techniques, many of which can benefit from ML. For example, ML is applied in advanced combustion engine design \cite{JANAKIRAMAN2016304}.
Hybrid electric vehicles, which are more efficient than combustion engines alone, rely on power management methods that can be improved with ML \cite{en11030476}. Aerodynamic efficiency improvements need turbulence modeling that is often computationally intensive and relies heavily on ML-based surrogate models \cite{YONDO201823}. Aerodynamic improvements can not only be made by vehicle design but also by rearranging load. Lai \et\cite{doi:10.1243/09544097JRRT92} use computer vision to detect aerodynamically inefficient loading on freight trains.  
Additive manufacturing (3-D printing) can produce lighter parts in vehicles, such as road vehicles and aircraft, that reduce energy consumption \cite{kaack2018decarbonizing, Hertwich_2019}. ML is applied to improve those processes, for example through failure detection \cite{Scime2018114, Shevchik2018598} or material design \cite{Gu2018939}. 

\paragraph*{Autonomous vehicles}\Gap\textbf{\HighRisk}\mbox{}\\Machine learning is essential in the development of autonomous vehicles (AVs), including in such basic tasks as following the road and detecting obstacles \cite{DBLP:journals/corr/BojarskiTDFFGJM16}.\footnote{Providing details on the general role of ML for AVs is beyond the scope of this paper.} While AVs could reduce energy consumption -- for example, by reducing traffic congestion and inducing efficiency through eco-driving -- it is also possible that AVs will lead to an increase in overall road traffic that nullifies efficiency gains. (For an overview of possible energy impacts of AVs see \cite{WADUD20161, Brown2014} and for broader impacts on mobility see \cite{Hancock7684}.)  Two advantages of AVs in the freight sector promise to cut GHG emissions: 
First, small autonomous vehicles, such as delivery robots and drones, could reduce the energy consumption of last-mile delivery \cite{samaras2018drones}, though they come with regulatory challenges \cite{marks2019law}. Second, trucks can reduce energy consumption by \emph{platooning} (driving very close together to reduce air resistance), thereby alleviating some of the challenges that come with electrifying long-distance road freight \cite{Guttenberg_2017}. Platooning relies on autonomous driving and communication technologies that allow vehicles to brake and accelerate simultaneously.

ML can help to develop AV technologies specifically aimed at reducing energy consumption. For example, Wu \et\cite{Wu2017EmergentBI,wu2017flow} develop AV controllers based on reinforcement learning to smooth out traffic involving non-autonomous vehicles, reducing congestion-related energy consumption. ML methods can also help to understand driving practices that are more energy efficient. For example, Jim\'enez et al.~\cite{en11020412} use data from smart phone sensors to identify driving behavior that leads to higher energy consumption in electric vehicles.

\subsection{Alternative fuels and electrification}\label{sec:TFuels}

\paragraph*{Electric vehicles}\Gap\textbf{\Rec}\mbox{}\\\label{sec:transport-evs}Electric vehicle (EV) technologies -- using batteries, hydrogen fuel cells, or electrified roads and railways -- are regarded as a primary means to decarbonize transport. EVs can have very low GHG emissions -- depending, of course, on the carbon intensity of the electricity. ML is vital for a range of different problems related to EVs. Rigas et al.~\cite{7000557} detail methods by which ML can improve charge scheduling, congestion management, and vehicle-to-grid algorithms. ML methods have also been applied to battery energy management (for example charge estimation \cite{HANSEN2005351} or optimization in hybrid EVs \cite{en11030476}), and to detect faults and lateral misalignment in wireless charging of EVs \cite{8096493}. 

As more people drive EVs, understanding their use patterns will become more important. Modeling charging behavior will be useful for grid operators looking to predict electric load. For this application, it is possible to analyze residential EV charging behavior from aggregate electricity load (\emph{energy disaggregation}, see also \S\ref{sec:use}) \cite{WANG20192611}. Also, in-vehicle sensors and communication data are increasingly becoming available and offer an opportunity to understand travel and charging behavior of EV owners, which can for example inform the placement of charging stations \cite{TAO2018735}. 

Battery electric vehicles are typically not used for more than a fraction of the day, allowing them to act as energy storage for the grid 
at other times, where charging and discharging is controlled for example by price signals \cite{doi:10.1002/wene.56} (see \S\ref{sec:electricity-variable},\ref{sec:electricity-currentSystemImpact}). There is much potential for ML to improve such vehicle-to-grid technology, for example with reinforcement learning \cite{VAZQUEZCANTELI20191072}, which can reduce GHG emissions from electricity generation. Vehicle-to-grid technology comes with private and social financial benefits. However, consumers are expected to be reluctant to agree to such services, as they might not want to compromise their driving range \cite{HIDRUE201567}. 

Finally, ML can also play a role in the research and development of batteries, a decisive technology for EV costs and usability. 
Work in this area has focused on predicting battery state, degradation, and remaining lifetime using supervised learning techniques, fuzzy logic, and clustering \cite{wu2016review, waag2014critical, si2011remaining, severson2019data, ellis2018new, hu2016advanced, kauwe2019data, buteau2019}.
However, many models developed in academia are based on laboratory data that do not account for real-world factors such as environmental conditions \cite{wu2016review, waag2014critical, si2011remaining}. By contrast, industry lags behind in ML modeling, but real-world operational data are readily available. Merging these two perspectives could yield significant benefits for the field.

\paragraph*{Alternative fuels}\Gap \textbf{\Longterm}\mbox{}\\Much of the transportation sector is highly dependent on liquid fossil fuels. Aviation, long-distance road transportation, and ocean shipping require fuels with high energy density and thus are not conducive to electrification \cite{Daviseaas9793}. Electrofuels \cite{BRYNOLF20181887}, solar fuels \ref{sec:electricity-materials}, biofuels \cite{biofuelsIEA}, hydrogen \cite{doeHydrogen, cano2018fuelCells}, and perhaps natural gas \cite{tong2015gas} offer alternatives, but the use of these fuels is constrained by factors such as cost, land-use, and (for hydrogen and natural gas) incompatibility with current infrastructure \cite{Daviseaas9793}.  
Electrofuels and biofuels have the potential to serve as low-carbon drop-in fuels that retain the properties of fossil fuels, such as high energy density, while retaining compatibility with the existing fleet of vehicles and the current fuel infrastructure \cite{kaack2018decarbonizing}. 
Fuels such as electrofuels and hydrogen can be produced using electricity-intensive processes and can be stored at lower cost than electricity. Thus, as a form of energy storage, these fuels could provide services to the electricity grid by enabling flexible power use and balancing variable electricity generators (\S\ref{sec:electricity-variable}). Given their relative long-term importance and early stage of development, they present a critical opportunity to mitigate climate change.
ML techniques may present opportunities for improvement at various stages of research and development of alternative fuels (similar to applications in  \S\ref{sec:electricity-variable}).

\subsection{Modal shift}\label{sec:modalshift}
Shifting passengers and freight to low carbon-intensity modes is one of the most important means to decarbonize transport. 
This \emph{modal shift} in passenger transportation can for example involve providing people with public transit, which requires analyzing mode choice and travel demand data. ML can also make low-carbon freight modes more competitive by helping to coordinate intermodal transport.

\paragraph*{Passenger preferences }\Gap\mbox{}\\ML can improve our understanding about passengers' travel mode choices, which in turn informs transportation planning, such as where public transit should be built. 
Some recent studies have shown that supervised ML based on survey data can improve passenger mode choice models \cite{omrani2015predicting, nam2017model, HAGENAUER2017273}. 
Seo \et propose to conduct long-term travel surveys with online learning, which reduces the demand on respondents, while obtaining high data quality \cite{SEO2017}. 
Sun \et \cite{SUN201896} use SVMs and neural networks for analyzing preferences of customers traveling by high speed rail in China. 
There is also work on inferring people's travel modes and destinations from social media or various mobile phone sensors such as GPS (\emph{transportation mode detection}), e.g.~\cite{DABIRI2018360, doi:10.1080/13658816.2017.1356464}.  Also in the freight sector, ML has been applied to analyze modal trade-offs, for example by imputing data on counterfactual mode choices \cite{doi:10.1080/03081060.2011.600092}.

\paragraph*{Enabling low-carbon options}\Gap\textbf{\Rec}\label{sec:LCcompetitive}\mbox{}\\In order to incentivize more users to choose low-carbon transport modes, their costs and service quality can be improved.
Many low-carbon modes must be integrated with other modes of transportation to deliver the same level of service. For example, when traveling by train, the trip to and from the station will often be by car, taxi, bus, or bike. There are many opportunities for ML to facilitate a better integration of modes, both in the passenger and freight sectors.
ML can also help to improve the operation of 
low-carbon modes, for example by reducing the operations and maintenance costs of rail \cite{JAMSHIDI2018185} and predicting track degradation \cite{doi:10.1177/0954409716657849}. 

Bike sharing and electric scooter services can offer low-carbon alternatives for urban mobility that do not require ownership and integrate well with public transportation. ML studies help to understand how usage patterns for bike stations depend on their immediate urban surroundings \cite{HYLAND201871}.
ML can also help solve the bike sharing rebalancing problem, where shared bikes accumulate in one location and are lacking in other locations, by improving forecasts of bike demand and inventory \cite{REGUE2014192}. Singla \et  \cite{Singla:2015:IUB:2887007.2887108} propose a pricing mechanism based on online learning to provide monetary incentives for bike users to help rebalancing. By producing accurate travel time estimates, ML can provide tools that help to integrate bike shares with other modes of transportation \cite{8005582}.
Many emerging bike and scooter sharing services are dockless, which means that they are parked anywhere in public space and can block sidewalks \cite{anderson2019governing}. ML has been applied to monitor public sentiment about such bike shares via tweets \cite{doi:10.1177/0361198119838982}. ML could also provide tools and information for regulators to ensure that public space can be used by everyone \cite{citylab2019race}. 

Coordination between modes resulting in faster and more reliable transit times could increase the amount of people or goods traveling on low-carbon modes such as rail. 
ML algorithms could be applied to make public transportation faster and easier to use. For example, there is a rich literature exploring ML methods to predict bus arrival times and their uncertainty \cite{altinkaya2013bus, MAZLOUMI2011534}.
Often freight is packaged so that it can switch between different modes of transport easily. Such \emph{intermodal} transportation relies on low-carbon modes such as rail and water for part of the journey \cite{kaack2018decarbonizing}.
ML can contribute by improving predictions of the estimated time of arrival (for example of freight trains \cite{BARBOUR2018211}) or the weight or volume of expected freight (for example for roll-on/roll-off transport -- often abbreviated as Ro-Ro \cite{doi:10.1111/itor.12337}). Intelligent transport systems of different modes could be combined and enable more efficient multimodal freight transportation \cite{kaack2018decarbonizing}.

Some modes with high GHG emissions, such as trucks, can be particularly cost-competitive in regions with lax enforcement of regulation, as they can benefit from overloading and not obeying labor or safety rules \cite{kaack2018decarbonizing}. 
ML can assist public institutions with enforcing their regulations. For example, image recognition can help law enforcement detect overloading of trucks \cite{2018AIPC.1955d0038Z}.

\subsection{Discussion}
Decarbonizing transport is essential to a low-carbon society, and there are numerous applications where ML can make an impact. This is because transportation causes a large share of GHG emissions, but reducing them has been slow and complex. Solutions are likely very technical, are highly dependent on existing infrastructure, and require detailed understanding of passengers' and freight companies' behavior. ML can help decarbonize transportation by providing data, gaining knowledge from data, planning, and automation. Moreover, ML is fundamental to shared mobility, AVs, EVs, and smart public transit, which, with the right incentives, can be used to enable significant reductions in GHG emissions.

 \newpage

\section{Buildings \& Cities\texorpdfstring{\hfill\textit{by Nikola Milojevic-Dupont and Lynn H.~Kaack}}{}}
\label{sec:buildings-cities}
Buildings offer some of the lowest-hanging fruit when it comes to reducing GHG emissions. While the energy consumed in buildings is responsible for a quarter of global energy-related emissions \cite{ipcc_global_2018}, a combination of easy-to-implement fixes and state-of-the-art strategies\footnote{The IPCC classifies mitigation actions in buildings into four categories: \textit{carbon efficiency} (switching to low-carbon fuels or to natural refrigerants); \textit{energy efficiency} (reducing energy waste through insulation, efficient appliances, better heating and ventilation, or other similar measures); \textit{system and infrastructure efficiency} (e.g.~passive house standards, urban planning, and district cooling and heating); and \textit{service demand reduction} (behavioral and lifestyle changes) \cite{lucon_buildings_2014}.} could reduce emissions for existing buildings by up to 90\% \cite{urge2013energy}. It is possible today for buildings to consume almost no energy \cite{OLSTHOORN20191382}.\footnote{There are even high-rise buildings, e.g. the Tower Raiffeisen-Holding N\"O-Vienna office, or large university buildings, e.g.~the Technical University also in Vienna, that achieve such performance.} Many of these energy efficiency measures actually result in overall cost savings \cite{STEPHENSON20106120} and simultaneously yield other benefits, such as cleaner air for occupants. This potential can be achieved while maintaining the services that buildings provide -- and even while extending them to more people, as climate change will necessitate. For example, with the changing climate, more people will need access to air conditioning in regions where deadly heat waves will become common \cite{mora2017twenty,mora2017global}.

Two major challenges are heterogeneity and inertia. Buildings vary according to age, construction, usage, and ownership, so optimal strategies vary widely depending on the context. For instance, buildings with access to cheap, low-carbon electricity may have less need for expensive features such as intelligent light bulbs. Buildings also have very long lifespans; thus, it is necessary both to create new, energy-efficient buildings, and to retrofit old buildings to be as efficient as possible \cite{creutzig_urban_2016}. Urban planning and public policy can play a major role in reducing emissions by providing infrastructure, financial incentives, or energy standards for buildings.\footnote{For example, see the case of New York City, which mandated that building owners collectively reduce their emissions by 40\% by 2040: \url{https://www.nytimes.com/2019/04/17/nyregion/nyc-energy-laws.html}.}

\begin{figure}
    \centering
    \includegraphics[width=\textwidth]{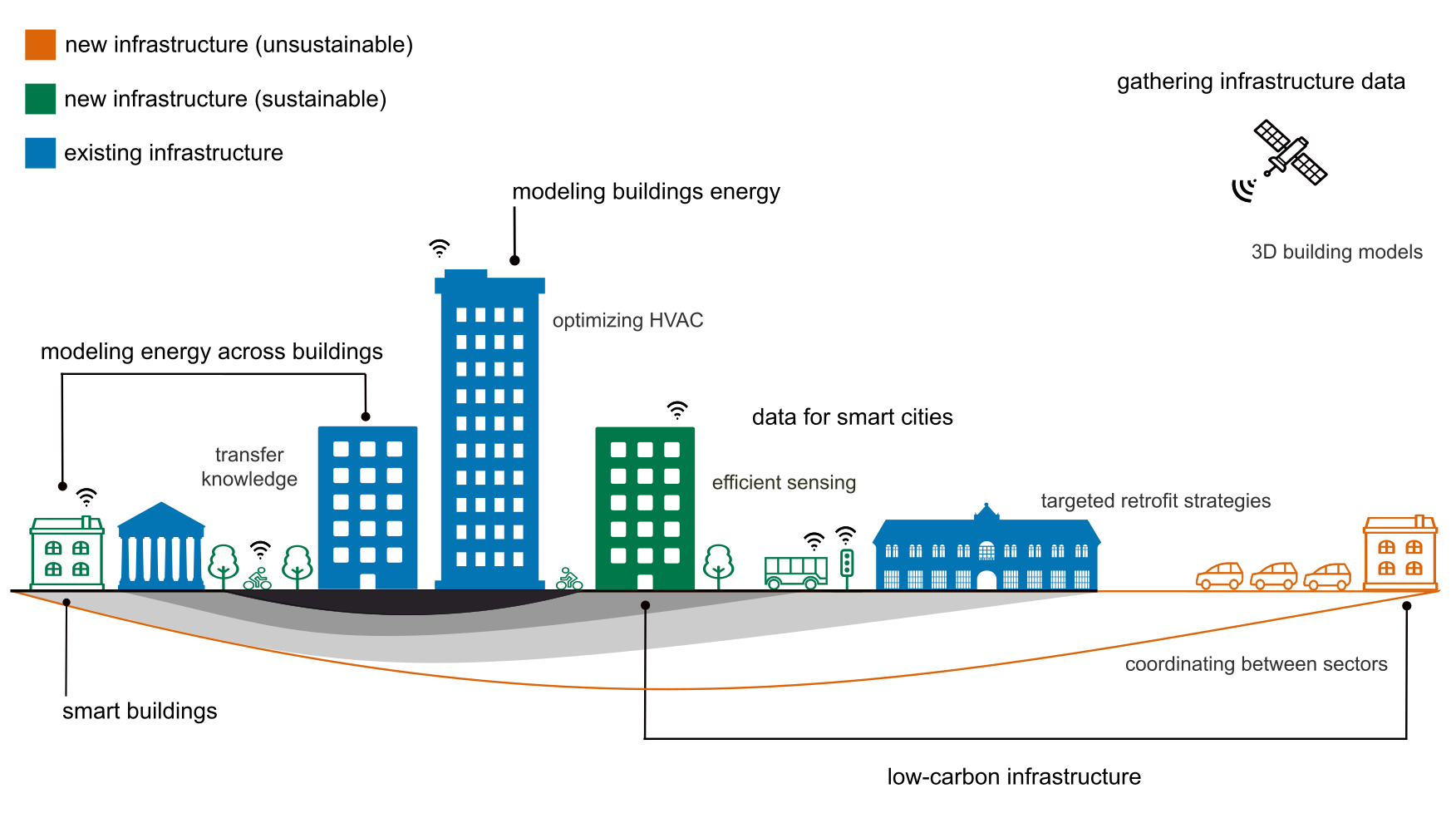}
    \caption{Selected strategies to mitigate GHG emissions from buildings and cities using machine learning.}
    \label{fig:buildingsncities}
\end{figure}

Machine learning provides critical tools for supporting both building managers and  policy makers in their efforts to reduce GHG emissions (Fig.~\ref{fig:buildingsncities}). At the level of building management, ML can help select strategies that are tailored to individual buildings, and can also contribute to implementing those strategies via smart control systems (\S\ref{sec:indv}). At the level of urban planning, ML can be used to gather and make sense of data to inform policy makers (\S\ref{sec:distr}). Finally, we consider how ML can help cities as a whole to transition to low-carbon futures (\S\ref{sec:cities}).  

\subsection{Optimizing buildings}\label{sec:indv}
In designing new buildings and improving existing ones, there are numerous technologies that can reduce GHG emissions, often saving money in the process \cite{lucon_buildings_2014,urge2013energy,OLSTHOORN20191382,STEPHENSON20106120,Gershenfeld:2010aa}. ML can accelerate these strategies by (i) modeling data on energy consumption and (ii) optimizing energy use (in \emph{smart buildings}). 
 
\paragraph{Modeling building energy}\mbox{}\\\label{sec:use}An essential step towards energy efficiency is making sense of the increasing amounts of data produced by meters and home energy monitors (see for example \cite{sense}). This can take the form of energy demand forecasts for specific buildings, which are useful for power companies (\S\ref{sec:electricity-variable}) and in evaluating building design and operation strategies \cite{AMASYALI20181192}. Traditionally, energy demand forecasts are based on models of the physical structure of a building that are essentially massive thermodynamics computations. ML has the potential to speed up these computations greatly, either by ignoring physical knowledge of the building entirely \cite{kreider1995building,paterakis2017deep}, by incorporating it into the computation \cite{dong2016hybrid}, or by learning to approximate the physical model to reduce the need for expensive simulation (\emph{surrogate models}) \cite{VANGELDER2014245}. Learning how to transfer the knowledge gained from modeling one building to another can make it easier to render precise estimations of more buildings. For instance, Mocanu \et \cite{mocanu_unsupervised_2016} modeled building load profiles with reinforcement learning and deep belief networks using data on commercial and residential buildings; they then used approximate reinforcement learning and transfer learning to make predictions about new buildings, enabling the transfer of knowledge from commercial to residential buildings, and from gas- to power-heated buildings.

Within a single building, understanding which appliances drive energy use (\emph{energy disaggregation}) is crucial for targeting efficiency measures, and can motivate behavioral changes. Promising ML approaches to this problem include hidden Markov models \cite{kolter2012approximate}, sparse  coding  algorithms for structured prediction
\cite{kolter2010energy}, harmonic analysis that picks out the ``signatures'' of individual appliances \cite{Srinivasan2006neural}, and deep neural networks \cite{Kelly:2015:NND:2821650.2821672}.

To verify the success or failure of energy efficiency interventions, statistical ML offers methods for causal inference. For example, Burlig \et \cite{burlig2017machine} used Lasso regression on hourly electricity consumption data from schools in California to find that energy efficiency interventions fall short of the expected savings. Such problems could represent a useful application of deep learning methods for counterfactual prediction \cite{hartford2017deep}.

\paragraph{Smart buildings}\Gap\textbf{\Rec}\mbox{}\\\label{sec:bldgopt}Intelligent control systems in buildings can decrease the carbon footprint both by reducing the energy consumed and by providing means to integrate lower-carbon sources into the electricity mix \cite{Gershenfeld1086}. Specifically, ML can reduce energy usage by allowing devices and systems to adapt to usage patterns. Further, buildings can respond to signals from the electricity grid, providing flexibility to the grid operator and lowering costs to the consumer (\S\ref{sec:dispatchDR}). 

Many critical systems inside buildings can be made radically more efficient. While this is also true for small appliances such as refrigerators and lightbulbs, we use the example of heating and cooling (HVAC) systems, both because they are notoriously inefficient and because they account for more than half of the energy consumed in buildings \cite{lucon_buildings_2014}. There are several promising ways to enhance HVAC operating performance, each providing substantial opportunities for using ML: forecasting what temperatures are needed throughout the system, better control to achieve those temperatures, and fault detection. Forecasting temperatures, as with modeling energy use in buildings, has traditionally been performed using detailed physical models of the system involved; however, ML approaches such as deep belief networks can potentially increase accuracy with less computational expense \cite{afroz2018modeling,fu2018deep} (see also \S\ref{sec:demandresponse}). For control, Kazmi \et \cite{kazmi_gigawatt-hour_2018} used deep reinforcement learning to achieve a scalable 20\% reduction of energy while requiring only three sensors: air temperature, water temperature, and energy use (see also \S\ref{sec:demandresponse} for similarly substantial gains in datacenter cooling). Finally, ML can automate building diagnostics and maintenance through fault-detection. For example, the energy efficiency of cooling systems can degrade if refrigerant levels are low \cite{KIM20121805}; ML approaches are well-suited to detect faults in these systems. Wang \et \cite{wang_fault_2017} treated HVAC fault-detection as a one-class classification problem, using only temperature readings for their predictions. Deep autoencoders can be used to simplify information about machine operation so that deep neural networks can then more easily predict multiple kinds of faults \cite{jia2016deep}.

Many systems within buildings -- such as lights and heating -- can also adjust how they operate based on whether a building or room is occupied, thereby improving both occupant comfort and energy use \cite{park2019lightlearn}. ML can help these systems dynamically adapt to changes in occupancy patterns \cite{rashidi2009keeping}. Moreover, occupancy detection itself represents an opportunity for ML algorithms, ranging from decision trees \cite{d2015occupancy,zhao_occupant_2014} to deep neural networks \cite{zou2018towards} that take input from occupancy sensors \cite{d2015occupancy}, WiFi signals \cite{zou2018towards,zou2019unsupervised}, or appliance power consumption data \cite{zhao_occupant_2014}.

In \S\ref{sec:dispatchDR}, we discussed how using variable low-carbon energy can mean that the supply and price of electricity varies over time. Thus, energy flexibility in buildings is increasingly useful to schedule consumption when supply is high \cite{riekstin2018time}.  For this, automated demand-side response \cite{hu2013hardware} can respond to electricity prices, smart meter signals, or learned user preferences \cite{JIN20171583}. Edge computing can be used to process data from distributed sensors and other \emph{Internet of Things} devices, and deep reinforcement learning can then use this data to efficiently schedule energy use \cite{liu2019intelligent}.

While smart building technologies have the capability to significantly increase efficiency, we should note that there are potential drawbacks \cite{hittinger2019internet}. First, smart building devices and connection networks, like wireless sensor networks, consume energy themselves; however, deep neural networks can be used to monitor and optimize their operations \cite{ateeq2019multi}.  Second, rebound effects are likely to happen in certain cases \cite{azevedo2014consumer}, leading to additional building energy consumption typically ranging between 10 and 20\% \cite{rau2018cross}. If control systems optimize for costs, interventions do not necessarily translate into energy efficiency measures or GHG reductions. Therefore, public policies are needed to mandate, support and complement the actions of individual building managers \cite{lucon_buildings_2014}.
Another concern in the case of widespread adoption of smart meters is the impact on mineral use and embodied energy use arising from their production \cite{sias2017characterization}. 
Finally, smart home applications present security and privacy risks \cite{couldry2019data} that require adequate regulation.

\subsection{Urban planning}\label{sec:distr}
For many impactful mitigation strategies -- such as district heating and cooling, neighborhood planning, and large-scale retrofitting of existing buildings -- coordination at the district and city level is essential. Policy makers use instruments such as building codes, retrofitting subsidies, investments in public utilities, and public-private partnerships in order to reduce GHG emissions without compromising equity.  
Where energy-use data on individual buildings exist, ML can be used to derive higher-level patterns. However, many regions of the world have almost no energy consumption data, which can make it difficult to design targeted mitigation strategies. ML is uniquely capable of predicting energy consumption and GHG mitigation potential at scale from other types of available data. 

\paragraph{Modeling energy use across buildings}\label{sec:bldg3d}\mbox{}\\
\emph{Urban Building Energy Models} provide simplified information on the energy use of all buildings across a city. These are different from individual-building models, which model energy use of only specific buildings, but with finer details and temporal granularity (\S\ref{sec:indv}). While UBEMs have yet to be adopted at scale, they are expected to become fundamental for enabling localized action by city planners \cite{reinhart_urban_2016}.\footnote{The startup nam.R is developing a database of all school buildings in France to help inform retrofitting decisions, harmonizing vast amounts of open and proprietary data with ML \cite{namr}.}  UBEMs can for example be used for planning and operating \emph{district heating and cooling}, where a central plant supplies many households in a district. In turn, district heating and cooling reduces HVAC energy consumption and can provide flexible load \cite{VANDERMEULEN2018103}, but it needs large amounts of data at the district level for implementation and operation.

UBEMs include features such as the location, geometries, and various other attributes of interest like building footprint, usage, material, roof type, immediate surroundings etc. ML can be used to held predict energy consumption from such features. For example, Kolter and Ferreira used Gaussian process regression to predict energy use from features such as property class or the presence of central AC \cite{kolter_large-scale_2011}. Based on energy data disclosed by residents of New York City, Kontokosta and colleagues used various ML methods to predict the energy use of the city's 1.1 million buildings \cite{kontokosta2017data}, analyzed the effect of energy disclosure on the demand \cite{papadopoulos2018pattern}, and developed a system for ranking buildings based on energy efficiency \cite{PAPADOPOULOS2019244}. Zhang \et \cite{ZHANG2018162} matched residential energy consumption survey data with public use microdata samples to estimate residential energy consumption at the neighborhood level. Using five commonly accessible features of buildings and climate, Robinson et al.~predict commercial building energy use across large American cities \cite{robinson2017machine}.

Beyond energy prediction, buildings' features can be used by ML algorithms to pinpoint which buildings have the highest retrofit potential. Simple building characteristics and surrounding environmental factors -- both potentially available at scale -- can be used \cite{khayatian2017building, bocher2018geoprocessing}.

There have also been attempts to upscale individual-building energy models to the district scale. Using deep neural networks for hybrid ML-physical modelling, Nutkiewicz \et provided precise energy demand forecasts that account for inter-building energy dynamics and urban microclimate factors for all buildings on a campus \cite{nutkiewicz2018data}.

\paragraph{Gathering infrastructure data}\Gap\textbf{\Rec}\mbox{}\\
\label{sec:bldginfrastructure}Specifics about building infrastructure can often be predicted using ML techniques.  
Remote sensing is key to inferring infrastructure data \cite{esch2017breaking,microsoftbuildings,yu2018deepsolar,LU2014134,Henn2012,blaha2016large} as satellite data\footnote{See \cite{zhu2017deep} for a review of different sources of data and deep learning methods for processing them.} present a source of information that is globally available and largely consistent worldwide. For example, using remote sensing data, Gei\ss \ \et \cite{Gei__2011} clustered buildings into types to assess the potential of district heat in a German town. 

The resolution of infrastructure data ranges from coarse localization of all buildings at the global scale \cite{esch2017breaking}, to precise 3D reconstruction of a neighborhood \cite{blaha2016large}. It is possible to produce a global map of human settlement footprints at meter-level resolution from satellite radar images \cite{esch2017breaking}. For this, Esch \et used highly automated learners, which make classification at such scale possible by retraining locally. Segmentation of high-resolution satellite images can now generate exact building footprints at a national scale \cite{microsoftbuildings}. Energy-relevant building attributes, such as the presence of photovoltaic panels, can also be retrieved from these images \cite{yu2018deepsolar} (see \S\ref{sec:electricity-variable}). 
To generate 3D models, LiDAR data is often used to retrieve heights or classify buildings at city scale \cite{LU2014134,Henn2012}, but its collection is expensive. 
Recent research showed that heights can be predicted even without such elevation data, as demonstrated by \cite{BILJECKI20171}, who predicted these from real estate records and census data. 
Studies, which for now are small scale, aim for complete 3D reconstruction with class labels for different components of buildings \cite{blaha2016large}. 

\subsection{The future of cities}\label{sec:cities}
Since most of the resources of the world are ultimately channeled into cities, municipal governments have a unique opportunity to mitigate climate change. City governments regulate (and sometimes operate) transportation, buildings, and economic activity. They handle such diverse issues as energy, water, waste, crime, health, and noise. Recently, data and ML have become more common for improving efficiency in such areas, giving rise to the notion of \emph{smart city}. While the phrase \emph{smart city} encompasses a wide array of technologies \cite{neirotti2014current}, here we discuss only applications that are relevant to reducing GHG emissions.

\paragraph{Data for smart cities}\Gap\Rec\mbox{}\\
\label{sec:data_pol}Increasingly, important aspects of city life come with digital information that can make the city function in a more coordinated way. Habibzadeh \et \cite{8291118} differentiate between \emph{hard-sensing}, i.e., fixed-location-dedicated sensors like traffic cameras, and \emph{soft-sensing}, for example from mobile devices. Hard sensing is the primary data collection paradigm in many smart city applications, as it is adapted to precisely meet the application requirements. However, there is a growing volume of data coming from soft sensing, due to the widespread adoption of personal devices like smartphones that can provide movement data and geotagged pictures.\footnote{Note that management of any such private data, even if they are anonymized, poses challenges \cite{creutzig_leveraging_2019}.} Urban computing \cite{urban-computing} is an emerging field looking at data analytics in urban spaces, and aiming to yield insights for data-driven policies. For example, clustering anonymized credit card payments makes it possible to model different communities and lifestyles -- of which the sustainability can be assessed \cite{di2018sequences}. Jiang \et provides a review of urban computing from mobile phone traces \cite{jiang2013review}.\footnote{See \url{https://www.microsoft.com/en-us/research/project/urban-computing/} for more applications of urban computing.} Relevant information on the urban space can also be learned from social media activity, e.g.~on Twitter, as reviewed in \cite{ilieva2018social, ruths2014social}.
There are, however, numerous challenges in making sense of this wealth of data (see \cite{mosannenzadeh2017identifying}), and privacy considerations are of paramount importance when collecting or working with many of these data sources.

First, cities need to obtain relevant data on activities that directly or indirectly consume energy. Data are often proprietary. To obtain these data, the city of Los Angeles now requires all \emph{mobility as a service} providers, i.e.~vehicle-sharing companies, to use an open-source API. Data on location, use, and condition of all those vehicles, which can be useful in guiding regulation, are thus transmitted to the city \cite{cityLAgit}. ML can also distill information on urban issues related to climate change through web-scraping and text-mining, e.g.~\cite{doi:10.1177/0361198119838982}. As discussed above (\S\ref{sec:distr}), ML can also be used to infer infrastructure data.

Second, smart city applications must transmit high volumes of data in real-time. ML is key to preprocessing large
amounts of data in large sensor networks, allowing only what is relevant to be transmitted, instead of all the raw
data that is being collected \cite{li2016geospatial, valerio2016hypothesis, ravi2017deep}. Similar techniques also help to reduce the amount of energy consumed during
transmission itself \cite{muhammad2019intelligent}.

Third, urban policy-making based on intelligent infrastructure faces major challenges with data management \cite{Giest_2017}.
Smart cities require the integration of multiple large and heterogeneous sources of data, for which ML can be a valuable tool, which includes data matching \cite{bordes2014semantic,doan2004ontology}, data fusion \cite{methodologies-for-cross-domain}, and ensemble learning \cite{krawczyk2017ensemble}.

\paragraph{Low-emissions infrastructure}\Gap\mbox{}\\\label{sec:smart_cities}When smart city projects are properly integrated into urban planning, they can make cities more sustainable and foster low-carbon lifestyles (see \cite{wu2016big,o2019smart,muhammad2019intelligent} for extensive reviews on this topic). Different types of infrastructure interact, meaning that planning strategies should be coordinated to achieve mitigation goals. For instance, urban sprawl influences the energy use from transport, as wider cities tend to be more car-oriented \cite{ewing_does_2017,creutzig_global_2015,ding2018applying}. ML-based analysis has shown that the development of efficient public transportation is dependent on both the extent of urban sprawl and the local development around transportation hubs \cite{silva2018scenario,monajem2015evaluation}. 

Cities can reduce GHG emissions by coordinating between infrastructure sectors and better adapting services to the needs of the inhabitants. ML and AI can help, for example, to coordinate district heating and cooling networks, solar power generation, and charging stations for electric vehicles and bikes \cite{o2019smart}, and can improve public lighting systems by regulating light intensity based on historical patterns of foot traffic \cite{de2016intelligent}. Due to inherent variability in energy demand and supply, there is a need for uncertainty estimation, e.g.~using Markov chain Monte Carlo methods or Gaussian processes \cite{o2019smart}.

Currently, most smart city projects for urban climate change mitigation are implemented in wealthier regions such as the United States, China, and the EU.\footnote{See for example the European Union H2020 smart cities project  \url{https://ec.europa.eu/inea/en/horizon-2020/smart-cities-communities}.} 
The literature on city-scale mitigation strategies is also strongly biased towards the Global North \cite{lamb2019learning}, while key mitigation challenges are expected to arise from the Global South \cite{nagendra2018urban}. Infrastructure models described in \S\ref{sec:distr} could be used to plan low-carbon neighborhoods without relying on advanced smart city technologies. To transfer strategies across cities, it is possible to cluster similar cities based on climate-relevant dimensions  \cite{han_1_nodate,louf_typology_2014}. Creutzig \et \cite{creutzig_global_2015} related the energy use of 300 cities worldwide to historical structural factors such as fuel taxes (which have a strong impact on urban sprawl).  Other relevant applications include groupings of transportation systems \cite{han_1_nodate} using a latent class choice model, or of street networks \cite{louf_typology_2014} to identify common patterns in urban development using hierarchical clustering.

\subsection{Discussion}
We have shown many different ways that ML can help to reduce GHG emissions from buildings and cities. A central challenge in this sector is the availability of high-quality data for training the algorithms, which rarely go beyond main cities or represent the full spectrum of building types. Techniques for obtaining these data, however, can themselves be an important application for ML (e.g.~via computer vision algorithms to parse satellite imagery). Realizing the potential of data-driven urban infrastructure can advance mitigation goals while improving the well-being of citizens \cite{urge2013energy,bai2018six,creutzig_urban_2016}.


 \newpage
 
\section{Industry\texorpdfstring{\hfill\textit{by Anna Waldman-Brown}}{}}
\label{sec:industry}
Industrial production, logistics, and building materials are leading causes of difficult-to-eliminate GHG emissions \cite{Daviseaas9793}. Fortunately for ML researchers, the global industrial sector spends billions of dollars annually gathering data on factories and supply chains \cite{Gualtieri2016} -- aided by improvements in the cost and accessibility of sensors and other data-gathering mechanisms (such as QR codes and image recognition). The availability of large quantities of data, combined with affordable cloud-based storage and computing, indicates that industry may be an excellent place for ML to make a positive climate impact. 

\begin{figure}[bpht]
    \centering
    \includegraphics[width=\textwidth]{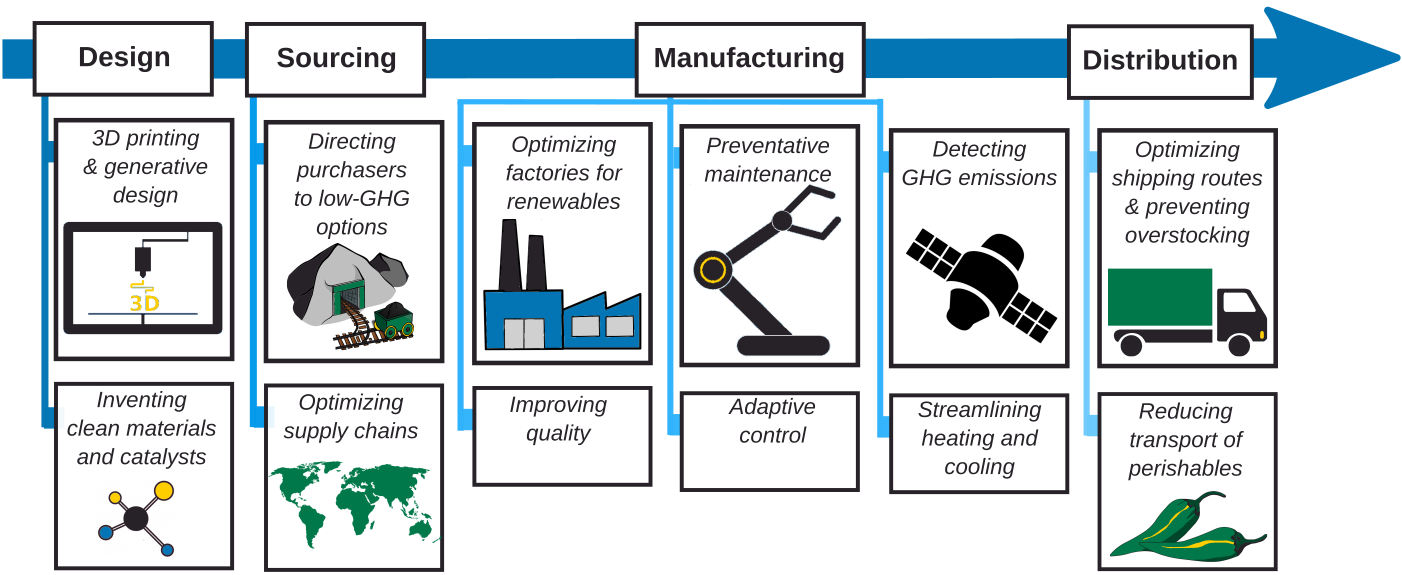}
    \caption{Selected opportunities to use machine learning to reduce greenhouse gas emissions in industry.}
    \label{fig:industry}
\end{figure}

ML demonstrates considerable potential for reducing industrial GHG emissions under the following circumstances:
\begin{itemize}
    \item When there is enough accessible, high-quality data around specific processes or transport routes.
    \item When firms have an incentive to share their proprietary data and/or algorithms with researchers and other firms.
    \item When aspects of production or shipping can be readily fine-tuned or adjusted, and there are clear objective functions.
    \item When firms' incentives align with reducing emissions (for example, through efficiency gains, regulatory compliance, or high GHG prices).
\end{itemize}
In particular, ML can potentially reduce global emissions (Fig.~\ref{fig:industry}) by helping to streamline supply chains, improve production quality, predict machine breakdowns, optimize heating and cooling systems, and prioritize the use of clean electricity over fossil fuels \cite{kazi2017dreamsketch, evans2016deepmind, Zhang2016, Berral2010}.
However, it is worth noting that greater efficiency may increase the production of goods and thus GHG emissions (via the Jevons paradox) unless industrial actors have sufficient incentives to reduce overall emissions \cite{sorrell2009jevons}.

\subsection{Optimizing supply chains}
\label{sec:supplychains}
In 2006, at least two Scottish seafood firms flew hundreds of metric tons of shrimp from Scotland to China and Thailand for peeling, then back to Scotland for sale -- because they could save on labor costs \cite{Cramb2006}. This indicates the complexity of today's globalized \emph{supply chains}, i.e., the organizational processes and shipping networks that are required to bring a product from producer to final consumer. ML can help reduce emissions in supply chains by intelligently predicting supply and demand, identifying lower-carbon products, and optimizing shipping routes. (For details on shipping and delivery optimization, see \S\ref{sec:transportation}.) However, for many of these applications to reduce emissions, firms' financial incentives must also align with climate change mitigation through carbon pricing or other policy mechanisms.

\paragraph*{Reducing overproduction} \Gap \HighRisk\mbox{}\\\label{sec:reducingexcess}The production, shipment, and climate-controlled warehousing of excess products is a major source of industrial GHG emissions, particularly for time-dependent goods such as perishable food or retail goods that quickly fall out of fashion \cite{wang2016energy}. 
Global excess inventory in 2011 amounted to about \$8 trillion worth of goods, according to the Council of Supply Chain Management Professionals \cite{winston2011}. This excess may be in part due to mis-estimation of demand, as the same organization noted that corporate sales estimates diverged from actual sales by an average of 40\% \cite{winston2011}. ML may be able to mitigate these issues of overproducing and/or overstocking goods by improving demand forecasting \cite{akyuz2017ensemble, tsoumakas2019survey}. For example, the clothing industry sells an average of only 60\% of its wares at full price, but some brands can sell up to 85\% due to just-in-time manufacturing and clever intelligence networks \cite{SCMGlobe2016}. As online shopping and just-in-time manufacturing become more prevalent and websites offer more product types than physical storefronts, better demand forecasts will be needed on a regional level to efficiently distribute inventory without letting unwanted goods travel long distances only to languish in warehouses \cite{rizet2010ghg}. Nonetheless, negative side effects can be significant depending on the type of product and regional characteristics; just-in-time manufacturing and online shopping are often responsible for smaller and faster shipments of goods, mostly on road, that lack the energy efficiency of freight aggregation and slower shipping methods such as rail \cite{ugarte2016lean, rizet2010ghg}.

\paragraph*{Recommender systems}\Gap \mbox{}\\\label{sec:recommendersystems}Recommender systems can potentially direct consumers and purchasing firms toward climate-friendly options, as long as one can obtain information about GHG emissions throughout the entire life-cycle of some product. The challenge here lies in hunting down usable data on every relevant material and production process from metal ore extraction through production, shipping, and eventual use and disposal of a product \cite{Hawkins2013,rebitzer2004life}. One must also convince companies to share proprietary data to help other firms learn from best practices. If these datasets can be acquired, ML algorithms could hypothetically assist in identifying the cleanest options.

\paragraph*{Reducing food waste}\Gap \Rec\mbox{}\\\label{sec:foodwaste}Globally, society loses or wastes 1.3 billion metric tons of food each year, which translates to \emph{one-third} of all food produced for human consumption \cite{faofood}. In developing countries, 40\% of food waste occurs between harvest and processing or retail, while over 40\% of food waste in industrialized nations occurs at the end of supply chains, in retail outlets, restaurants, and consumers' homes \cite{faofood}. ML can help reduce food waste by optimizing delivery routes and improving demand forecasting at the point of sale (see \S\ref{sec:supplychains}), as well as improving refrigeration systems \cite{meneghetti2015greening} (see \S\ref{sec:adaptivecontrol}). ML can also potentially assist with other issues related to food waste, such as helping develop sensors to identify when produce is about to spoil, so it can be sold quickly or removed from a storage crate before it ruins the rest of the shipment \cite{fuertes2016intelligent}.

\subsection{Improving materials}
\label{sec:materialsandconstruction}

\paragraph*{Climate-friendly construction}\Gap \Rec\Longterm\mbox{}\\\label{sec:construction}Cement and steel production together account for over 10\% of all global GHG emissions \cite{fischedick2014industry}; the cement industry alone emits more GHGs than every country except the US and China \cite{lehne2018making}. ML can help minimize these emissions by reducing the need for carbon-intensive materials, by transforming industrial processes to run on low-carbon energy, and even by redesigning the chemistry of structural materials. To reduce the use of cement and steel, researchers have combined ML with generative design to develop structural products that require less raw material, thus reducing the resulting GHG emissions \cite{kazi2017dreamsketch}. Novel manufacturing techniques such as 3D printing allow for the production of unusual shapes that use less material but may be impossible to produce through traditional metal-casting or poured concrete; ML and finite element modeling have been used to simulate the physical processes of 3D printing in order to improve the quality of finished products \cite{baturynska}.

Assuming future advances in materials science, ML research could potentially draw upon open databases such as the Materials Project \cite{jain2013commentary} and the UCI Machine Learning Repository \cite{ge2019accelerated} to invent new, climate-friendly materials \cite{ward2016general}. Using semi-supervised generative models and concrete compression data, for example, Ge et al.~proposed novel, low-emission concrete formulas that could satisfy desired structural characteristics \cite{ge2019accelerated}.

\paragraph*{Climate-friendly chemicals}\Gap \Rec\Longterm\mbox{}\\\label{sec:chemicals}Researchers are also experimenting with supervised learning and thermal imaging systems to rapidly identify promising catalysts and chemical reactions \cite{rizkin2019, coley2019graph}, as described in \S\ref{sec:electricity-materials}. Firms are unlikely to adopt new materials or change existing practices without financial incentives, so widespread adoption might require subsidies for low-carbon alternatives or penalties for high GHG emissions.

Ammonia production for fertilizer use relies upon natural gas to heat up and catalyze the reaction, and accounts for around 2\% of global energy consumption \cite{montoya2015challenge}. To develop cleaner ammonia, chemists may be able to invent electrochemical strategies for lower-temperature ammonia production \cite{montoya2015challenge, wood2004review}. Given the potential of ML for predicting chemical reactions \cite{coley2019graph}, ML may also be able to help with the discovery of new materials for electrocatalysts and/or proton conductors to facilitate ammonia production.

\subsection{Production and energy}
\label{sec:demandresponse}
ML can potentially assist in reducing overall electricity consumption; streamlining factories' heating, ventilation, and air conditioning (HVAC) systems; and redesigning some types of industrial processes to run on low-carbon energy instead of coal, oil, or gas. Again, the higher the incentives for reducing carbon emissions, the more likely that firms will optimize for low-carbon energy use. New factory equipment can be very expensive to purchase and set up, so firms' cost-benefit calculations may dissuade them from retrofitting existing factories to run using low-carbon electricity or to save a few kilowatts \cite{gillingham2018cost, plambeck2012reducing, tao2010innovation}. Given the heterogeneity across industrial sectors and the secrecy of industrial data, firms will also need to tailor the requisite sensors and data analysis systems to their individual processes. ML will become a much more viable option for industry when factory workers can identify, develop, implement, and monitor their own solutions internally instead of relying upon outside experts \cite{helper2019profits}. The ML community can assist by building accessible, customizable industry tools tailored for people without a strong background in data science. 

\paragraph*{Adaptive control}\Gap \Rec\mbox{}\\\label{sec:adaptivecontrol}On the production side, ML can potentially improve the efficiency of HVAC systems and other industrial control mechanisms---given necessary data about all relevant processes. To reduce GHG emissions from HVAC systems, researchers have suggested combining optimization-based control algorithms with ML techniques such as image recognition, regression trees, and time delay neural networks \cite{aftab2017automatic, DRGONA2018199} (see also \ref{sec:bldgopt}). DeepMind has used reinforcement learning to optimize cooling centers for Google's internal servers by predicting and optimizing the \emph{power usage effectiveness (PUE)}, thus lowering HFC emissions and reducing cooling costs \cite{evans2016deepmind, gao2014}. Deep neural networks could also be used for adaptive control in a variety of industrial networking applications \cite{ahmed2016green}, enabling energy savings through self-learning about devices' surroundings.

\paragraph*{Predictive maintenance} \Gap\mbox{}\\\label{sec:predictive}ML could also contribute to predictive maintenance by more accurately modelling the wear and tear of machinery that is currently in use, and interpretable ML could assist factory owners in developing a better understanding of how best to minimize GHG emissions for specific equipment and processes. For example, creating a \emph{digital twin} model of some industrial equipment or process could enable a manufacturer to identify and prevent undesirable scenarios, as well as virtually test out a new piece of code before uploading it to the actual factory floor -- thus potentially increasing the GHG efficiency of industrial processes \cite{Glaessgen2012, Tao2018}. Digital twins can also reduce production waste by identifying broken or about-to-break machines before the actual factory equipment starts producing damaged products. Industrial systems can employ similar models to predict which pipes are liable to spring leaks, in order to minimize the direct release of GHGs such as HFCs and natural gas.

\paragraph*{Using cleaner electricity}\Gap \Rec\mbox{}\\\label{sec:electritydemand}ML may be particularly useful for enabling more flexible operation of industrial electrical loads, through optimizing a firm's \emph{demand response} to electricity prices as addressed in \S\ref{sec:electricity-systems}. Such optimization can contribute to cutting GHG emissions as long as firms have a financial incentive to optimize for minimal emissions, maximal low-carbon energy, or minimum overall power usage. Demand response optimization algorithms can help firms adjust the timing of energy-intensive processes such as cement crushing \cite{Zhang2016} and powder-coating \cite{Rockwell} to take advantage of electricity price fluctuations, although published work on the topic has to date used relatively little ML. Online algorithms for optimizing demand response can reduce overall power usage for computer servers by dynamically shifting the internet traffic load of data providers to underutilized servers, although most of this research, again, has focused on minimizing costs rather than GHG emissions \cite{buchbinder2011online, Horner2016}. Berral et al.~proposed a framework that demonstrates how such optimization algorithms might be combined with RL, digitized controls, and feedback systems to enable the autonomous control of industrial processes \cite{Berral2010}.

\subsection{Discussion}
\label{sec:industrydiscussion}
Given the globalized nature of international trade and the urgency of climate change, decarbonizing the industrial sector must become a key priority for both policy makers and factory owners worldwide.  As we have seen, there are a number of highly impactful applications where ML can help reduce GHG emissions in industry, with several caveats. First, incentives for cleaner production and distribution are not always aligned with reduced costs, though policies can play a role in aligning these incentives. Second, despite the proliferation of industrial data, much of the information is proprietary, low-quality, or very specific to individual machines or processes; practitioners estimate that 60-70\% of industrial data goes unused \cite{Coffey2019, Gualtieri2016}. Before investing in extensive ML research, researchers should be sure that they will be able to eventually access and clean any data needed for their algorithms. Finally, misjudgments can be very costly for manufacturers and retailers, leading most managers to adopt risk-averse strategies towards relatively untested technologies such as ML \cite{helper2019profits}. For this reason, ML algorithms that determine industrial activities should be robust enough to guarantee both performance and safety, along with providing both interpretable and reproducible results \cite{henderson2018deep}.

 \newpage
 
\section{Farms \& Forests\texorpdfstring{\hfill\textit{by Alexandre Lacoste}}{}} \label{sec:afolu}
Plants, microbes, and other organisms have been drawing \cd from the atmosphere for millions of years. Most of this carbon is continually broken down and recirculated through the carbon cycle, and some is stored deep underground as coal and oil, but a large amount of carbon is sequestered in the biomass of trees, peat bogs, and soil. Our current economy encourages practices that are freeing much of this sequestered carbon through deforestation and unsustainable agriculture. On top of these effects, cattle and rice farming generate methane, a greenhouse gas far more potent than \cd itself. Overall, land use by humans is estimated to be responsible for about a quarter of global GHG emissions \cite{ipcc2014summary} (and this may be an underestimate \cite{mahowald2017impacts}). In addition to this direct release of carbon through human actions, the permafrost is now melting, peat bogs are drying, and forest fires are becoming more frequent as a consequence of climate change itself -- all of which release yet more carbon \cite{Climate_change_feedback}.

\begin{figure}[hbpt]
    \centering
    \includegraphics{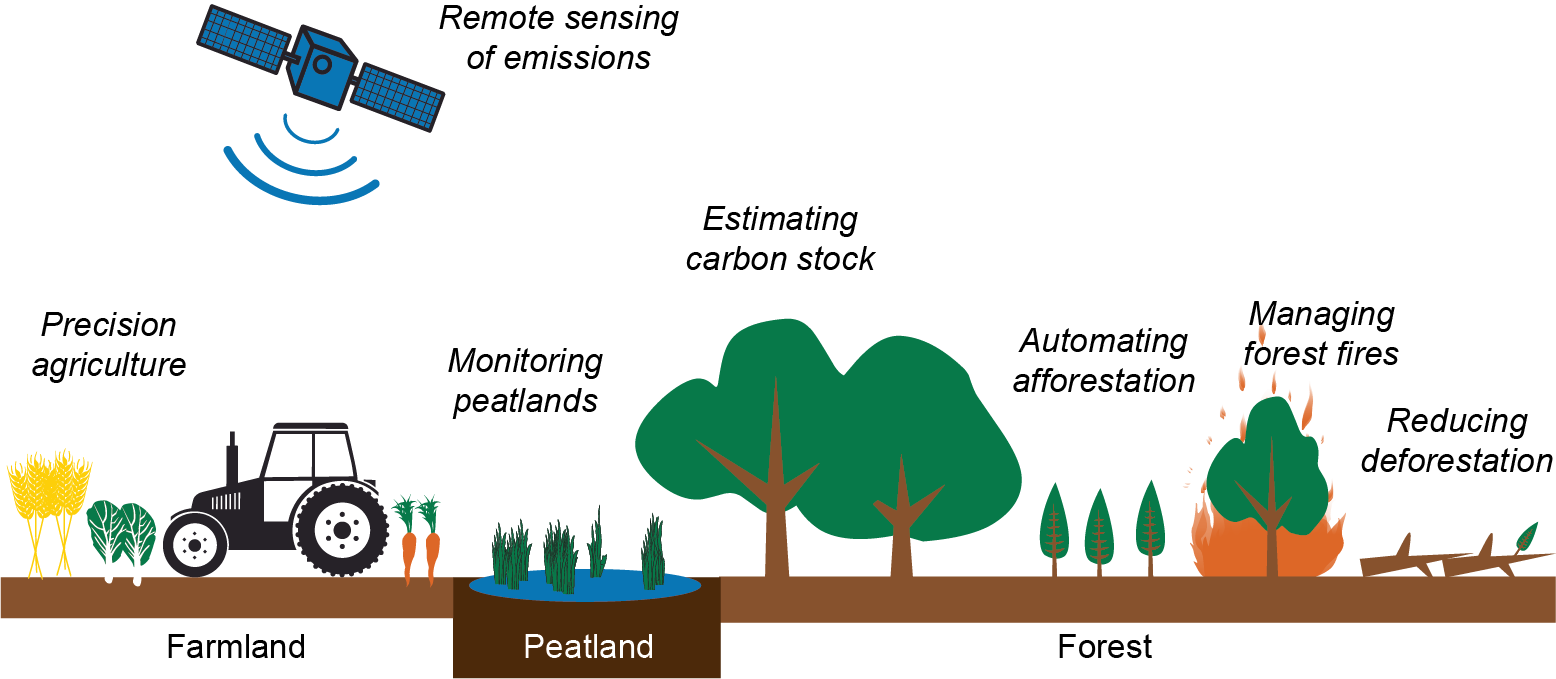}
    \caption{Selected strategies to mitigate GHG emissions from land use using machine learning.}
    \label{fig:agriculture}
\end{figure}

The large scale of this problem allows for a similar scale of positive impact. According to one estimate \cite{hawken2017drawdown}, about a third of GHG emissions reductions could come from better land management and agriculture. ML can play an important role in some of these areas. Precision agriculture could reduce carbon release from the soil and improve crop yield, which in turn could reduce the need for deforestation. Satellite images make it possible to estimate the amount of carbon sequestered in a given area of land, as well as track GHG emissions from it. ML can help monitor the health of forests and peatlands, predict the risk of fire, and contribute to sustainable forestry (Fig.~\ref{fig:agriculture}). These areas represent highly impactful applications, in particular, of sophisticated computer vision tools, though care must be taken in some cases to avoid negative consequences via the Jevons paradox.

\subsection{Remote sensing of emissions\Gap \Rec} \label{sec:emissions-detection}
Having real-time maps of GHGs could help us quantify emissions from agriculture and forestry practices, as well as monitor emissions from other sectors (\S \ref{sec:electricity-currentSystemImpact}).
 
Such information would be valuable in guiding regulations or incentives that could lead to better land use practices. For example, data on emissions make it possible to set effective targets, and pinpointing the sources of emissions makes it possible to enforce regulations.

While greenhouse gases are invisible to our eyes, they must by definition interact with sunlight. This means that we can observe these compounds with hyperspectral cameras \cite{Hyperspectral, scafutto2018detection}. These cameras can record up to several hundred wavelengths (instead of simply RGB), providing information on the interaction between light and individual chemicals. Many satellites are equipped with such cameras and can perform, to some extent, estimations of \cd, CH$_4$ (methane), H$_2$O, and N$_2$O (nitrous oxide) emissions \cite{jacob2016satellite, bernath2017near}. While extremely useful for studying climate change, most of these satellites have very coarse spatial resolution and large temporal and spatial gaps, making them unsuitable for precise tracking of emissions. Standard satellite imagery provides RGB images with much higher resolution, which could be used in an ML algorithm to fill the gaps in hyperspectral data and obtain more precise information about emissions\footnote{Microsatellites with higher resolution hyperspectral cameras are expected to launch over the coming years, including a proposal by Bluefield Technologies that would provide methane detection at 20-meter spatial resolution with daily refresh. Even once this technology comes online, ML will remain useful to cover the daily gaps and to estimate emissions of other GHGs.}. Some preliminary work \cite{jacob2016satellite} has studied this possibility, but there are no clear results as of yet. This is therefore an open problem with high potential impact.

\subsection{Precision agriculture\Gap \Rec
\HighRisk}\label{sec:agriculture}
Agriculture is responsible for about 14\% of GHG emissions \cite{ipcc2014summary}. This might come as a surprise, since plants take up \cd from the air. However, modern industrial agriculture involves more than just growing plants. First, the land is stripped of trees, releasing carbon sequestered there. Second, the process of tilling exposes topsoil to the air, thereby releasing carbon that had been bound in soil aggregates and disrupting organisms in the soil that contribute to sequestration. Finally, because such farming practices strip soil of nutrients, nitrogen-based fertilizers must be added back to the system. Synthesizing these fertilizers consumes massive amounts of energy, about 2\% of global energy consumption \cite{montoya2015challenge} (see \S\ref{sec:chemicals}). Moreover, while some of this nitrogen is absorbed by plants or retained in the soil, some is converted to nitrous oxide,\footnote{Some fertilizer additionally often ends up in waterways, which can contaminate drinking water and induce blooms of toxic algae \cite{robertson2009nitrogen}.} a greenhouse gas that is about 300 times more potent than \cd.  

Such industrial agriculture approaches are ultimately based on making farmland more uniform and predictable. This allows it to be managed at scale using basic automation tools like tractors, but can be both more destructive and less productive than approaches that work with the natural heterogeneity of land and crops. Increasingly, there is demand for sophisticated tools which would allow farmers to work at scale, but adapt to what the land needs. This approach is often known as ``precision agriculture.'' 

Smarter robotic tools can help enable precision agriculture. RIPPA \cite{sukkarieh2017mobile}, a robot under development at the University of Sydney, is equipped with a hyperspectral camera and has the capacity to perform mechanical weeding, targeted pesticide application, and vacuuming of pests. It can cover 5 acres per day on solar energy and collect large datasets \cite{bender2019ladybird} for continual improvement. Many other robotic platforms\footnote{Examples include \url{sagarobotics.com}, \url{ecorobotix.com}, and \url{farm.bot}.} likewise offer opportunities for developing new ML algorithms. There remains significant room for development in this space, since current robots still sometimes get stuck, are optimized only for certain types of crops, and rely on ML algorithms that may be highly sensitive to changes of environment. 

There are many additional ways in which ML can contribute to precision agriculture. Intelligent irrigation systems can save large amounts of water while reducing pests that thrive under excessive moisture \cite{hawken2017drawdown}. ML can also help in disease detection, weed detection, and soil sensing \cite{wahabzada2016plant, liakos2018machine, rossel2016soil}. ML can guide crop yield prediction \cite{you2017deep} and even macroeconomic models that help farmers predict crop demand and decide what to plant at the beginning of the season \cite{ma2019interpretable}. These problems often have minimal hardware requirements, as devices such as Unmanned Aerial Vehicles (UAVs) with hyperspectral cameras can be used for all of these tasks.

Globally, agriculture constitutes a \$2.4 trillion industry \cite{agricultureTrillionIndustry}, and there is already a significant economic incentive to increase efficiency. However, efficiency gains do not necessarily translate into reduced GHG emissions (e.g.~via the Jevons paradox). Moreover, significantly reducing emissions may require a shift in agricultural paradigms -- for example, widespread adoption of regenerative agriculture, silvopasture, and tree intercropping \cite{hawken2017drawdown}. ML tools for policy makers and agronomists \cite{agronomy9030142} could potentially encourage climate-positive action: for example, remote sensing with UAVs and satellites could perform methane detection and carbon stock estimation, which could be used to incentivize farmers to sequester more carbon and reduce emissions. 

\subsection{Monitoring peatlands\Gap\Rec}\label{sec:peatlands}
Peatlands (a type of wetland ecosystem) cover only 3\% of the Earth's land area, yet hold twice the total carbon in all the world's forests, making peat the largest source of sequestered carbon on Earth \cite{parish2008assessment}. When peat dries, however, it releases carbon through decomposition and also becomes susceptible to fire \cite{parish2008assessment,peatfire}. A single peat fire in Indonesia in 1997 is reported to have released emissions comparable to 20-50\% of global fossil fuel emissions during the same year \cite{page2002amount}. 

Monitoring peatlands and protecting them from artificial drainage or droughts is essential to preserve the carbon sequestered in them \cite{joosten2012peatlands, holden2004artificial}. In \cite{minasny2018open}, ML was applied to features extracted from remote sensing data to estimate the thickness of peat and assess the carbon stock of tropical peatlands. A more precise peatlands map is expected to be made by 2020 using specialized satellites \cite{peatlandMap}. Advanced ML could potentially help develop precise monitoring tools at low cost and predict the risk of fire.

\subsection{Managing forests}\label{sec:forests}

\paragraph{Estimating carbon stock}\Gap\textbf{\Rec}\mbox{}\\
Modeling (and pricing) carbon stored in forests requires us to assess how much is being sequestered or released across the planet. Since most of a forest's carbon is stored in above-ground biomass \cite{rodriguez2017quantifying}, tree species and heights are a good indicator of the carbon stock. 

The height of trees can be estimated fairly accurately with LiDAR devices mounted on UAVs, but this technology is not scalable and many areas are closed to UAVs. To address this challenge, ML can be used to predict the LiDAR's outcome from satellite imagery \cite{planetCarbonStock, rodriguez2017quantifying}. From there, the learned estimator can perform predictions at the scale of the planet. Despite progress in this area, there is still significant room for improvement. For example, LiDAR data is often not equally distributed across regions or seasons. Hence domain adaptation and transfer learning techniques may help algorithms to generalize better. 

\paragraph{Automating afforestation}\Gap\textbf{\Longterm\HighRisk}\mbox{}\\
Planting trees, also called \emph{afforestation}, can be a means of sequestering \cd over the long term. According to one estimate, up to 0.9 billion hectares of extra canopy cover could theoretically be added \cite{Bastin76} globally.  However, care must be taken when planting trees to ensure a positive impact. Afforestation that comes at the expense of farmland (or ecosystems such as peat bogs) could result in a net increase of GHG emissions. Moreover, planting trees without regard for local conditions and native species can reduce the climate impact of afforestation as well as negatively affecting biodiversity.  

ML can be helpful in automating large-scale afforestation by locating appropriate planting sites, monitoring plant health, assessing weeds, and analyzing trends. Startups like BioCarbon Engineering\footnote{\url{www.biocarbonengineering.com}} and Droneseed\footnote{\url{www.droneseed.co}} are even developing UAVs that are capable of planting seed packets more quickly and cheaply than traditional methods \cite{biocarbonengineeringInterview}.

\paragraph{Managing forest fires}\Gap\mbox{}\\
Besides their potential for harming people and property, forest fires release \cd into the atmosphere (which in turn increases the rate of forest fires \cite{westerling2016increasing}). On the other hand, small forest fires are part of natural forest cycles. Preventing them causes biomass to accumulate on the ground and increases the chances of large fires, which can then burn all trees to the ground and erode top soil, resulting in high \cd emissions, biodiversity loss, and a long recovery time \cite{montgomery2014agent}.
Drought forecasting \cite{rhee2016drought} is helpful in predicting regions that are more at risk, as is estimating the water content in the tree canopy \cite{brodrick2019forest}. In \cite{ganapathi2018using, ganapathi2018combining}, reinforcement learning is used to predict the spatial progression of fire. This helps firefighters decide when to let a fire burn and when to stop it \cite{houtman2013allowing}. With good tools to evaluate regions that are more at risk, firefighters can perform controlled burns and cut select areas to prevent the progression of fires. 

\paragraph{Reducing deforestation}\Gap 
\textbf{\Rec}\mbox{}\\
Only 17\% of the world's forests are legally protected \cite{macdicken2016global}. The rest are subject to deforestation, which contributes to approximately 10\% of global GHG emissions \cite{ipcc2014summary} as vegetation is burned or decays. While some deforestation is the result of expanding agriculture or urban developments, most of it comes from the logging industry. Clearcutting, which has a particularly ruinous effect upon ecosystems and the carbon they sequester, remains a widespread practice across the world.

Tools for tracking deforestation can provide valuable data for informing policy makers, as well as law enforcement in cases where deforestation may be conducted illegally. ML can be used to differentiate selective cutting from clearcutting using remote sensing imagery \cite{hethcoat2019machine, lippitt2008mapping, baccini2012estimated, defries2002carbon}. Another approach is to install (old) smartphones powered by solar panels in the forest; ML can then be used to detect and report chainsaw sounds within a one-kilometer radius \cite{rfcx}.

Logistics and transport still dominate the cost of wood harvesting, which often motivates clearcutting. Increasingly, ML tools \cite{silviaterra} are becoming available to help foresters decide when to harvest, where to fertilize, and what roads to build. However, once more, the Jevons paradox is at play; making forestry more efficient can have a negative effect by increasing the amount of wood harvested. On the other hand, developing the right combination of tools for regulation and selective cutting could have a significant positive impact.

\subsection{Discussion}
Farms and forests make up a large portion of global GHG emissions, but reducing these emissions is challenging. The scope of the problem is highly globalized, but the necessary actions are highly localized. Many applications also involve a diversity of stakeholders. Agriculture, for example, involves a complex mix of large-scale farming interests, small-scale farmers, agricultural equipment manufacturers, and chemical companies. Each stakeholder has different interests, and each often has access to a different portion of the data that would be useful for impactful ML applications. Interfacing between these different stakeholders is a practical challenge for meaningful work in this area.

 \newpage

\section{Carbon Dioxide Removal\texorpdfstring{\hfill\textit{by Andrew S.~Ross and Evan D.~Sherwin}}{}}
\label{sec:ccs}
Even if we could cut emissions to zero today, we would still face significant climate consequences from greenhouse gases already in the atmosphere. 
Eliminating emissions entirely may also be tricky, given the sheer diversity of sources (such as airplanes and cows). Instead, many experts argue that to meet critical climate goals, global emissions must become net-negative---that is, we must remove more \carbon~from the atmosphere than we release \cite{fuss_betting_2014,gasser2015negative}. Although there has been significant progress in negative emissions research \cite{board2019negative,icef2018roadmap,minx2018negative,fuss2018negative,nemet2018negative}, the actual \carbon~removal industry is still in its infancy. As such, many of the ML applications we outline in this section are either speculative or in the early stages of development or commercialization.

Many of the primary candidate technologies for \carbon~removal directly harness the same natural processes which have (pre-)historically shaped our atmosphere.
One of the most promising methods is simply allowing or encouraging more natural uptake of \carbon~by plants (whose ML applications we discuss in \textsection\ref{sec:afolu}).
Other plant-based methods include bioenergy with carbon capture and \emph{biochar}, where plants are grown specifically to absorb \carbon~and then burned in a way that sequesters it (while creating energy or fertilizer as a useful byproduct) \cite{creutzig2015bioenergy,robledo2017bioenergy,board2019negative}. Finally, the way most of Earth's \carbon~has been removed over geologic timescales is the slow process of mineral weathering, which also initiates further \carbon~absorption in the ocean due to alkaline runoff \cite{schuiling2006enhanced}. These processes can both be massively accelerated by human activity to achieve necessary scales of \carbon~removal \cite{board2019negative}. However, although these biomass, mineral, and ocean-based methods are all promising enough as techniques to merit mention, they may have drawbacks in terms of land use and potentially serious environmental impacts, and (more relevantly for this paper) they would not likely benefit significantly from ML.

\subsection{Direct air capture\Gap \Longterm}
\label{subsec:dac}
Another approach is to build facilities to extract \carbon~from power plant exhaust, industrial processes, or even ambient air \cite{rubin_cost_2015}. While this ``direct air capture'' (DAC) approach faces technical hurdles, it requires little land and has, according to current understanding, minimal negative environmental impacts \cite{creutzig2019mutual}. The basic idea behind DAC is to blow air onto \carbon~sorbents (essentially like sponges, but for gas), which are either solid or in solution, then use heat-powered chemical processes to release the \carbon~in purified form for sequestration \cite{board2019negative, icef2018roadmap}. Several companies have recently been started to pilot these methods.\footnote{\url{https://carbonengineering.com/}}\footnote{\url{https://www.climeworks.com/}}\footnote{\url{https://globalthermostat.com/}}

While \carbon~sorbents are improving significantly \cite{zelevnak2008amine,cashin2018surface}, issues still remain with efficiency and degradation over time, offering potential (though still speculative) opportunities for ML. ML could be used (as in \S\ref{sec:electricity-materials}) to accelerate materials discovery and process engineering workflows \cite{raccuglia2016machine,liu2017materials,gomez2018automatic,butler2018machine} to maximize sorbent reusability and \carbon~uptake while minimizing the energy required for \carbon~release. ML might also help to develop corrosion-resistant components capable of withstanding high temperatures, as well as optimize their geometry for air-sorbent contact (which strongly impacts efficiency \cite{holmes2012air}).

\subsection{Sequestering \carbon \Gap \Rec
\Longterm \HighRisk}
\label{subsubsec: sequestrativervin}
Once \carbon~is captured, it must be sequestered or stored, securely and at scale, to prevent re-release back into the atmosphere.
The best-understood form of \carbon~sequestration is direct injection into geologic formations such as saline aquifers, which are generally similar to oil and gas reservoirs \cite{board2019negative}.
A Norwegian oil company has successfully sequestered \carbon~from an offshore natural gas field in a saline aquifer for more than twenty years \cite{zoback_earthquake_2012}.
Another promising option is to sequester \carbon~in volcanic basalt formations, which is being piloted in Iceland \cite{snaebjornsdottir2016co2}.

Machine learning may be able to help with many aspects of \carbon~sequestration.
First, ML can help identify and characterize potential storage locations. Oil and gas companies have had promising results using ML for subsurface imaging based on raw seismograph traces \cite{araya2018deep}. These models and the data behind them could likely be repurposed to help trap \carbon~rather than release it.
Second, ML can help monitor and maintain active sequestration sites. Noisy sensor measurements must be translated into inferences about subsurface \carbon~flow and remaining injection capacity \cite{celia2015status}; recently, \cite{mo2019deep} found success using convolutional image-to-image regression techniques for uncertainty quantification in a global \carbon~storage simulation study. Additionally, it is important to monitor for \carbon~leaks \cite{moriarty2014rapid}. ML techniques have recently been applied to
monitoring potential \carbon~leaks from wells \cite{chen2018geologic}; computer vision approaches for emissions detection (see \cite{wang2019machine} and \textsection\ref{sec:emissions-detection}) may also be applicable.

\subsection{Discussion}
Given limits on how much more \carbon~humanity can safely emit and the difficulties associated with eliminating emissions entirely, \carbon~removal may have a critical role to play in tackling climate change.
Promising applications for ML in  \carbon~removal include informing research and development of novel component materials, characterizing geologic resource availability, and monitoring underground \carbon~in sequestration facilities.
Although many of these applications are speculative, the industry is growing, which will create more data and more opportunities for ML approaches to help.

 \newpage
 \part*{Adaptation}
 \addcontentsline{toc}{part}{Adaptation}

\section{Climate Prediction\texorpdfstring{\hfill\textit{by Kelly Kochanski}}{}}
\label{sec: climate prediction}

The first global warming prediction was made in 1896, when Arrhenius estimated that burning fossil fuels could eventually release enough CO$_2$ to warm the Earth by $5^\circ$C.
The fundamental physics underlying those calculations has not changed, but our predictions have become far more detailed and precise. The predominant predictive tools are climate models, known as \emph{General Circulation Models (GCMs)} or \emph{Earth System Models (ESMs)}\footnote{Learn about climate modeling from \url{climate.be/textbook} \cite{Goosse} or Climate Literacy, \url{youtu.be/XGi2a0tNjOo}}. These models inform local and national government decisions (see IPCC reports \cite{ipcc2014summary,IPCC2014,ipcc_global_2018}), help people calculate their climate risks (see \textsection\ref{sec:tools-individuals} and \textsection\ref{sec:societal-impacts}) and allow us to estimate the potential impacts of solar geoengineering (see \textsection\ref{sec:geoengineering}).

\begin{figure}[bhpt]
    \centering
    \includegraphics[width=8.7cm]{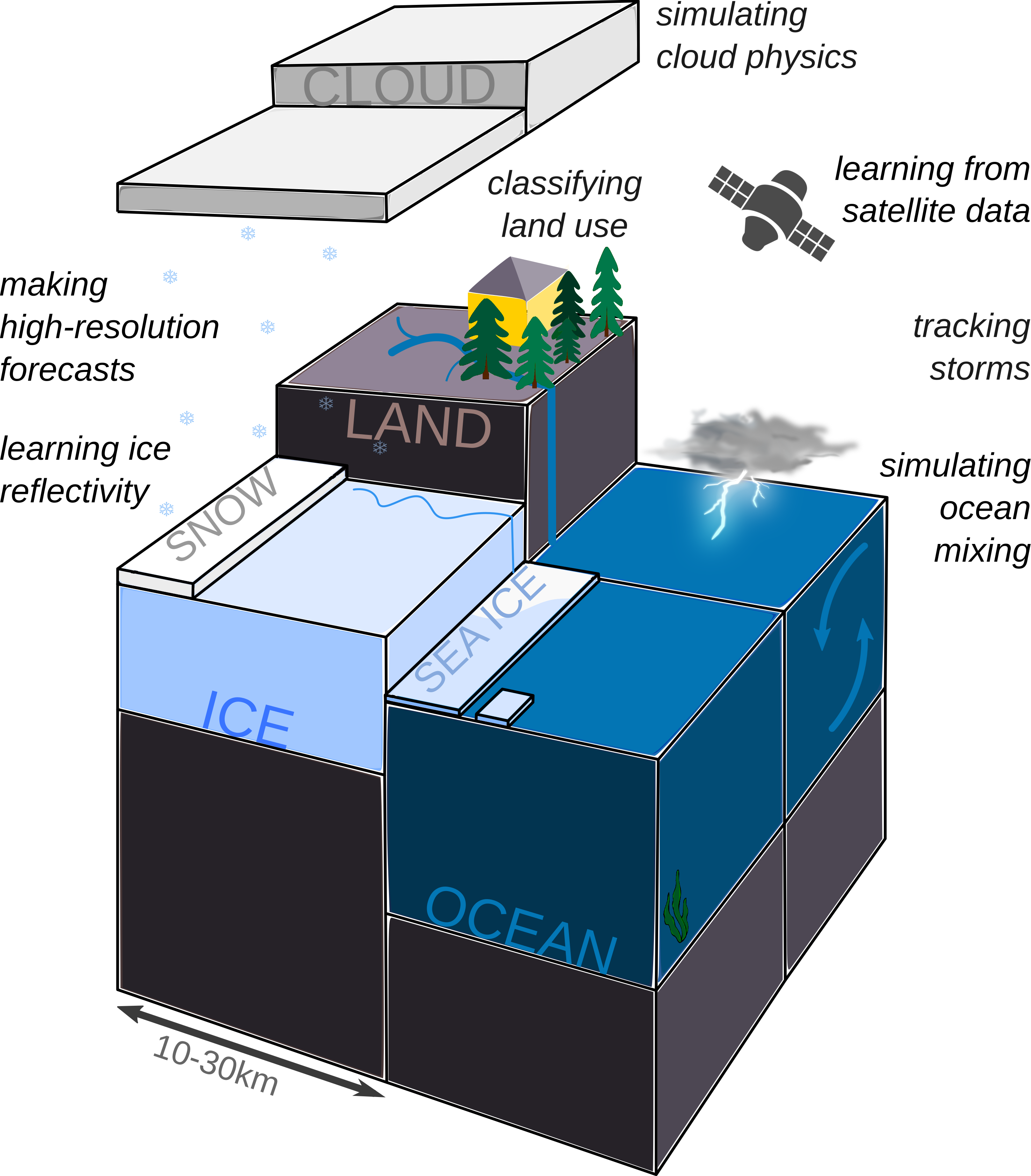}
    \caption{Schematic of a climate model, with selected strategies to improve climate change predictions using machine learning.}
    \label{fig:climateModels}
\end{figure}

Recent trends have created opportunities for ML to advance the state-of-the-art in climate prediction (Fig.~\ref{fig:climateModels}). First, new and cheaper satellites are creating petabytes of climate observation data\footnote{e.g.~NASA's Earth Science Data Systems program, \url{earthdata.nasa.gov}, and ESA's Earth Online, \url{earth.esa.int}}. Second, massive climate modeling projects are generating petabytes of simulated climate data\footnote{e.g.~the Coupled Model Intercomparison Project, \url{cmip.llnl.gov} \cite{Taylor2012, Eyring2016} and Community Earth System Model Large Ensemble \cite{Kay2015}}. Third, climate forecasts are computationally expensive \cite{Carman2017} (the simulations in \cite{Kay2015} took three weeks to run on NCAR supercomputers), while ML methods are becoming increasingly fast to train and run, especially on next-generation computing hardware. As a result, climate scientists have recently begun to explore ML techniques, and are starting to team up with computer scientists to build new and exciting applications.

\subsection{Uniting data, ML, and climate science}
\label{sec:climate-models-params}
Climate models represent our understanding of Earth and climate physics. We can learn about the Earth by collecting data. To turn that data into useful predictions, we need to condense it into coherent, computationally tractable models.
ML models are likely to be more accurate or less expensive than other models where:
(1) there is plentiful data, but it is hard to model systems with traditional statistics, or
(2) there are good models, but they are too computationally expensive to use in production.

\subsubsection{Data for climate models}
When data are plentiful, 
climate scientists build data-driven models. In these areas, ML techniques may solve many problems that were previously challenging.
These include black box problems, for instance sensor calibration \cite{Lary2009}, and classification of observational data, for instance classifying crop cover or identifying pollutant sources in satellite imagery \cite{Lary2015,Kussul2017}.
More applications like these are likely to appear as satellite databases grow. 
The authors of \cite{Monteleoni2013chapter} describe many opportunities for data scientists to assimilate data from diverse field and remote sensing sources, many of which have since been explored by climate informatics researchers.

Numerous authors, such as \cite{Gil2019}, have identified geoscience problems that would be aided by the development of benchmark datasets. Efforts to develop such datasets include
EnviroNet \cite{Mukkavilli2019}, the IS-GEO benchmark datasets \cite{Ebert2017}, and ExtremeWeather \cite{Racah2017}. 
We expect the collection of curated geoscience datasets to continue to grow;
this process might even be accelerated by ML optimizations in data collection systems \cite{Gil2019}.
We strongly encourage modellers to dive into the data in collaboration with domain experts.
We also recommend that modellers who seek to learn directly from data see \cite{Hourdin2017} for specific advice on fitting and over-fitting climate data.

\subsubsection{Accelerating climate models}
Many climate prediction problems are irremediably data-limited. No matter how many weather stations we construct, how many field campaigns we run, or how many satellites we deploy, the Earth will generate at most one year of new climate data per year. Existing climate models deal with this limitation by relying heavily on physical laws, such as thermodynamics.
These models are structured in terms of coupled partial differential equations that represent physical processes like cloud formation, ice sheet flow, and permafrost melt.
ML models provide new techniques for solving such systems efficiently.

\paragraph*{Clouds and aerosols}\Gap\textbf{\Rec}\mbox{}\\
Recent work has shown how deep neural networks could be combined with existing thermodynamics knowledge to fix the largest source of uncertainty in current climate models: clouds. Bright clouds block sunlight and cool the Earth; dark clouds catch outgoing heat and keep the Earth warm \cite{IPCC2014, Sherwood2014}. These effects are controlled by small-scale processes such as cloud convection and atmospheric aerosols (see uses of aerosols for cloud seeding and solar geoengineering in \textsection\ref{sec:geoengineering}). Physical models of these processes are far too computationally expensive to include in global climate models --- but ML models are not. Gentine et al.~trained a deep neural network to emulate the behavior of a high-resolution cloud simulation, and found that the network gave similar results for a fraction of the cost \cite{Gentine2018} and was stable in a simplified global model \cite{Rasp2018}. Existing scientific model structures do not always offer great trade-offs between cost and accuracy. Neural networks trained on those scientific models produce similar predictions, but offer an entirely new set of compromises between training cost, production cost, and accuracy. Replacing select climate model components with neural network approximators may thus improve both the cost and the accuracy of global climate models. Additional work is needed to identify more climate model components that could be replaced by neural networks (we highlight other impactful components below), to optimize those models, and to automate their training workflows (see examples in \cite{Reichstein2019}).

\paragraph{Ice sheets and sea level rise}\Gap\textbf{\Rec}\mbox{}\\
The next most important targets for climate model improvements are ice sheet dynamics and sea level rise. The Arctic and Antarctic are warming faster than anywhere else on Earth, and their climates control the future of global sea level rise and many vulnerable ecosystems \cite{ipcc2014summary, ipcc_global_2018}. Unfortunately, these regions are dark and cold, and until recently they were difficult to observe. In the past few years, however, new satellite campaigns have illuminated them with hundreds of terabytes of data\footnote{See e.g.~\url{icebridge.gsfc.nasa.gov} and \url{pgc.umn.edu/data/arcticdem}.}. These data could make it possible to use ML to solve some of the field's biggest outstanding questions. In particular, models of mass loss from the Antarctic ice-sheet are highly uncertain \cite{Kopp2017} and models of the extent of Antarctic sea ice do not match reality well \cite{Gagne2015}. The most uncertain parts of these models, and thus the best targets for improvement, are snow reflectivity, sea ice reflectivity, ocean heat mixing and ice sheet grounding line migration rates \cite{Hourdin2017,Kopp2017,Hanna2013}. Computer scientists who wish to work in this area could build models that learn snow and sea ice properties from satellite data, or use new video prediction techniques to predict short-term changes in the sea ice extent.

\subsubsection{Working with climate models}
ML could also be used to identify and leverage relationships between climate variables. Pattern recognition and feature extraction techniques could allow us to identify more useful connections in the climate system, and regression models could allow us to quantify non-linear relationships between connected variables. For example, Nowack et al.~demonstrated that ozone concentrations could be computed as a function of temperature, rather than physical transport laws, which led to considerable computational savings \cite{Nowack2018}.

The best climate predictions are synthesized from ensembles of 20+ climate models \cite{Tebaldi2007}. Making good ensemble predictions is an excellent ML problem. Monteleoni et al.~proposed that online ML algorithms could create better predictions of one or more target variables in a multi-model ensemble of climate models \cite{Monteleoni2011}; this idea has been refined in \cite{Mcquade2012,Strobach2015}. More recently, Anderson and Lucas used random forests to make high-resolution predictions from a mix of high- and low-resolution models, which could reduce the costs of building multi-model ensembles \cite{Anderson2018}.

In the further future, the Climate Modeling Alliance has proposed to build an entirely new climate model that learns continuously from data and from high-resolution simulations \cite{Schneider2017}. The proposed model would be written in Julia, in contrast to existing models which are mostly written in C++ and Fortran. At the cost of a daunting translation workload, they aim to build a model that is more accessible to new developers and more compatible with ML libraries.

\subsection{Forecasting extreme events}
\label{sec:models-extreme-events}
For most people, extreme event prediction means the local weather forecast and a few days' warning to stockpile food, go home, and lock the shutters.
Weather forecasts are shorter-term than climate forecasts, but they produce abundant data. Weather models are optimized to track the rapid, chaotic changes of the atmosphere; since these changes are fast, tomorrow's weather forecast is made and tested every day. Climate models, in contrast, are chaotic on short time scales, but their long-term trends are driven by slow, predictable changes of ocean, land, and ice (see \cite{Shukla1998})\footnote{This is one of several reasons why climate models produce accurate long-term predictions in spite of atmospheric chaos.}. As a result, climate model output can only be tested against long-term observations (at the scale of years to decades).  Intermediate time scales, of weeks to months, are exceptionally difficult to predict, although Cohen et al.~\cite{Cohen2018} argue that machine learning could bridge that gap by making good predictions on four to six week timescales \cite{Hwang2019}. Thus far, however, weather modelers have had hundreds of times more test data than climate modelers, and began to adopt ML techniques earlier. Numerous ML weather models are already running in production. For example, Gagne et al.~recently used an ensemble of random forests to improve hail predictions within a major weather model \cite{Gagne2017}. 

A full review of the applications of ML for extreme weather forecasting is beyond the scope of this article. Fortunately, that review has already been written: see \cite{McGovern2017using}.
The authors describe ML systems that correct bias, recognize patterns, and predict storms. Moving forward, they envision human experts working alongside automated forecasts.

\subsubsection{Storm tracking}
Climate models cannot predict the specific dates of future events, but they can predict changes in long-term trends like drought frequency and storm intensity. Information about these trends helps individuals, corporations and towns make informed decisions about infrastructure, asset valuation and disaster response plans (see also \S\ref{subsub:crisis}). Identifying extreme events in climate model output, however, is a classification problem with a twist: all of the available data sets are strongly skewed because extreme events are, by definition, rare. ML has been used successfully to classify some extreme weather events.
Researchers have used deep learning to classify \cite{Liu2016}, detect \cite{Racah2017} and segment \cite{Kurth2018} cyclones and atmospheric rivers, as well as tornadoes \cite{Lakshmanan2010}, in historical climate datasets.
Tools for more event types would be useful, as would online tools that work within climate models, labelled datasets for predicting future events, and statistical tools that quantify the uncertainty in new extreme event forecasts.

\subsubsection{Local forecasts \Gap \Rec}
Forecasts are most actionable if they are specific and local. ML is widely used to make local forecasts from coarse 10--100 km climate or weather model predictions; various authors have attempted this using support vector machines, autoencoders, Bayesian deep learning, and super-resolution convolutional neural networks (e.g.~\cite{Li2019}). Several groups are now working to translate high-resolution climate forecasts into risk scenarios. For example, ML can predict localized flooding patterns from past data \cite{perignon2018patterns}, which could inform individuals buying insurance or homes. Since ML methods like neural networks are effective at predicting local flooding during extreme weather events \cite{sit2019decentralized}, these could be used to update local flood risk estimates to benefit individuals. The start-up Jupiter Intelligence
is working to make climate predictions more actionable by translating climate forecasts into localised flood and temperature risk scores.

\subsection{Discussion}
\label{sec: climate models - ml+science}
 ML may change the way that scientific modeling is done. The examples above have shown that many components of large climate models can be replaced with ML models at lower computational costs. 
 From an ML standpoint, learning from an existing model has many advantages: modelers can generate new training and test data on-demand, and the new ML model inherits some community trust from the old one. This is an area of active ML research.
 Recent papers have explored data-efficient techniques for learning dynamical systems \cite{Raissi2018}, including physics-informed neural networks \cite{Raissi2017} and neural ordinary differential equations \cite{chen2018neural}. In the further future, researchers are developing ML approaches for a wide range of scientific modeling challenges, including crash prediction \cite{Lucas2013}, adaptive numerical meshing \cite{Jiang2016}, uncertainty quantification \cite{Ling2015, Lakshminarayanan2017} and performance optimization \cite{Thiagarajan2018}. If these strategies are effective, they may solve some of the largest structural challenges facing current climate models.
 
New ML models for climate will be most successful if they are closely integrated into existing scientific models. This has been emphasized, again and again, by authors who have laid future paths for artificial intelligence within climate science \cite{Lary2018,Gil2019,Rasp2018,McGovern2017using,Reichstein2019,Schneider2017}. New models need to leverage existing knowledge to make good predictions with limited data. In ten years, we will have more satellite data, more interpretable ML techniques, hopefully more trust from the scientific community, and possibly a new climate model written in Julia. For now, however, ML models must be creatively designed to work within existing climate models. The best of these models are likely to be built by close-knit teams including both climate and computational scientists.

     \newpage

\section{Societal Impacts\texorpdfstring{\hfill\textit{by Kris Sankaran}}{}}
\label{sec:societal-impacts}

Changes in the atmosphere have impacts on the ground.
The expected societal impacts of climate change include prolonged ecological and
socioeconomic stresses as well as brief, but severe, societal disruptions. For example, impacts could include both gradual decreases in crop yield and localized food shortages.
If we can anticipate climate impacts well enough, then we can prepare for
them by asking:

\begin{itemize}
\item How do we reduce vulnerability to climate impacts?
\item How do we support rapid recovery from climate-induced disruptions?
\end{itemize}

\begin{figure}[hppt]
    \centering
    \includegraphics[width=\textwidth]{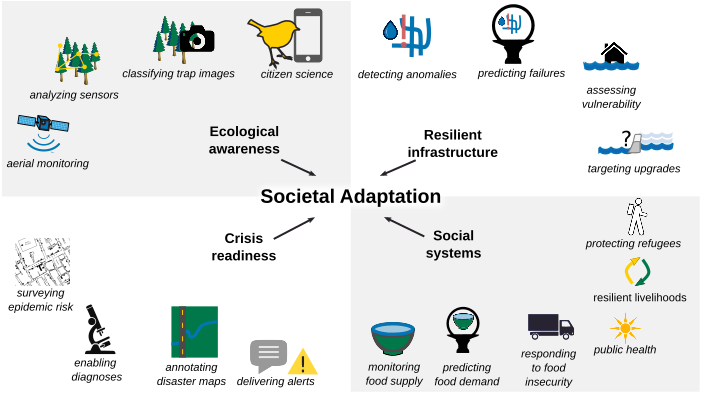}
    \caption{Selected strategies to accelerate societal adaptation to climate change using machine learning.}
    \label{fig:society}
\end{figure}

A wide variety of strategies have been put forward, from robust power grids to
food shortage prediction (Fig.~\ref{fig:society}), and while this is good news for society, it can be
overwhelming for an ML practitioner hoping to contribute. Fortunately, a few
critical needs tend to recur across strategies -- it is by meeting
these needs that ML has the greatest potential to support societal adaptation
\cite{ford2016opinion, quinn2014computational, kelling2018computational}.
From a high level, these involve
\begin{itemize}
\item Sounding alarms: Identifying and prioritizing the areas of highest risk, by using evidence of risk from historical data.
\item Providing annotation: Extracting actionable information or labels from unstructured raw data.
\item Promoting exchange: Making it easier to share resources and information to pool and reduce risk.
\end{itemize}

These unifying threads will appear repeatedly in the sections below, where we review strategies to help ecosystems, infrastructure, and societies adapt to climate change, and explain how ML supports each strategy (Fig.~\ref{fig:society}).

We note that the projects involved vary in scale from local to global, from infrastructure upgrades
and crisis preparedness planning to international ecosystem monitoring and
disease surveillance. Hence, we anticipate valuable contributions by
researchers who have the flexibility to formulate experimental
approaches, by industrial engineers and entrepreneurs who have the expertise to
translate prototypes into wide-reaching systems, and by civil servants who lead
many existing climate adaptation efforts.

\subsection{Ecology}
\label{subsub:ecology}

Changes in climate are increasingly affecting the distribution and composition of ecosystems. This has profound implications for global biodiversity, as well as agriculture, disease, and natural resources such as wood and fish.
ML can  help by supporting efforts to monitor ecosystems and biodiversity.

\paragraph*{Monitoring ecosystems}\Gap\textbf{\Rec}\mbox{}\\ To preserve
ecosystems, it is important to know which are most at risk. This has
traditionally been done via manual, on-the-ground observation, but the process
can be accelerated by annotation of remote sensing data
\cite{potter2008terrestrial,boriah2008land,malkin2018label,bragilevsky2017deep} (see also \S\ref{sec:emissions-detection}). For example,
tree cover can be automatically extracted from aerial imagery to characterize
deforestation \cite{mcdowell2015global, huynh2018annotation}. At the scale of
regions or biomes, analysis of large-scale simulations can illuminate the
evolution of ecosystems across potential climate futures
\cite{klausmeyer2009climate, feng2018improving}. A more direct source of data is
offered by environmental sensor networks, made from densely packed but low-cost
devices \cite{hart2006environmental, dietterich2009machine, hut2012tahmo}. To
monitor ocean ecosystems, marine robots are useful, because they can be used to
survey large areas on demand \cite{griffiths1998towards, dunbabin2012robots}.

For a system to have the most real-world impact, regardless of the underlying
data source, it is necessary to ``personalize'' predictions across a range of
ecosystems. A model trained on the Sahara would almost certainly fail if
deployed in the Amazon. Hence, these applications may motivate ML researchers
interested in heterogeneity, data collection, transfer learning, and rapid
generalization. In sensor networks, individual nodes fail frequently, but are
redundant by design -- this is an opportunity for research into anomaly
detection and missing data imputation \cite{dereszynski2012probabilistic,hill2010anomaly}. In marine robotics, improved techniques for sampling regions
to explore and automatic summarization of expedition results would both provide
value \cite{das2015data, flaspohler2017feature}. Finally, beyond aiding
adaptation by prioritizing at-risk environments, the design of effective methods
for ecosystem monitoring will support the basic science necessary to shape
adaptation in the long-run \cite{faghmous2014big,gomes2009computational,marotzke2017climate}.

\paragraph*{Monitoring biodiversity }\Gap\textbf{\Rec}\mbox{}\\
Accurate estimates of species populations are the foundation on which conservation efforts are built. Camera traps and aerial imagery have increased the richness and coverage of sampling efforts. ML can help infer biodiversity counts from image-based sensors.
For instance, camera traps take photos automatically whenever a motion sensor is activated -- computer vision can be used to classify the species that pass by, supporting a real-time, less labor-intensive species count \cite{zamba, beery2019synthetic, norouzzadeh2018automatically}. It is also possible to use aerial imagery to estimate the size of large herds \cite{van2014nature} or count birds \cite{ghioca2008assessing}. In underwater ecosystems, ML has been used to identify plankton automatically from underwater cameras \cite{faillettaz2016imperfect} and to infer fish populations from the structure of coral reefs \cite{young2018convolutional}.

Citizen science can also enable dataset collection at a scale impossible in individual studies \cite{sullivan2009ebird, plantsnap, branchini2015using, menon2016animal}. For example, by leveraging public enthusiasm for birdwatching, eBird has logged more than 140 million observations \cite{sullivan2009ebird}, which have been used for population and migration studies \cite{kelly2016novel}. Computer vision algorithms that can classify species from photographs have furthered such citizen science efforts by making identifications easier and more accurate \cite{van2015building, ralls2018systems}, though these face challenges such as class imbalances in training data \cite{van2017devil}. Work with citizen science data poses the additional challenge that researchers have no control over where samples come from. To incentivize observations from undersampled regions, mechanisms from game theory can be applied \cite{xue2016avicaching}, and even when sampling biases persist, estimates of dataset shift can minimize their influence \cite{chen2018bias}.

Monitoring biodiversity may be paired with interventions to protect rare species or control invasive pests. Machine learning is providing new solutions to assess the impact of ecological interventions \cite{rana2018machine, albers2018role, lydakiscomputing} and prevent poaching \cite{xue2016avicaching}.

\subsection{Infrastructure} \label{subsub:infrastructure}

Physical infrastructure is so tightly woven into the fabric of everyday life -- like the buildings we inhabit and lights we switch on -- that it is easy to forget that it exists (see \S\ref{sec:buildings-cities}). The fact that something so basic will have to be rethought in order to adapt to climate change can be unsettling, but viewed differently, the sheer necessity of radical redesign can inspire creative thinking.

We first consider the impacts of climate change on the built environment. Shifts in weather patterns are likely to put infrastructure under more persistent stress. Heat and wind damage roads, buildings, and power lines. Rising water tables near the coast will lead to faults in pipelines. Urban heat islands will be exacerbated and it is likely that there will be an increased risk of flooding caused by heavy rain or coastal inundations, resulting in property damage and traffic blockages\cite{Pachauri2014}.

A clear target is construction of physical defenses -- for example, ``climate proofing'' cities with new coastal embankments and increased storm drainage capacity.  However, focusing solely on defending existing structures can stifle proactive thinking about urban and social development -- for example, floating buildings are being tested in Rotterdam -- and one may alternatively consider resilience and recovery more broadly \cite{pelling2010adaptation, shi2016roadmap}. From this more general perspective  of improving social processes, ML can support two types of activities: Design and maintenance.

\paragraph*{Designing infrastructure}\Gap\textbf{\Longterm}\mbox{}\\ How can infrastructure be (re)designed to dampen climate impacts?
In road networks, it is possible to incorporate flood hazard and traffic
information in order to uncover vulnerable stretches of road, especially those with few
alternative routes \cite{gupta2018infrastructure}. If traffic data are not
directly available, it is possible to construct  proxies from mobile
phone usage and city-wide CCTV streams -- these are promising in rapidly developing
urban centers \cite{frias2012estimation, jain2012road}. Beyond drawing from
flood hazard maps, it is possible to use data from real-world flooding events
\cite{pastor2014flooding}, and to send localized predictions to those at risk 
\cite{wiesel2018ml}.
For electrical, water, and waste collection networks, the same
principle can guide investments in resilience -- using proxy or historical data
about disruptions to anticipate vulnerabilities 
\cite{oshri2018infrastructure, muharemi2019machine, nateghi2018multi, panteli2015grid}. 
Robust components can replace those at risk; for example, \emph{adaptive islands}, parts of an energy grid that continue to provide power even when disconnected from the network, 
prevent cascading outages in power distribution \cite{fang2012smart}.

Infrastructure is long-lived, but the future is uncertain, and planners must weigh immediate resource costs against future societal risks \cite{fletcher2019learning}. One area that urgently needs adaptation strategies is the consistent access to drinking water, which can be jeopardized by climate variability \cite{delpla2009impacts, unihe}. Investments in water infrastructure can be optimized; for example, a larger dam might cost more up front, but would have a larger storage capacity, giving a stronger buffer against drought. To delay immediate decisions, infrastructure can be upgraded in phases -- the technical challenge is to discover policies that minimize a combination of long-term resource and societal costs under plausible climate futures, with forecasts being updated as climates evolve \cite{quinn2017direct, giuliani2015curses, shen2017trans}.

\paragraph*{Maintaining infrastructure}\Gap\textbf{\Rec}\mbox{}\\ What types of systems can keep infrastructure functioning
well under increased stress? Two strategies for efficiently managing limited
maintenance resources are predictive maintenance and anomaly detection;
both can be applied to electrical, water, and transportation
infrastructure. In predictive maintenance, operations are prioritized according
to the predicted probability of a near-term breakdown \cite{rudin2012machine,
nguyen2018automatic, srivastava2018design, dragomir2009review}. For anomaly detection, failures are discovered as soon as
they occur, without having to wait for inspectors to show up, or complaints to
stream in \cite{baig2011use, difallah2013scalable}. 

The systems referenced here have required the manual curation of data streams,
structured and unstructured. The data are plentiful, just difficult to glue
together. Ideas from the missing data, multimodal data, and AutoML communities
have the potential to resolve some of these issues.

\subsection{Social systems}
\label{subsub:social_systems}

While less tangible, the social systems we construct are just as critical to the
smooth functioning of society as any physical infrastructure, and it is
important that they adapt to changing climate conditions. First, consider
what changes these systems may encounter. Decreases in crop yield, due to drought and other factors, will pose a threat to food
security, as already evidenced by long periods of drought in North America, West
Africa and East Asia \cite{ipcc_foodsec, dai2011drought}. More generally, communities
dependent on ecosystem resources will find their livelihoods at risk, and this
may result in mass migrations, as people seek out more supportive environments.

At first, these problems may seem beyond the reach of algorithmic thinking,
but investments in \textit{social} infrastructure can increase
resilience. ML can amplify the reach and effectiveness of this infrastructure. See also \S{\ref{sec:toolsforsociety}} for
perspective on how ML can support the function and analysis of complex
social environments.

\paragraph*{Food security}\Gap\textbf{\Rec}\mbox{}\\
Data can be used to monitor the risk of food insecurity in real time, to forecast
near-term shortages, and to identify areas at risk in the long-term, all of which
can guide interventions. For real-time and near-term systems, it is
possible to distill relevant signals from mobile phones, credit card
transactions, and social media data \cite{decuyper2014estimating,
pulse2015using, kim2017nowcasting}. These have emerged as low-cost, high-reach
alternatives to manual surveying. The idea is to train models that link these
large, but decontextualized, data with ground truth consumption or survey
information, collected on small representative samples. This process of 
developing proxies to link small, rich datasets with large, coarse ones 
can be viewed as a type of semi-supervised learning, and is fertile ground
for research.

For longer-term warnings, spatially localized crop yield predictions are needed.
These can be generated by aerial imagery or meteorological data (see \S\ref{sec:agriculture}), if they can be
linked with historical yield data \cite{Chakraborty2011, wang2018deep}. 
On the ground, it is possible to perform crop-disease
identification from plant photos -- this can alert communities to disease
outbreaks, and enhance the capacity of agricultural inspectors. For even
longer-run risk evaluation, it is possible to simulate crop yield via biological
and ecological models \cite{tebaldi2008towards,rosenzweig2014assessing,konduri_kumar_hoffman_bhatia_gouthier_ganguly},
presenting another opportunity for blending large scale simulation with
ML \cite{paganini2018accelerating, welling2015ml}.

Beyond sounding alarms, ML can improve resilience of food supply chains. As
detailed in \S\ref{sec:industry}, ML can reduce waste along these chains;
we emphasize that for adaptation, it is important that supply chains also be made
robust to unexpected disruptions \cite{rancourt2015tactical,
prasad_vuyyuru_gupta, drivendata_supply, mwebaze2010causal}.

\paragraph*{Resilient livelihoods}\Gap\mbox{}\\Individuals whose
livelihoods depend on one activity, and who have less access to community
resources, are those who are most at risk \cite{agrawal2009climate,
rodima2012social}. Resilient livelihoods can be promoted through increased diversification, cooperation, and exchange, all of which can be facilitated by ML systems. 
For example, they can guide equipment and information sharing in farming cooperatives,
via growers' social networks \cite{assefa}. Mobile money efforts can increase access to
liquid purchasing power; they can also be used to monitor economic health \cite{un_global_pulse_2013, frias2012computing}.
Skill-matching programs and online training are often driven by data, with some
programs specifically aiming to benefit refugees \cite{marivate2017employment, bansak2018improving, un_global_pulse_2017_jakarta} (see also \S{\ref{sec:education}}).

\paragraph*{Supporting displaced people}\Gap\textbf{\Longterm\HighRisk}\mbox{}\\ Human populations move in response to threats and opportunities, and ML can be used to predict large-scale migration patterns. Work in this area has relied on
accessible proxies, like social media, where users' often self-report location
information, or aerial imagery, from which the extent of informal settlement can
be gauged \cite{zagheni2017leveraging, isaacman2017climate,
blumenstock2012inferring, quinn2018humanitarian}. More than
quantifying migration patterns, there have been efforts directly aimed at
protecting refugees, either through improving rescue operations
\cite{pham2018data, lomonaco2018intelligent} or monitoring negative public
sentiment \cite{un_global_pulse_2017}.
It is worth cautioning that immigrants and refugees are
vulnerable groups, and systems that surveil them can easily be exploited by bad
actors. Designing methodology and governance mechanisms that allow vulnerable
populations to benefit from such data, without putting them at additional risk,
should be a research priority.

\paragraph*{Assessing health risks}\Gap\mbox{}\\Climate change will affect exposure to
health hazards, and machine learning can play a role in measuring and mitigating
their impacts across subpopulations. Two of the most relevant expected shifts
are (1) heat waves will become more frequent and (2) outdoor and indoor air quality will
deteriorate \cite{haines2006climate, sarofim2016impacts}. These exposures have
either direct or indirect effects on health. For example, prolonged heat
episodes both directly cause heat stroke and can trigger acute episodes in chronic
conditions, like heart or respiratory disease \cite{schwartz2004hospital,
dominici2006fine}.

Careful data collection and analysis have played a leading role in
epidemiology and public health efforts for generations. It should be no surprise that ML has emerged as
an important tool in these disciplines, supporting a variety of research
efforts, from increasing the efficiency of disease simulators to supporting the
fine-grained measurement of exposures and their health impacts
\cite{khoury2013transforming, salathe2012digital}.

These disciplines are increasingly focused on the risks posed by climate change
specifically. For example, new sources of data have enabled detailed sensing of
urban heat islands \cite{clinton2013modis, ho2014mapping, voelkel2016peer},
water quality \cite{hafeez2019comparison, koditala2018water}, and air pollution
\cite{di2018machine, chen2018op}. Further, data on health indicators, which are
already collected, can quantitatively characterize observed impacts across
regions as well as illuminate which populations are most at risk to
climate-change induced health hazards \cite{watts2017lancet}. For example, it is
known that the young, elderly, and socially isolated are especially vulnerable
during heat waves, and finer-grained risk estimates could potentially drive
outreach \cite{pentland2009using, rose2013mortality}.

Across social applications, there are worthwhile research challenges -- guiding
interventions based on purely observational, potentially unrepresentative
data poses risks. In these contexts, transparency is necessary, and ideally,
causal effects of interventions could be estimated, to prevent feedback loops in
which certain subgroups are systematically ignored from policy interventions.

\subsection{Crisis}
\label{subsub:crisis}

Perhaps counterintuitively, natural disasters and health crises are not entirely
unpredictable -- they can be prepared for, risks can be reduced, and
coordination can be streamlined. Furthermore, while crises may be some of the
most distressing consequences of climate change, disaster response and public
health are mature disciplines in their own right, and have already benefited
extensively from ML methodology \cite{meier2013human, castillo2016big,
yasnoff2000public}.

\paragraph*{Managing epidemics}\Gap\mbox{}\\ 
Climate change will increase the range of vector and water-borne
diseases, elevating the likelihood that these new environments experience epidemics \cite{haines2006climate}.
Disease surveillance and outbreak forecasting systems can be
built from web data and specially-designed apps, in addition to traditional
surveys \cite{pervaiz2012flubreaks, lampos2010flu, johansson2016evaluating}.
While non-survey proxies are observational and self-reported, current research
attempts to address these issues \cite{lazer2014parable, nuti2014use}. Beyond
surveillance, point-of-care diagnostics have enjoyed a renaissance, thanks in
part to ML \cite{quinn2014computational, onu2017ubenwa}. These are tools that
allow health workers to make diagnoses when specialized lab equipment is
inaccessible. An example is malaria diagnosis based on photos of prepared
pathology slides taken with a mobile phone \cite{quinn2014automated}. Ensuring
that these systems reliably and transparently augment extension workers, guiding
data collection and route planning when appropriate, are active areas of study
\cite{robertson2010agile, brunskill2010routing}.

\paragraph*{Disaster response}\Gap\textbf{\Rec}\mbox{}\\ In disaster preparation and response, two types of ML
tasks have proven useful: creating maps from aerial imagery and performing
information retrieval on social media data. Accurate and well-annotated maps can
inform evacuation planning, retrofitting campaigns, and delivery of relief
\cite{doshi2018satellite, bastani2018machine}. Further, this imagery can assist 
damage assessment, by comparing scenes immediately pre- and  post-disaster 
\cite{voigt2007satellite, gupta2019creating}. Social media data can contain kernels of insight -- 
places without water, clinics without supplies -- which can inform relief efforts. 
ML can help properly surface these insights, compressing large volumes of social media 
data into the key takeaways, which can be acted upon by disaster managers 
\cite{olteanu2014crisislex, imran2015processing, castillo2016big}.

\subsection{Discussion}

Climate change will have profound effects on the planet, and
the ML community can support efforts to minimize the damage it does to
ecosystems and the harm it inflicts on people. This section has suggested areas
of research that may help societies adapt more effectively to these ever
changing realities. We have identified a few recurring themes, but also
emphasized the role of understanding domain-specific needs. The use of ML to
support societal resilience would be a noble goal at any time, but the need for
tangible progress towards it may never have been so urgent as it is today, in the
face of the wide-reaching consequences of climate change.

     \newpage

\section{Solar Geoengineering\texorpdfstring{\hfill\textit{by Andrew S.~Ross}}{}}
\label{sec:geoengineering}

Airships floating through the sky, spraying aerosols; robotic boats crisscrossing the ocean, firing vertical jets of spray; arrays of mirrors carefully positioned in space, micro-adjusted by remote control: these images seem like science fiction, but they are actually real proposals for solar radiation management, commonly called solar geoengineering \cite{keith2000geoengineering,shepherd2009geoengineering,irvine2016overview,keith2018science}. Solar geoengineering, much like the greenhouse gases causing climate change, shifts the balance between how much heat the Earth absorbs and how much it releases. The difference is that it is done deliberately, and in the opposite direction. The most common umbrella strategy is to make the Earth more reflective, keeping heat out, though there are also methods of helping heat escape (besides \carbon~removal, which we discuss in \textsection\ref{sec:afolu} and \textsection\ref{sec:ccs}).

Solar geoengineering generally comes with a host of potential side effects and governance challenges. Moreover, unlike \carbon~removal, it cannot simply reverse the effects of climate change (average temperatures may return to pre-industrial levels, but location-specific climates still change), and also comes with the risk of \emph{termination shock} (fast, catastrophic warming if humanity undertakes solar geoengineering but stops suddenly) \cite{parker2018risk}. Because of these and other issues, it is not within the scope of this paper to evaluate or recommend any particular technique. However, the potential for solar geoengineering to moderate some of the most catastrophic hazards of climate change is well-established \cite{irvine2019halving}, and it has received increasing attention in the wake of societal inaction on mitigation. Although \cite{keith2018science} argue that the ``hardest and most important problems raised by solar geoengineering are non-technical,'' 
there are still a number of important technical questions that machine learning may be able to help us study.

\paragraph*{Overview}\Gap\mbox{}\\
The primary candidate methods for geoengineering are marine cloud brightening \cite{jones2009climate} (making low-lying clouds more reflective), cirrus thinning \cite{storelvmo2014cirrus} (making high-flying clouds trap less heat), and stratospheric aerosol injection \cite{rasch2008overview} (which we discuss below). Other candidates (which are either less effective or harder to implement) include ``white-roof'' methods \cite{akbari2012long} and even launching sunshades into space \cite{angel2006feasibility}.

Injecting sulfate aerosols into the stratosphere is considered a leading candidate for solar geoengineering both because of its economic and technological feasibility \cite{mcclellan2012cost,smith2018production} and because of a reason that should resonate with the ML community: we have data. (This data is largely in the form of temperature observations after volcanic eruptions, which release sulfates into the stratosphere when sufficiently large \cite{robock2013studying}.) Once injected, sulfates circulate globally and remain aloft for 1 to 2 years. As a result, the process is reversible, but must also be continually maintained. Sulfates come with a well-studied risk of ozone loss \cite{eastham2018quantifying}, and they make sunlight slightly more diffuse, which can impact agriculture \cite{proctor2018estimating}.

\subsection{Understanding and improving aerosols}
\label{subsub:better-aerosols}

\paragraph*{Design}\Gap\mbox{\Longterm}\\
The effects and side-effects of aerosols in the stratosphere (or at slightly lower altitudes for cirrus thinning \cite{doi:10.1029/2018JD029815}) vary significantly with their optical and chemical properties. Although sulfates are the best understood due to volcanic eruption data, many others have been studied, including zirconium dioxide, titanium dioxide, calcite (which preserves ozone), and even synthetic diamond \cite{dykema2016improved}. However, the design space is far from fully explored. Machine learning has had recent success in predicting or even optimizing for specific chemical and material properties \cite{raccuglia2016machine,liu2017materials,gomez2018automatic,butler2018machine}. Although speculative, it is conceivable that ML could accelerate the search for aerosols that are chemically nonreactive but still reflective, cheap, and easy to keep aloft.

\paragraph*{Modeling}\Gap\mbox{}\\
One reason that sulfates have been the focus for aerosol research is that atmospheric aerosol physics is not perfectly captured by current climate models, so having natural data is important for validation. Furthermore, even if current aerosol models are correct, their best-fit parameters must still be determined (using historical data), which comes with uncertainty and computational difficulty. ML may offer tools here, both to help quantify and constrain uncertainty, and to manage computational load. As a recent example, \cite{fletcher2018quantifying} use Gaussian processes to emulate climate model outputs based on nine possible aerosol parameter settings, allowing them to establish plausible parameter ranges (and thus much better calibrated error-bars) with only 350 climate model runs instead of $>$100,000. Although this is important progress, ideally we want uncertainty-aware aerosol simulations with a fraction of the cost of one climate model run, rather than 350. ML may be able to help here too (see \textsection\ref{sec: climate prediction} for more details).

\subsection{Engineering a planetary control system\Gap \Rec \Longterm \HighRisk}
\label{subsub:planetary-control}
Efficient emulations and error-bars will be essential for what MacMartin and Kravitz \cite{macmartin2018engineering} call ``The Engineering of Climate Engineering.'' According to \cite{macmartin2018engineering}, any practical deployment of geoengineering would constitute ``one of the most critical engineering design and control challenges ever considered: making real-time decisions for a highly uncertain and nonlinear dynamic system with many input variables, many measurements, and a vast number of internal degrees of freedom, the dynamics of which span a wide range of timescales.'' Bayesian and neural network-based approaches could facilitate the fast, uncertainty-aware nonlinear system identification this challenge might require. Additionally, there has been recent progress in reinforcement learning for control \cite{moe2018machine,boczar2018finite,amos2018differentiable}, which could be useful for fine-tuning geoengineering interventions such as deciding where and when to release aerosols. For an initial attempt at analyzing stratospheric aerosol injection as a reinforcement learning problem (using a neural network climate model emulator), see \cite{de2019stratospheric}.

\subsection{Modeling impacts \Gap \Longterm}
\label{subsub:impact-models}
Of course, optimizing interventions requires defining objectives, and the choices here are far from clear. Although it is possible to stabilize global mean temperature and even regional temperatures through geoengineering, it is most likely impossible to preserve all relevant climate characteristics in all locations. Furthermore, climate model outputs do not tell the full story; ultimately, the goal of climate engineering is to minimize harm to people, ecosystems, and society. It is therefore essential to develop robust tools for estimating the extent and distribution of these potential harms.  There has been some recent work in applying ML to assess the impacts of geoengineering. For example, \cite{di2016assessing} use deep neural networks to estimate the effects of aerosols on human health, while \cite{crane2018using} use them to estimate the effects of solar geoengineering on agriculture. References \cite{burke2015global,diffenbaugh2019global} use relatively simple local and polynomial regression techniques but applied to extensive empirical data to estimate the past and future effects of temperature change on economic production. More generally, the field of \emph{Integrated Assessment Modeling}  \cite{kelly1999integrated,weyant2017some} aims to map the outputs of a climate model to societal impacts; for a general discussion of potential opportunities for applying ML to IAMs, see \S\ref{sec:decisionmaking}.

\subsection{Discussion}

Any consideration of solar geoengineering raises many moral questions. 
It may help certain regions at the expense of others, introduce risks like termination shock, and serve as a ``moral hazard'': widespread awareness of its very possibility may undermine mainstream efforts to cut emissions \cite{lin2013does}. Because of these issues, there has been significant debate about whether it is ethically responsible to research this topic \cite{preston2013ethics,keith2017toward}. However, although it creates new risks, solar geoengineering could actually be a moderating force against the terrifying uncertainties climate change already introduces \cite{macmartin2015solar,irvine2019halving}, and ultimately many environmental groups and governmental bodies have come down on the side of supporting further research.\footnote{ \url{https://www.edf.org/climate/our-position-geoengineering}}\footnote{\url{https://www.nrdc.org/media/2015/150210}}\footnote{\url{https://www.ucsusa.org/sites/default/files/attach/2019/gw-position-Solar-Geoengineering-022019.pdf}}
In this section, we have attempted to outline some of the technical challenges in implementing and evaluating solar geoengineering. We hope the ML community can help geoengineering researchers tackle these challenges.

 \newpage
 \part*{Tools for Action}
\addcontentsline{toc}{part}{Meta tools}

\section{Individual Action\texorpdfstring{\hfill\textit{by Natasha Jaques}}{}}
\label{sec:tools-individuals}
Individuals may worry that they are powerless to affect climate change, or lack clarity on which of their behaviors are most important to change. In fact, there are actions which can meaningfully reduce each person's carbon footprint, and, if widely adopted, could have a significant impact on mitigating global emissions \cite{ccneedsbehaviorchange,hawken2017drawdown}. AI can help to identify those behaviors, inform individuals, and provide constructive opportunities by modeling individual behavior. 

\subsection{Understanding personal carbon footprint}
\label{sec:personal_carbon_footprint}
We as individuals are constantly confronted with decisions that affect our carbon footprint, but we may lack the data and knowledge to know which decisions are most impactful. Fortunately, ML can help determine an individual's carbon footprint from their personal and household data\footnote{See e.g.~\url{https://www.tmrow.com/}}. For example, natural language processing can be used to extract the flights a person takes from their email, or determine specific grocery items purchased from a bill, making it possible to predict the associated emissions. Systems that combine this information with data obtained from the user's smartphone (e.g.~from a ride-sharing app) can then help consumers who wish to identify which behaviors result in the highest emissions. Given such a ML model, counterfactual reasoning can potentially be used to demonstrate to consumers how much their emissions would be reduced for each behavior they changed. As a privacy-conscious alternative, emissions estimates could be directly incorporated into grocery labels \cite{supermarketfuture} or interfaces for purchasing flights. Such information can empower people to understand how they can best help mitigate climate change through behavior change.

Residences are responsible for a large share of GHG emissions \cite{ipcc_global_2018} (see also \S\ref{sec:buildings-cities}). 
A large meta-analysis found that significant residential energy savings can be achieved \cite{ehrhardt2010advanced}, by targeting the right interventions to the right households \cite{albert2016predictive, allcott2011social, allcott2014short}. ML can predict a household's emissions in transportation, energy, water, waste, foods, goods, and services, as a function of its characteristics \cite{jones2011quantifying}. These predictions can be used to tailor customized interventions for high-emissions households \cite{jones2014spatial}.
Changing behavior both helps mitigate climate change and benefits individuals; studies have shown that many carbon mitigation strategies also provide cost savings to consumers \cite{jones2011quantifying}. 

Household energy disaggregation breaks down overall electricity consumption into energy use by individual appliances (see also \S\ref{sec:indv}) \cite{armel2013disaggregation}, which can help facilitate behavior change \cite{sundramoorthy2011domesticating}. For example, it can be used to inform consumers of high-energy appliances of which they were previously unaware. This alone could have a significant impact, since many devices consume a large amount of electricity even when not in use; standby power consumption accounts for roughly 8\% of residential electricity demand \cite{mackay2008sustainable}. A variety of ML techniques have been used to effectively disaggregate household energy, such as spectral clustering, Hidden Markov Models, and neural networks  \cite{armel2013disaggregation}.

ML can also be used to predict the marginal emissions of energy consumption in real time, on a scale of hours\footnote{\url{https://www.watttime.org/}}, potentially allowing consumers to effectively schedule activities such as charging an electric vehicle when the emissions (and prices \cite{klenert2018making}) will be lowest
\cite{olivierElectricitymap}. Combining these predictions with disaggregated energy data allows for the efficient automation of household energy consumption, ideally through products that present interpretable insights to the consumer (e.g.~\cite{strbac2008demand, schweppe1989algorithms}). Methods like reinforcement learning can be used to learn how to optimally schedule household appliances to consume energy more efficiently and sustainably \cite{mocanu2018line, remani2018residential}. Multi-agent learning has also been applied to this problem, to ensure that groups of homes can coordinate to balance energy consumption to keep peak demand low \cite{ramchurn2011agent2, ygge1999homebots}.

\subsection{Facilitating behavior change \Gap \Rec}
\label{sec:behavior_change}
ML is highly effective at modeling human preferences, and this can be leveraged to help mitigate climate change. Using ML, we can model and cluster individuals based on their climate knowledge, preferences, demographics, and consumption characteristics (e.g.~\cite{beiser2018assessing,carr2011exploring,de2018graph,gabe2016householders,yang2013review}), and thus predict who will be most amenable to new technologies and sustainable behavior change. Such techniques have improved the enrollment rate of customers in an energy savings program by 2-3x \cite{albert2016predictive}. Other works have used ML to predict how much consumers are willing to pay to avoid potential environmental harms of energy consumption \cite{de2013willingness}, finding that some groups were totally insensitive to cost and would pay the maximum amount to mitigate harm, while other groups were willing to pay nothing. Given such disparate types of consumers, targeting interventions toward particular households may be especially worthwhile; all the more so because data show that the size and composition of household carbon footprints varies dramatically across geographic regions and demographics \cite{jones2011quantifying}.

Citizens who would like to engage with policy decisions, or explore different options to reduce their personal carbon footprint, can have difficulty understanding existing laws and policies due to their complexity. They may benefit from tools that make policy information more manageable and relevant to the individual (e.g.~based on where the individual lives).
There is the potential for natural language processing to derive understandable insights from policy texts for these applications, similar to automated compliance checking \cite{doi:10.1061/(ASCE)CP.1943-5487.0000427, bell2016systems}.

Understanding individual behavior can help signal how it can be nudged. For example, path analysis has shown that an individual's \textit{psychological distance} to climate change (on geographic, temporal, social, and uncertainty dimensions) fully mediates their level of climate change concern \cite{jones2017future}. This suggests that interventions minimizing psychological distance to the effects of climate change may be most effective. Similarly, ML has revealed that cross-cultural support for international climate programs is not reduced, even when individuals are exposed to information about other countries' climate behavior  \cite{beiser2019commitment}.
To make the effects of climate change more real for consumers, and thus help motivate those who wish to act, image generation techniques such as CycleGANs have been used to visualize the potential consequences of extreme weather events on houses and cities \cite{schmidt2019visualizing}. 
Gamification via deep learning has been proposed to further allow individuals to explore their personal energy usage \cite{konstantakopoulos2019deep}. All of these programs may be an incredibly cost-effective way to reduce energy consumption; behavior change programs can cost as little as 3 cents to save a kilowatt hour of electricity, whereas generating one kWh would cost 5-6 cents with a coal or wind power plant, and 10 cents with solar \cite{peerpressureOPower, forbesEnergyCost}. 

\subsection{Discussion}
While individuals can sometimes feel that their contributions to climate change are dwarfed by other factors, in reality individual actions can have a significant impact in mitigating climate change. ML can aid this process by empowering consumers to understand which of their behaviors lead to the highest emissions, automatically scheduling energy consumption, and providing insights into how to facilitate behavior change.

 \newpage

\section{Collective Decisions\texorpdfstring{\hfill\textit{by Tegan Maharaj and Nikola Milojevic-Dupont}}{}}
\label{sec:toolsforsociety}

Addressing climate change requires swift and effective decision-making by groups at multiple levels -- communities, unions, NGOs, businesses, governments, intergovernmental organizations, and many more. Such collective decision-making encompasses many kinds of action -- for example, negotiating international treaties to reduce GHG emissions, designing carbon markets, building resilient infrastructure, and establishing community-owned solar farms.
These decisions often involve multiple stakeholders with different goals and priorities, requiring difficult trade-offs. The economic and societal systems involved are often extremely complex, and the impacts of climate-related decisions can play out globally across long time horizons. 
To address some of these challenges, researchers are using empirical and mathematical methods from fields such as policy analysis, operations research, economics, game theory, and
computational social science; there are many opportunities for ML to support and supplement these methods.

\subsection{Modeling social interactions}\label{sec:coordination}

When designing climate change strategies, it is critical to understand how organizations and  individuals act and interact in response to different incentives and constraints. Agent-based models (ABMs) \cite{de2014agent, 10.2307/j.ctt7rxj1} represent one approach used in simulating the actions and interactions of \emph{agents} (people, companies, etc.) in their environment. ABMs have been applied to a multitude of problems relevant to climate change, in particular to study low-carbon technology adoption \cite{RAI2015163, Haelg_2018, zhang2012agent, NOORI2016215}. 
For example, when modeling solar PV adoption \cite{Zhang2016abm}, agents may represent individuals who act based on factors such as financial interest and the behavior of their peers \cite{rai2016overcoming, bollinger2012peer}; the goal is then to study how these agents interact in response to different conditions, such as electricity rates, subsidy programs, and geographical considerations. 
Other applications of ABMs include modeling how behavior under social norms changes with external pressures \cite{doi:10.1098/rspb.2015.2431}, how the economy and climate may evolve given a diversity of political and economic beliefs \cite{geisendorf}, and how individuals may migrate in response to environmental changes \cite{thober}. While agent and environment models in ABMs are often hand-designed by experts, ML can help integrate data-driven insights into these models \cite{Zhang2019}, for example by learning rules or models for agents based on observational data \cite{Zhang2016abm, gunaratne2018evolutionary}, or by using unsupervised methods such as VAEs or GANs to discover salient features useful in modeling a complex environment. While the hope of learning or tuning behavior from data is promising for generalization, many data-driven approaches lose the interpretability for which ABMs are valued; work in interpretable ML methods could potentially help with this.

In addition to ABMs, techniques from game theory can be valuable in modeling behavior, e.g.~to explore cooperation in the face of a depleting resource \cite{hilbe2018evolution}. Multi-agent reinforcement learning can also be applied to understand the behavior of groups of agents who need to cooperate; see \cite{panait2005MARL} for an overview and \cite{lee2019imitate, jaques2018motivation} for recent examples. Combined with mechanism design\footnote{Mechanism design is often called ``inverse game theory'' -- rather than determining optimal strategies for players, mechanism design seeks to design games such that certain strategies are incentivized.}, such approaches can be used to design methods for cooperation that lead to mutually beneficial outcomes, for example when formalizing procedures around international climate agreements \cite{mechclim, Procaccia:2013:CCJ:2483852.2483870}.

\subsection{Informing policy}
\label{sec:decisionmaking}

The actions required to address climate change, both in mitigation and adaptation, require making policies\footnote{\emph{Policy} can refer, for example, to laws, measures, standards, or best practices.} at the local, national, and international levels \cite{sterner2019}. Various institutions act as policy-makers:~for instance, governments, international organizations, non-governmental organizations, standards committees, and professional institutions.  Tools from \emph{policy analysis} -- the process of evaluating the outcomes of past policies and assessing future policy alternatives\footnote{The former is often referred to as \textit{ex-post policy analysis} and the latter as \textit{ex-ante policy analysis}.} -- can help inform the choices these institutions make.
Policy analysis uses quantitative tools from statistics, economics, and operations research such as cost-benefit analysis, uncertainty analysis, and multi-criteria decision making to inform the policy-making process; see \cite{morgan_2017, patton2015policy} for an introduction.
ML can provide data for policy analysis, help improve existing tools for assessing policy options, and provide new tools for evaluating the effects of policies.

\paragraph*{Gathering data}\Gap\mbox{\Rec}\\
When creating policies, decision-makers must often negotiate fundamental uncertainties in the underlying data.
ML can help alleviate some of this uncertainty by providing data. For instance, as detailed elsewhere in this paper, ML can help pinpoint sources of emissions (\S\ref{sec:electricity-methane},\ref{sec:emissions-detection}), approximate traffic patterns (\S\ref{sec:transport-data}), identify infrastructure at risk (\S\ref{subsub:infrastructure}), and mine information from companies' financial disclosures (\S\ref{sec:climate-analytics}).
Natural language processing, network analysis, and clustering techniques can also be used to analyze social media data to understand public opinions and discourse around climate change \cite{veltri2017climate, williams2015network, kirilenko2014public}.
These data can then be used to identify areas of intervention, compute the benefits and costs of a project, or evaluate the effectiveness of a policy after it has been implemented. 

\paragraph*{Assessing policy options}\Gap\mbox{}\\
Decision-makers often construct mathematical models to help them assess or trade off between different policy alternatives. ML is particularly relevant to approaches that model large and complex socio-economic systems to assess outcomes of particular strategies, as well as optimization-based tools that help with navigating the decision.

Policy-makers often wish to analyze how different policy alternatives may contribute to achieving a particular objective. Computational approaches such as simulation and (partial) equilibrium models can be used to compare different policy options, assess the effects of underlying assumptions, or propose strategies that are consistent with the objectives of decision-makers.
Of particular relevance to climate change mitigation are \emph{integrated assessment models} (IAMs), which incorporate economic models, climate models, and policy information (see \cite{10.1093/reep/rew018} for an overview). IAMs are used to explore future societal pathways that are consistent with climate goals (e.g.~1.5$^{\circ}$C mean global temperature increase), and play a prominent role in the IPCC assessments \cite{moss2010iams}. 
While these models can simulate interactions between many variables in great detail, this comes at the cost of computational complexity and presents opportunities for machine learning.
Much as with Earth system models (\S\ref{sec: climate prediction}), ML can be applied within any of the various sub-models that make up an IAM. One set of applications involves deriving results at the appropriate spatial resolution, since different components of an IAM operate at different scales. Outputs with high resolution may be aggregated via clustering methods to provide insights \cite{dietrich2013reducing}, while at coarser resolution, statistical \emph{downscaling} can help to disaggregate data to an appropriate spatial resolution, as seen in applications such as crop yield \cite{folberth2019spatio}, wind speed \cite{li2019geographically} or surface temperature \cite{li2019evaluation}.
ML also has the potential to help with sensitivity and uncertainty analysis \cite{JAXAROZEN2018245}, with finding numerical solutions for computational expensive submodels \cite{scheidegger2019machine, duarte2018}, and assessing the validity of the models \cite{Mori2018}.

In addition to assessing the outcomes of various policies, policy-makers may also employ optimization-based tools to figure out what decisions to make.
For example, combinatorial optimization is a powerful tool used widely for decision-making in operations research. See \cite{2018arXiv181106128B} for a survey of how ML can be employed to help solve combinatorial optimization problems.

Tools from the field of \emph{multi-criteria decision-making} can also help policy-makers manage trade-offs between different policies by reconciling competing objectives and minimizing negative side-effects; in particular, in cases where policy objectives and constraints can be mathematically formalized, \emph{multi-objective optimization} can provide a pragmatic approach to making decisions. 
Here, a decision-maker would formulate their decision-making process as an optimization problem by combining multiple optimization objectives subject to physical or other types of constraints;
the goal is to then find a solution (or set of solutions) that is \emph{Pareto-optimal} with respect to all of the objective functions.
However, finding these solutions is often computationally expensive.
Practitioners have applied bio-inspired algorithms such as particle swarm, genetic, or evolutionary algorithms to search for or compute Pareto-optimal solutions that satisfy the constraints.
This approach has been applied in a number of climate change-related fields, including energy and infrastructure planning \cite{pohekar, mattiussi, atabaki, shi_dendritic, wu2018efficiently, jhhan}, industry \cite{hassine, chaabane}, land use \cite{lakicevic, varma}, and more \cite{gutierrez, miniciardi, chen, lithuania}. 
Previous work has also employed parallel surrogate search, assisted by ML, to efficiently solve multi-objective optimization problems \cite{akhtar2019efficient}.
Optimization algorithms which have been successful in the context of hyperparameter tuning (e.g. Bayesian optimization \cite{shahriari2015taking, Snoek:2012:PBO:2999325.2999464}) or guided search algorithms (e.g. tree search algorithms \cite{alphazero}) could also potentially be applied to this problem.

\paragraph*{Evaluating policy effects}\Gap\mbox{\textbf{\Rec}}\\
When creating new policies, decision-makers may wish to understand previous policies (e.g.~from other jurisdictions) and how these policies performed.
ML can help analyze previous policy actions automatically and at scale by improving computational text analysis. In particular, natural language processing methods are already used in the field of political science to analyze political texts and legislation \cite{grimmer_stewart_2013}; these approaches could be promising for systematically studying climate change policies.
Causal inference techniques can also help assess the effect of a particular policy or climate-related event from observed outcomes.
ML can play a role in causal inference \cite{pearl2019, athey2019machine, doi:10.1080/09332480.2019.1579578}, including in the context of policy problems \cite{kreif2019m, Athey483} and in climate-relevant scenarios such as estimating the effects of temperature on human mortality \cite{hovdahl2019use} and the effects of World Bank projects on vegetative cover \cite{10.1007/978-3-319-71273-4_17}.

\subsection{Designing markets}
\label{subsec:markets}

In economics, GHG emissions can be seen as a \emph{negative externality}:~while a changing climate results in a cost for society, this cost is often not reflected in the market price of goods or services that cause GHG emissions. This is problematic, since organizations and individuals making decisions solely on the basis of market prices will tend to favor cheaper goods, even if those goods emit a large amount of GHGs. Market-based tools\footnote{For background on market-based strategies, see \cite{stiglitz2017report,stern2008economics,ellerman2010pricing}.} such as carbon taxes aim to enforce prices reflecting the societal cost of GHGs and thus encourage socially beneficial behavior through market forces. ML can help in understanding the impacts of market instruments; assessing their effectiveness at reducing emissions; and supporting a swift, effective and fair implementation.\footnote{For a review on ML for energy economics and finance, see \cite{ghoddusi2019machine}.}

\paragraph*{Predicting carbon prices}\Gap\mbox{}\\
There are several approaches to pricing GHG emissions. Carbon taxes and quotas aim to influence the behavior of organizations by shaping supply and demand within an existing market. By contrast, cap-and-trade approaches such as those within the European Union involve a completely new market, an \emph{Emissions Trading Scheme}, within which companies can buy and sell a limited number of GHG emissions permits.
Prices within such cap-and-trade markets are highly sensitive to control elements such as the number of permits released at a given time. 
ML can be used to analyze prices within these markets, for example by predicting prices via supervised learning \cite{zhu2017forecasting,sun2018analysis,zhu2018novel,wei} or analyzing the main drivers of prices via hierarchical clustering \cite{zhu2015carbon}.

\paragraph*{Non-carbon markets}\Gap\mbox{}\\
Market design can influence GHG emissions even in settings where such emissions are not directly penalized. For instance, dynamic pricing in electricity markets -- varying the price of electricity to consumers based on, e.g.,  how much wind power is available -- can shape demand for low-carbon energy sources (see \S\ref{sec:dispatchDR}). 
Following seminal research on modeling pricing in markets as a bandit problem \cite{rothschild}, many works have applied bandit and other reinforcement learning (RL) algorithms to determine prices or other market values. For example, RL has been applied to predict bids \cite{ragupathi} and market power \cite{nanduri} in electricity markets, and to set dynamic prices in more general settings 
\cite{maestre}. ML can also help solve auctions in supply chains \cite{Sandholm_very-large-scalegeneralized}.

\paragraph*{Assessing market effects}\Gap\mbox{}\\
When designing market-based strategies, it is necessary to understand how effectively each strategy will reduce emissions, as well as how the underlying socio-technical system may be affected. 
Studies have considered effects of carbon pricing on economic growth and energy intensity \cite{fang2017investigating,fang2018optimize}, or on electricity prices \cite{nagapurkar2019techno}. 
Effects of pricing mechanisms can also be indirect, as companies' strategic decisions can have longer-term effects. ML can be useful in analyzing these effects. For example, self-organizing maps have been used to analyze how R\&D investment in green technologies changes in response to fuel prices \cite{barbieri2016fuel}, while a game theoretical framework using neural networks has been used to study the optimal production strategies for companies under carbon quotas \cite{zheng2018optimal}.

To ensure that market-based strategies are effective and equitable, it is important to understand their distributional effects, as certain social groups or classes of stakeholders may be affected more than others. For example, a flat carbon tax on gasoline will have a larger effect on lower-income populations, as fuel expenses are a bigger share of their total budget. Here, clustering can help identify permit allocation schemes that maximize social welfare \cite{wu2019research}, and supervised learning has been used to predict winners and losers from changing electricity tariff schemes \cite{granell2014predicting}. \emph{Hedonic pricing} can also help identify how much different consumers may be willing to pay for a environmental good or a service, which is a noisy measure for the monetary value of that good or service;  these values are typically inferred using regression or ML techniques on historical market data \cite{hedonic_residential, hedonic_park, hedonic_theory,hedonic_air}.
It is also important to analyze which organizations or individuals can actually participate in a given market.
For example, carbon markets can be more flexible if viable offsets exist, including those offered by landowners who sequester carbon through forest conservation and management; ML has been used to examine the factors influencing the financial viability of such projects \cite{kerchner2015california}.

\subsection{Discussion}
The complexity, scale, and fundamental uncertainty inherent in the problems of climate change can pose challenges for collective decision-making.
ML can help supplement existing mathematical frameworks that are employed to alleviate some of these challenges, including agent-based models, integrated assessment models, multi-objective optimization, and market design. 
Interpretable and fair ML techniques may be of particular importance in this context, as they may enable decision-makers to more effectively and equitably employ insights from ML models.
While these quantitative assessment tools can provide useful input to the decision-making process, it is worth noting that decisions regarding climate change may ultimately depend on qualitative discussions around norms, values, or equity considerations that may not be captured in quantitative models.

 \newpage

\section{Education\texorpdfstring{\hfill\textit{by Alexandra Luccioni}}{}}
\label{sec:education}

Access to quality education is a key part of sustainable development, with significant benefits for climate and society at large. Education contributes to improving quality of life, helps individuals make informed decisions, and trains the next generation of innovators. Education is also paramount in helping people across societies understand and address the causes and consequences of climate change and provides the skills and tools necessary for adapting to its impacts. For instance, education can both improve the resilience of communities, particularly in developing countries that will be disproportionately affected by climate change~\cite{unesco2015}, and empower individuals, especially from developed countries, to adopt more sustainable lifestyles~\cite{fisher2016}. As climate change itself may diminish educational outcomes for some populations, due to its negative effects on agricultural productivity and household income~\cite{randell2016, randell2019}, this makes providing high-quality educational interventions globally all the more important.

\paragraph*{AI in Education}\Gap\textbf{\Longterm}\label{sec:aied}\mbox{}\\There are a number of ways that AI and ML can contribute to education and teaching -- for instance by improving access to educational opportunities, helping personalize the teaching process, and stepping in when teachers have limited time.
The field of AIED (Artificial Intelligence in EDucation) has existed for over 30 years, and until recently relied on explicitly modeling content, learners, and tutoring strategies based on psychological theories of learning.
However, AIED is increasingly incorporating data-driven insights derived from ML techniques.

One important area of AIED research has been Intelligent Tutoring Systems (ITSs), which can adapt their behavior in real time according to the needs of individuals or to support collaborative learning~\cite{chaplot2018}. 
While ITSs have traditionally been defined and constructed by hand, recent approaches have applied ML techniques such as multi-armed bandit techniques to adaptively personalize sequences of learning activities~\cite{clement2013}, LSTMs to generate questions to evaluate language comprehension~\cite{du2017}, and reinforcement learning to improve the strategies used within the ITS~\cite{iglesias2009,koedinger2013}. However, there remains much work to be done to bridge the performance gap between digital and human tutors, and ML-based approaches have an important role to play in this endeavor -- for example, via natural language processing techniques for creating conversational agents~\cite{gnewuch2018}, learner analytics for classifying student profiles,~\cite{romero2008}, and adaptive learning approaches to propose relevant educational activities and exercises ~\cite{svihla2012}.~\footnote{For further background on this area, see~\cite{nkambou2010,joksimovic2018,pinkwart2016}.}

While ITSs generally focus on individualized or small-group instruction, AIED can also help provide tools that improve educational outcomes at scale for larger groups of learners. For instance, scalable, adaptive online courses could give hundreds of thousands of learners access to learning resources that they would not usually have in their local educational facilities~\cite{roll2018}. Furthermore, giving teachers guidance derived from computational teaching algorithms or heuristics could help them design better educational curricula and improve student learning outcomes~\cite{dede2009}. In this context, AIED applications can be used either as a standalone tool for independent learners or as an educational resource that frees up teachers to have more one-on-one time with students. Key considerations for creating AIED tools that can be applied across the globe include adapting to local technological and cultural needs, addressing barriers such as access to electricity and internet~\cite{khandker2009welfare1, khandker2009welfare2}, and taking into account students' computing skills, language, and culture~\cite{cakmak2014,nye2015}. 

\paragraph*{Learning about climate}
\label{sec:climate-ed}\mbox{}\\
Research has shown that educational activities centered on climate change and carbon footprints can engage learners in understanding the connection between personal and collective actions and their impact on global climate, and can enable individuals to make climate-friendly lifestyle choices such as reducing energy use~\cite{cordero2008}. There have also been proposals for interactive websites explaining climate science as well as educational interventions focusing on local and actionable aspects of sustainable development~\cite{anderson2012}. In these contexts, ML can help create personalized educational tools, for instance by generating images of future impacts of extreme weather events based on a learner's address ~\cite{schmidt2019visualizing} or by anchoring an individual's learning experience in a digital replica of their real-life location and allowing them to explore the way that climate change will impact a specific location~\cite{angel2015}.

 \newpage

\section{Finance\texorpdfstring{\hfill\textit{by Alexandra Luccioni}}{}}
\label{sec:finance}

The rise and fall of financial markets is linked to many events, both sporadic (e.g.~the 2008 global financial crisis) and cyclical (e.g.~the price of gas over the years), with profits and losses that can be measured in the billions of dollars and can have global consequences. Climate change poses a substantial financial risks to global assets measured in the trillions of dollars \cite{dietz2016}, and it is hard to forecast where, how, or when climate change will impact the stock price of a given company, or even the debt of an entire nation. While financial analysts and investors focus on pricing risk and forecasting potential earnings, the majority of the current financial system is based on quarterly or yearly performance. This fails to incentivize the prediction of medium or long-term risks, which include most climate change-related exposures such as physical impacts on assets or distribution chains, legislative impacts on profit generation, and indirect market consequences such as supply and demand\footnote{For further reading regarding the impact of climate change on financial markets, see~\cite{boissinot2016, battiston2017,campiglio2018}.}.

\paragraph*{Climate investment}
\label{sec:climate-investment}\mbox{}\\
\emph{Climate investment}, the current dominant approach in climate finance, involves investing money in low-carbon assets~\cite{eyraud2013}. The dominant ways in which major financial institutions take this approach are by creating ``green'' financial indexes that focus on low-carbon energy, clean technology, and/or environmental services~\cite{diaz2017} or by designing carbon-neutral investment portfolios that remove or under-weight companies with relatively high carbon footprints \cite{gianfrate2018}. This investment strategy is creating major shifts in certain sectors of the market (e.g.~utilities and energy) towards renewable energy alternatives, which are seen as having a greater growth potential than traditional energy sources such as oil and gas~\cite{bergmann2006}. While this approach currently does not utilize ML directly, we see the potential in applying deep learning both for portfolio selection (based on features of the stocks involved) and investment timing (using historical patterns to predict future demand), to maximize both the impact and scope of climate investment strategies.

\paragraph*{Climate analytics}\Gap\Rec\label{sec:climate-analytics}\mbox{}\\
The other main approach to climate finance is \emph{climate analytics}, which aims to predict the financial effects of climate change, and is still gaining momentum in the mainstream financial community~\cite{eyraud2013}.
Since this is a predictive approach to addressing climate change from a financial perspective, it is one where ML can potentially have greater impact. 
Climate analytics involves analyzing investment portfolios, funds, and companies in order to pinpoint areas with heightened risk due to climate change, such as timber companies that could be bankrupted by wildfires or water extraction initiatives that could see their sources polluted by shifting landscapes. Approaches used in this field include:~natural language processing techniques for identifying climate risks and investment opportunities in disclosures made by companies~\cite{stanny2008} as well as for analyzing the evolution of climate coverage in the media to dynamically hedge climate change risk~\cite{engle2019}; econometric approaches for developing arbitrage strategies that take advantage of the carbon risk factor in financial markets~\cite{andersson2016}; and ML approaches for forecasting the price of carbon in emission exchanges\footnote{Carbon pricing, e.g.~via CO$_2$~cap-and-trade or a carbon tax, is a commonly-suggested policy approach for getting firms to price future climate change impacts into their financial calculations. For an introduction to these topics, see \cite{pizer2006choosing} and also \S\ref{subsec:markets}.}~\cite{zhu2017, zhou2018}. 

To date, the field of climate finance has been largely neglected within the larger scope of financial research and analysis. This leaves many directions for improvement, such as (1) improving existing traditional portfolio optimization approaches; (2) in-depth modeling of variables linked to climate risk; (3) designing a statistical climate factor that can be used to project the variation of stock prices given a compound set of events; and (4) identifying direct and indirect climate risk exposure in annual company reports. ML plays a central role in these strategies, and can be a powerful tool in leveraging the financial sector to mitigate climate change and in reducing the financial impacts of climate change on society.

 \newpage

\part*{Conclusion}
\addcontentsline{toc}{part}{Conclusion}

Machine learning, like any technology, does not always make the world a better place --- but it can. In the fight against climate change, we have seen that ML has significant contributions to offer across domain areas. ML can enable automatic monitoring through remote sensing (e.g.~by pinpointing deforestation, gathering data on buildings, and assessing damage after disasters). It can accelerate the process of scientific discovery (e.g.~by suggesting new materials for batteries, construction, and carbon capture). ML can optimize systems to improve efficiency (e.g.~by consolidating freight, designing carbon markets, and reducing food waste). And it can accelerate computationally expensive physical simulations through hybrid modeling (e.g.~climate models and energy scheduling models). These and other cross-cutting themes are shown in Table \ref{tab:themes}. We emphasize that in each application, ML is only one part of the solution; it is a tool that enables other tools across fields.

Applying machine learning to tackle climate change has the potential both to benefit society and to advance the field of machine learning. Many of the problems we have discussed here highlight cutting-edge areas of ML, such as interpretability, causality, and uncertainty quantification. Moreover, meaningful action on climate problems requires dialogue with fields within and outside computer science and can lead to interdisciplinary methodological innovations, such as improved physics-constrained ML techniques.

The nature of climate-relevant data poses challenges and opportunities. For many of the applications we identify, data can be proprietary or include sensitive personal information. Where datasets exist, they may not be organized with a specific task in mind, unlike typical ML benchmarks that have a clear objective. Datasets may include information from  heterogeneous sources, which must be integrated using domain knowledge. Moreover, the available data may not be representative of global use cases. For example, forecasting weather or electricity demand in the US, where data are abundant, is very different from doing so in India, where data can be scarce. Tools from transfer learning and domain adaptation will likely prove essential in low-data settings. For some tasks, it may also be feasible to augment learning with carefully simulated data. Of course, the best option if possible is always more real data; we strongly encourage public and private entities to release datasets and to solicit involvement from the ML community.

\begin{table}
\begin{small}
\begin{center}
\begin{tabular}{l l l l l l l l l l }  \toprule
     \multicolumn{2}{l}{ }
         & \small{\rotatebox{90}{\parbox{2.4cm}{Accelerated\\experimentation}}}
         & \small{\rotatebox{90}{\parbox{2.4cm}{Control systems}}}
         & \small{\rotatebox{90}{\parbox{2.4cm}{Forecasting}}}
         & \small{\rotatebox{90}{\parbox{2.4cm}{Human\\interaction}}}
         & \small{\rotatebox{90}{\parbox{2.4cm}{Hybrid physical\\models}}}
         & \small{\rotatebox{90}{\parbox{2.4cm}{Predictive\\maintenance}}}
         & \small{\rotatebox{90}{\parbox{2.4cm}{Remote sensing}}}
         & \small{\rotatebox{90}{\parbox{2.4cm}{System\\optimization}}}
    \\ \midrule
    \rowcolor{ccai-yellow}
    \multicolumn{2}{l}{1 \hyperref[sec:electricity-systems]{Electricity systems}} 
         & 
         & 
         & 
         &
         & 
         & 
         & 
         &\\ 
    & \hyperref[sec:electricity-lowCarbon]{Enabling low-carbon electricity}
         & $\bullet$
         & $\bullet$
         & $\bullet$
         &
         & $\bullet$
         & $\bullet$
         & $\bullet$
         & $\bullet$\\
    & \hyperref[sec:electricity-currentSystemImpact]{Reducing current-system impacts}
         & 
         & 
         & $\bullet$
         &
         & 
         & $\bullet$
         & $\bullet$
         & \\
    & \hyperref[sec:electricity-developing]{Ensuring global impact}
         & 
         & 
         & $\bullet$
         &
         & $\bullet$
         & 
         & $\bullet$
         & \\
    \rowcolor{ccai-yellow}
    \multicolumn{2}{l}{2 \hyperref[sec:transportation]{Transportation}} 
         & 
         & 
         & 
         &
         & 
         & 
         & 
         & \\
    & \hyperref[sec:TReducing]{Reducing transport activity}
         & $\bullet$
         & $\bullet$
         & $\bullet$
         &
         & 
         & 
         & $\bullet$
         & $\bullet$\\
   & \hyperref[sec:TEfficient]{Improving vehicle efficiency}
         & $\bullet$
         & $\bullet$
         & 
         &
         & 
         & 
         & 
         & $\bullet$ \\ 
   & \hyperref[sec:TFuels]{Alternative fuels \& electrification}
         & $\bullet$
         & $\bullet$
         & $\bullet$
         &
         & 
         & 
         & 
         &$\bullet$\\ 
   & \hyperref[sec:modalshift]{Modal shift}
         & 
         & $\bullet$
         & $\bullet$
         & $\bullet$ 
         & 
         & $\bullet$
         & $\bullet$
         & $\bullet$\\ 
    \rowcolor{ccai-yellow}
    \multicolumn{2}{l}{3 \hyperref[sec:buildings-cities]{Buildings and cities}} 
         & 
         & 
         & 
         &
         &  
         & 
         & 
         &\\ 
    & \hyperref[sec:indv]{Optimizing buildings}
         & 
         & $\bullet$
         & $\bullet$
         &
         & $\bullet$
         & $\bullet$
         & 
         &$\bullet$\\ 
    & \hyperref[sec:distr]{Urban planning}
         & 
         & 
         & 
         &
         & 
         & 
         & $\bullet$
         &\\ 
    & \hyperref[sec:cities]{The future of cities}
         & 
         & 
         & 
         &
         & 
         & 
         & 
         &$\bullet$\\ 
    \rowcolor{ccai-yellow}
    \multicolumn{2}{l}{4 \hyperref[sec:industry]{Industry}} 
         & 
         & 
         & 
         &
         & 
         & 
         & 
         &\\ 
    & \hyperref[sec:supplychains]{Optimizing supply chains}
         & 
         & $\bullet$
         &$\bullet$ 
         &
         & 
         & 
         & 
         &$\bullet$\\ 
    & \hyperref[sec:materialsandconstruction]{Improving materials}
         & $\bullet$
         & 
         & 
         &
         & 
         & 
         & 
         &\\ 
    & \hyperref[sec:demandresponse]{Production \& energy}
         & 
         & $\bullet$
         & 
         &
         & 
         & $\bullet$
         & 
         &$\bullet$\\ 
    \rowcolor{ccai-yellow}
    \multicolumn{2}{l}{5 \hyperref[sec:afolu]{Farms \& forests}} 
         & 
         & 
         & 
         &
         & 
         & 
         &
         &\\ 
    & \hyperref[sec:emissions-detection]{Remote sensing of emissions}
         & 
         & 
         & 
         &
         & 
         & 
         &$\bullet$ 
         &\\ 
    & \hyperref[sec:agriculture]{Precision agriculture}
         & 
         & $\bullet$
         & $\bullet$
         &
         & 
         & 
         & $\bullet$
         &\\ 
    & \hyperref[sec:peatlands]{Monitoring peatlands}
         & 
         & 
         & 
         &
         & 
         & 
         &$\bullet$ 
         &\\ 
    & \hyperref[sec:forests]{Managing forests}
         & 
         & $\bullet$
         & $\bullet$
         &
         & 
         & 
         &$\bullet$ 
         &\\ 
    \rowcolor{ccai-yellow}
    \multicolumn{2}{l}{6 \hyperref[sec:ccs]{Carbon dioxide removal}}
         & 
         & 
         & 
         &
         & 
         & 
         & 
         &\\ 
    & \hyperref[sec:ccs]{Direct air capture}
         & $\bullet$
         & 
         & 
         &
         & $\bullet$
         & 
         & 
         &\\ 
    & \hyperref[subsubsec: sequestrativervin]{Sequestering~\cd}
         & 
         &
         & 
         &
         & $\bullet$
         & 
         & 
         & \\
    \rowcolor{ccai-yellow}\multicolumn{2}{l}{7 \hyperref[sec: climate prediction]{Climate prediction}}
         & 
         & 
         & 
         &
         & 
         & 
         & 
         & \\
    & \hyperref[sec:climate-models-params]{Uniting data, ML \& climate science}
         & 
         & 
         & $\bullet$
         &
         & $\bullet$
         & 
         & $\bullet$
         & \\
    & \hyperref[sec:models-extreme-events]{Forecasting extreme events}
         & 
         & 
         &$\bullet$ 
         &
         & $\bullet$
         & 
         & $\bullet$
         & \\
    \rowcolor{ccai-yellow}
    \multicolumn{2}{l}{8 \hyperref[sec:societal-impacts]{Societal impacts}} 
         & 
         & 
         & 
         &
         & 
         & 
         & 
         &\\ 
    & \hyperref[subsub:ecology]{Ecology}
         & 
         & 
         & 
         &
         & 
         & 
         & $\bullet$
         &\\ 
    & \hyperref[subsub:infrastructure]{Infrastructure}
         & 
         & 
         & 
         &
         & 
         & $\bullet$
         & 
         &$\bullet$\\ 
    & \hyperref[subsub:social_systems]{Social systems}
         & 
         & 
         & $\bullet$
         & $\bullet$
         & 
         & 
         & $\bullet$
         & $\bullet$\\
    & \hyperref[subsub:crisis]{Crisis}
         & 
         & 
         & $\bullet$
         &
         & 
         & 
         & $\bullet$
         & \\
    \rowcolor{ccai-yellow}
    \multicolumn{2}{l}{9 \hyperref[sec:geoengineering]{Solar geoengineering}} 
         & 
         & 
         & 
         &
         & 
         & 
         & 
         & \\
    & \hyperref[subsub:better-aerosols]{Understanding \& improving aerosols}
         & 
         & $\bullet$
         & 
         &
         & $\bullet$
         & 
         & 
         & \\
    & \hyperref[subsub:planetary-control]{Engineering a planetary control system}
         & 
         & $\bullet$
         & 
         &
         & $\bullet$
         & 
         & 
         & \\
    & \hyperref[subsub:impact-models]{Modeling impacts}
         & 
         & 
         & 
         &
         & $\bullet$
         & 
         & 
         & \\
    \rowcolor{ccai-yellow}
    \multicolumn{2}{l}{10 \hyperref[sec:tools-individuals]{Individual action}} 
         & 
         & 
         & 
         & 
         & 
         & 
         & 
         & \\
    & \hyperref[sec:personal_carbon_footprint]{Understanding personal footprint}
         & 
         & 
         & $\bullet$
         & $\bullet$
         & 
         & 
         & 
         & \\
    & \hyperref[sec:behavior_change]{Facilitating behavior change}
         & 
         & 
         & 
         & $\bullet$
         & 
         & 
         & 
         & \\
    \rowcolor{ccai-yellow}
    \multicolumn{2}{l}{11 \hyperref[sec:toolsforsociety]{Collective decisions}} 
         & 
         & 
         & 
         & 
         & 
         & 
         & 
         & \\
    & \hyperref[sec:coordination]{Modeling social interactions}
         & 
         & 
         & 
         & $\bullet$
         & 
         & 
         & 
         & \\
    & \hyperref[sec:decisionmaking]{Informing policy}
         & 
         & 
         & 
         & $\bullet$
         & 
         & 
         & 
         & \\
    & \hyperref[subsec:markets]{Designing markets}
         & 
         & 
         & $\bullet$
         &
         & 
         & 
         & 
         & $\bullet$\\
    \rowcolor{ccai-yellow}
    \multicolumn{2}{l}{12 \hyperref[sec:education]{Education}} 
         & 
         & 
         & 
         & $\bullet$
         & 
         & 
         & 
         & \\
    \rowcolor{ccai-yellow}
    \multicolumn{2}{l}{13 \hyperref[sec:finance]{Finance}} 
         & 
         & 
         & $\bullet$
         & $\bullet$
         & 
         & 
         & 
         & \\
    \bottomrule
\end{tabular}
\caption{Cross-cutting objectives that are relevant to many climate change domains.}
\label{tab:themes}
\end{center}
\end{small}
\end{table}

For those who want to apply ML to climate change, we provide a roadmap:
\begin{itemize}
    \item \textbf{Learn.} Identify how your skills may be useful -- we hope this paper is a starting point.
    \item \textbf{Collaborate.} Find collaborators, who may be researchers, entrepreneurs, established companies, or policy makers. Every domain discussed here has experts who understand its opportunities and pitfalls, even if they do not necessarily understand ML.
    \item \textbf{Listen.} Listen to what your collaborators and other stakeholders say is needed. Groundbreaking technologies have an impact, but so do well-constructed solutions to mundane problems.
    \item \textbf{Deploy.} Ensure that your work is deployed where its impact can be realized.
\end{itemize}
We call upon the machine learning community to use its skills as part of the global effort against climate change.

 \newpage

\section*{Acknowledgments}
\label{sec:acknowledgments}

\paragraph*{Electricity systems.} We thank James Kelloway (National Grid ESO), Jack Kelly (Open Climate Fix), Zico Kolter (CMU), and Henry Richardson (WattTime) for their help and ideas in shaping this section. We also thank Samuel Buteau (Dalhousie University) and Marc Cormier (Dalhousie University) for their inputs on accelerated science and battery storage technologies; Julian Kates-Harbeck (Harvard) and Melrose Roderick (CMU) for their extensive inputs and ideas on nuclear fusion; and Alasdair Bruce (formerly National Grid ESO) for inputs on emissions factor forecasting and automated dispatch. Finally, we thank Lea Boche (EPRI), Carl Elkin (DeepMind), Jim Gao (DeepMind), Muhammad Hasan (DeepMind), Guannan He (CMU), Jeremy Keen (CMU), Zico Kolter (CMU), Luke Lavin (CMU), Sanam Mirzazad (EPRI), David Pfau (DeepMind), Crystal Qian (DeepMind), Juliet Rothenberg (DeepMind), Sims Witherspoon (DeepMind) and Matt Wytock (Gridmatic, Inc.) for helpful comments and feedback.

\paragraph*{Transportation.}
We are grateful for advice from Alan T.~Jenn (UC Davis) and Prithvi S.~Acharya (CMU) on electric vehicles, Alexandre Jacquillat (CMU) on decarbonizing aviation, Michael Whiston (CMU) on hydrogen fuel cells, Evan Sherwin (CMU) on alternative fuels, and Samuel Buteau (Dalhousie University) on batteries.

\paragraph*{Buildings and Cities.} We thank \'Erika Mata (IVL - Swedish Environmental Research Institute, IPCC Lead Author Buildings section), Duccio Piovani (nam.R) and Jack Kelly (Open Climate Fix) for feedback and ideas.

\paragraph*{Industry.}
We appreciate all the constructive feedback from Angela Acocella (MIT), Kevin McCloskey (Google), and Bill Tubbs (University of British Columbia), and we are grateful to Kipp Bradford (Yale) for his recommendations around embodied energy and refrigeration. Thanks to Allie Schwertner (Rockwell Automation), Greg Kochanski (Google), and Paul Weaver (Abstract) for their suggestions around optimizing industrial processes for low-carbon energy.

\paragraph*{Farms \& Forests.}
We would like to give thanks to David Marvin (Salo) and Remi Charpentier (Tesselo) on remote sensing for land use. Max Nova (SilviaTerra) provided insight on forestry, Mark Crowley (University of British Columbia) on forest fire management, Benjamin Deleener (ChrysaLabs) on precision agriculture, and Lindsay Brin (Element AI) on soil chemistry.

\paragraph{Climate prediction.} We thank Ghaleb Abdulla (LLNL), Ben Kravitz (PNNL) and David John Gagne II (UCAR) for enlightening conversations; Goodwin Gibbins (Imperial College London) and Ben Kravitz (PNNL) for detailed editing and feedback; and Claire Monteleoni (CU Boulder) and Prabhat (LBL) for feedback which improved the quality of this manuscript.

\paragraph*{Societal adaptation.}
We thank Loubna Benabbou (UQAR), Mike Sch{\"a}fer (University of Zurich), Andrea Garcia Tapia (Stevens Tech), Slava Jankin Mikhaylov (Hertie School Berlin), and Sarah M.~Fletcher (MIT) for valuable conversations on the social aspects of climate change. 

\paragraph*{Solar geoengineering.}
We thank David Keith (Harvard), Peter Irvine (Harvard), Zhen Dai (Harvard), Colleen Golja (Harvard), Ross Boczar (UC Berkeley), Jon Proctor (UC Berkeley), Ben Kravitz (Indiana University),  Andrew Lockley (University College London), Trude Storelvmo (University of Oslo), and Simon Gruber (University of Oslo) for help and useful feedback.

\paragraph*{Individual action.} We thank Priyanka deSouza (MIT), Olivier Corradi (Tomorrow), Jack Kelly (Open Climate Fix), Ioana Marinescu (UPenn), and Aven Satre-Meloy (Oxford).

\paragraph*{Collective Decisions.}
We thank Sebastian Sewerin (ETH Z\"urich), D.~Cale Reeves (UT Austin), and Rahul Ladhania (UPenn).

\paragraph*{Education.} We appreciated the constructive feedback received by Jacqueline Bourdeau (T\'{E}LUQ University), who gave us valuable insights regarding the field of AIED.

\paragraph*{Finance.} We thank Himanshu Gupta (ClimateAI), and Bjarne Steffen (ETH Z\"urich) for constructive discussions and the valuable feedback.

\paragraph*{}The authors gratefully acknowledge support from National Science Foundation grant 1803547, the Center for Climate and Energy Decision Making through a cooperative agreement between the National Science Foundation and Carnegie Mellon University (SES-00949710), US Department of Energy contract DE-FG02-97ER25308, the Natural Sciences and Engineering Research Council of Canada, and the MIT Media Lab Consortium.

\small{
\bibliographystyle{unsrt}
\bibliography{references}
}

\end{document}